\documentclass[aps,prc,reprint,showpacs,superscriptaddress,eqsecnum]{revtex4-1}

\usepackage{amsmath}
\usepackage{amssymb}
\usepackage{graphicx}
\usepackage{bm}
\usepackage{enumitem}
\usepackage{tensor}
\usepackage{color}
\usepackage{combelow}
\usepackage{slashed}
\usepackage{epstopdf}

\def\feq{\ensuremath{f^{(\mathrm{eq})}}}

\def\omegabar{\overline{\omega}}

\begin{document}


\title{Study of transport coefficients in ultrarelativistic kinetic theory}


\author{Victor E. \surname{Ambru\cb{s}}}
\email[E-mail: ]{victor.ambrus@e-uvt.ro}
\affiliation{Department of Physics, West University of Timi\cb{s}oara,\\
Bd.~Vasile P\^arvan 4, 300223 Timi\cb{s}oara, Romania}


\date{\today}

\begin{abstract}
A spatially-periodic longitudinal wave is considered 
in relativistic dissipative hydrodynamics. 
At sufficiently small wave amplitudes, an analytic solution 
is obtained in the linearised limit of the 
macroscopic conservation equations within the first- and 
second-order relativistic hydrodynamics formulations. 
A kinetic solver is used to obtain the numerical solution of 
the relativistic Boltzmann equation for massless particles 
in the Anderson-Witting approximation for 
the collision term. It is found that, at small values of the 
Anderson-Witting relaxation time $\tau$, 
the transport coefficients emerging from the 
relativistic Boltzmann equation agree with those predicted 
through the Chapman-Enskog procedure, 
while the relaxation times of the heat flux and shear pressure 
are equal to $\tau$.
These claims are further strengthened by considering a 
moment-type approximation based on orthogonal 
polynomials under which the Chapman-Enskog results for the 
transport coefficients are exactly recovered.
\end{abstract}


\maketitle

\section{Introduction} \label{sec:intro}

The relativistic Boltzmann equation is known to reduce to the 
equations of relativistic hydrodynamics in the limit when the mean 
free path of the particle constituents is negligible compared to the typical 
length scales of the system \cite{cercignani02}. 
The transition from kinetic theory to relativistic hydrodynamics is traditionally 
performed following two approaches: the Chapman-Enskog procedure and Grad's 14
moments approximation \cite{cercignani02}. These two approaches 
yield different expressions for the transport coefficients 
appearing in the constitutive equations of the underlying hydrodynamic equations. 
While the non-relativistic limit of these expressions 
coincides between the two formulations, their ultrarelativistic limits differ.

In order to check which of the two approaches (Chapman-Enskog expansion 
or the Grad method) correctly predicts the transport coefficients 
of the hydrodynamic equations, a solution of the relativistic Boltzmann equation 
is required. Solving the Boltzmann equation requires an explicit expression for the 
collision term, which in general is an integral operator taking into account local 
binary collisions. A considerable simplification arises by employing a single 
relaxation time (SRT) approximation. The most common SRT approximations are the 
Marle \cite{marle69} and the Anderson-Witting \cite{anderson74a,anderson74b} models, which 
generalise the widely-used Bhatnagar-Gross-Krook (BGK) model introduced in Ref.~\cite{bhatnagar54}
for the non-relativistic case. Since it is known that the Marle model is not appropriate for 
the study of the flow of massless particles \cite{cercignani02,anderson74a}, 
only the Anderson-Witting model will be considered in this paper.

There has been recent evidence in the literature indicating that the 
transport coefficients predicted by the Chapman-Enskog method are closer to 
those recovered from solutions of the Boltzmann equation than those 
obtained through Grad's 14 moment approximation. 

Florkowski et al.~obtained a solution of the Anderson-Witting-Boltzmann (AWB) equation 
in the case of Bjorken flow \cite{bjorken83} at non-vanishing relaxation time,
written in integral form for the case when the distribution function depends only on proper 
time. This solution was restricted to the massless case in Refs.~\cite{florkowski13a,florkowski13b} and 
extended to the massive case in Ref.~\cite{florkowski14}. In Refs.~\cite{florkowski13a,florkowski13b},
it was shown that the numerical solution of the Israel-Stewart equations \cite{israel79}
leads to better agreement with the AWB solution (computed also numerically) when the Chapman-Enskog 
value for the shear viscosity $\eta$ is used compared to when the 14 moment approximation is used.
The same conclusion is reached in Ref.~\cite{florkowski15} for the case of Bose-Einstein 
and Fermi-Dirac statistics.

The solution of the AWB equation describing the Bjorken flow of massive particles 
obtained in Ref.~\cite{florkowski14} is used in Ref.~\cite{ryblewski15} to highlight 
that the second-order Chapman-Enskog expansion is in closest agreement to the 
solution of the AWB equation as compared to the Israel-Stewart and the 14 moment 
approaches introduced in Refs.~\cite{israel79} and \cite{denicol14}, respectively.

In Ref.~\cite{bhalerao14}, Bhalerao et al demonstrated that a Chapman-Enskog-like approximation
of the non-equilibrium distribution function yielded a closer agreement with the inviscid limit 
of the Bjorken flow than the Grad approximation.

The relativistic lattice Boltzmann (LB) method has also been used as a tool to 
solve the AWB equation \cite{mendoza10prl,romatschke11,hupp11,mohseni13,mendoza13prd,
blaga17prc,blaga17aip,gabbana17a,gabbana17b,coelho17}.
The propagation of planar shock waves of massless
\cite{mendoza10prl,hupp11,mohseni13,mendoza13prd,blaga17prc} and 
massive \cite{gabbana17a,gabbana17b} particles was investigated using the LB method 
and the results were validated by comparison with the data obtained using the 
Boltzmann Approach to Multi Parton Scattering (BAMPS) reported in 
Refs.~\cite{bouras09prl,bouras09nucl,bouras10}.
In order to compare the LB results obtained in the frame of the AWB equation and the 
BAMPS results, the shear viscosity to entropy ratio $\eta / s$ must be kept constant. 
Matching within the LB method 
the values of $\eta / s$ employed in the BAMPS simulations requires 
as an input the exact expression for the shear viscosity $\eta$, which can be 
obtained via Grad's 14 moment approach \cite{mendoza10prl,hupp11,mohseni13,mendoza13prd,gabbana17a} 
or the Chapman-Enskog procedure \cite{blaga17prc,gabbana17b}. 
The authors of Refs.~\cite{blaga17prc,gabbana17b} note that employing the Chapman-Enskog value 
for $\eta$ leads to better agreement with the BAMPS data than when the Grad value is 
employed.

Recently, the relativistic lattice Boltzmann model developed in Ref.~\cite{gabbana17a}
was used in Ref.~\cite{gabbana17b} to study the dissipative attenuation of the relativistic 
equivalent of the Taylor-Green vortices. This study allowed the authors of 
Ref.~\cite{gabbana17b} to demonstrate that the correct value of the shear viscosity 
is that given by the Chapman-Enskog procedure rather than the Grad method for a 
wide range of particle masses. 

In this paper, a study of the transport coefficients and second-order relaxation 
times arising in the Anderson-Witting model is presented, 
by considering the dissipative attenuation of a harmonic longitudinal wave. 
Three regimes of wave propagation will be analysed in this paper, 
corresponding to the cases when
the velocity $\beta$ along the wave propagation direction ({\it Case 1}), 
pressure $P = P_0 + \delta P$ ({\it Case 2a}) or density 
$n = n_0 + \delta n$ ({\it Case 2b}) are perturbed harmonically about $\beta = 0$, $n = n_0$ and 
$P = P_0$. In each case, the other two macrosopic variables are left unperturbed in the initial 
state.
In all cases, at initial time, the fluid is assumed to be in local thermodynamic 
equilibrium characterised by the Maxwell-J\"uttner distribution corresponding to 
the local values of 
$n$, $P$ and $\beta$. The analytic analysis of this system is restricted to 
the regime of small amplitudes $\beta_0$, $\delta n_0$ and $\delta P_0$, where the 
linearised form of the macroscopic equations can be solved exactly.

In the first-order description (i.e.~the five field approximation), the wave amplitude 
predicted by the analytic solution 
consists of a damped, oscillatory term (with respect to time) which allows the wave 
to propagate at approximately the speed of sound, for which the attenuation 
coefficient $\alpha_d$ is 
directly proportional to the shear viscosity $\eta$. The second term is non-oscillatory 
(evanescent) and its attenuation coefficient $\alpha_\lambda$ is directly proportional 
to the heat conductivity $\lambda$. Depending on the initial conditions, this system 
allows $\eta$ and $\lambda$ to be measured separately.
It can be shown that the wave corresponding to {\it Case 1} 
propagates adiabatically (with no heat flux being present), allowing $\eta$ to 
be measured independently.
Furthermore, the heat flux $q$ is purely evanescent (has no oscillatory contribution), 
such that its time
evolution is completely determined by $\alpha_\lambda$, being independent 
of $\alpha_d$ and hence of $\eta$.
{\it Cases 2a} and {\it 2b} will be therefore used 
to measure $\lambda$ independently of $\eta$.

A known fundamental limitation of the first-order theory is that it allows the non-causal instantaneous response
in the heat flux $q$ and shear pressure $\Pi$ induced by changes in the gradients of the fundamental variables $n$, 
$\beta$ and $P$. In particular, this theory does not allow the values of $q$ and $\Pi$ to be set at initial time 
$t = 0$ independently of the values of $n$, $\beta$ and $P$. This is incompatible with the initial local 
thermodynamic equilibrium state considered in this paper, 
in which $q = \Pi = 0$, such that the first-order theory prediction 
for the evolution of $q$ and $\Pi$ is not accurate for a duration of time proportional to 
the Anderson-Witting relaxation time $\tau$. In the second-order hydrodynamics approach,
both $q$ and $\Pi$ obey independent evolution equations which, for sufficiently small values of $\tau$, 
allow them to relax from arbitrary initial configurations to the first-order predictions on time scales 
given by the relaxation times $\tau_q$ and $\tau_\Pi$, respectively. These relaxation 
times will be studied for 
{\it Case 1} and {\it Case 2b}, where it will be demonstrated that $\tau_q \simeq \tau$ and 
$\tau_\Pi \simeq \tau$ at small values of $\tau$. 

In the analysis of {\it Case 2b}, a more subtle limitation of the first-order theory
was encountered. If the initial state 
is prepared as described in {\it Case 2b}, the numerical simulations indicate that
the pressure perturbation $\delta P$ and the shear pressure $\Pi$ 
remain zero throughout the evolution of the wave, for all tested values of the relaxation 
time, provided the initial perturbation $\delta n_0$ is small. Furthermore, 
the decay of the amplitudes of
$\delta n$, $\beta$ and $q$ is strictly exponential, with no oscillations. While the above 
behaviour is successfully recovered in the second-order theory, in the first-order theory, 
$\delta P$ and $\Pi$ are non-zero. Moreover, the amplitude of $\beta$ is oscillatory and the 
oscillation amplitude is of the same order of magnitude as the non-oscillatory $\beta$ predicted by the 
second-order theory.

Next, a moment-based approach similar to the one introduced in 
Refs.~\cite{romatschke11,blaga17prc,blaga17aip,denicol12} is considered, 
where the distribution function is expanded with respect to the Laguerre 
and Legendre polynomials, corresponding 
to the magnitude $p$ of the particle momentum and the $z$ component $\xi = p^z / p$ 
of its velocity, respectively.
Retaining zeroth and first-order terms with respect to the Laguerre polynomials and 
terms up to second-order with respect to the Legendre polynomials, a system of 
$6$ evolution equations is obtained for the density $n$, pressure $P$, 
velocity $\beta$, heat flux 
$q$, shear pressure $\Pi$ and an extra non-hydrodynamic variable. In the frame of this 
moment-based model, the Chapman-Enskog predictions for the 
shear viscosity $\eta$ and heat conductivity $\lambda$ are exactly recovered. 
As highlighted in 
Ref.~\cite{denicol12}, there is a fundamental difference between the above proposed 
moment-based method and Grad's 14 moment approach, due to the fact that the former 
is based on an expansion with respect to orthogonal polynomials, while the latter
relies on an expansion on polynomials in $p^\mu$, which do not constitute 
an orthogonal basis.
An analytic solution obtained within the above model is employed to 
show that at large values of the relaxation time $\tau$, the moment-based 
approach provides a better analytic description of the evolution of the heat 
flux compared to the second-order hydrodynamics result.

Finally, the ballistic regime is analysed, where the particle constituents stream freely. 
Since the flow is now collisionless, no dissipation occurs and the wave attenuation 
is no longer exponential. Instead, the dispersive regime sets in, since now the wave 
can be regarded as a packet which consists of non-interacting constituents which 
propagate in the longitudinal direction at different velocities $\xi$.
An analytic solution for the linearised limit of the ballistic regime is presented,
with the aid of which the capability of the numerical code employed in 
this paper to capture the free-streaming dynamics is demonstrated.

According to Refs.~\cite{blaga17prc,blaga17aip}, high order quadratures (i.e. large 
velocity sets) are required to obtain accurate simulation results at large values of 
$\tau$, when the flow is out of equilibrium and rarefaction effects 
become important. This is performed in a straightforward manner following the 
procedure described in Ref.~\cite{blaga17prc} and summarised in Appendix~\ref{app:num}. 
The numerical experiments presented in this paper were therefore conducted using the 
quadrature-based R-SLB models developed in Ref.~\cite{blaga17prc}.
While the analysis presented herein is 
restricted to the case of massless particles, it can be easily extended to the 
case of massive particles, e.g.~following Refs.~\cite{gabbana17a,gabbana17b,romatschke12}.

The paper is organised as follows. The general framework for the 
study of the propagation of longitudinal waves is introduced in 
Sec.~\ref{sec:fdyn} by linearising the relativistic hydrodynamics
equations with respect to the wave amplitude. 
The relativistic Boltzmann equation in the Anderson-Witting approximation 
for the collision term (the AWB equation) is also briefly presented, alongside a 
description of the Landau frame. In Secs.~\ref{sec:hydro1} and \ref{sec:hydro2}, 
the longitudinal wave problem is considered from the perspective of the first-
and second-order hydrodynamics theories, respectively, while in Sec.~\ref{sec:mom},
the same problem is considered using a moment-based approach. 
In all cases, numerical simulations are employed
to study the validity and applicability of these theories as the relaxation time is 
increased. In Sec.~\ref{sec:bal}, the propagation of the longitudinal wave 
is analysed analytically and numerically 
in the ballistic regime. A short description 
of the numerical method employed in this paper is provided in Appendix~\ref{app:num}.

Throughout this paper, the metric convention $\eta_{\mu\nu} = {\rm diag}(-1,1,1,1)$
is employed.
The non-dimensionalisation convention is presented in Appendix~\ref{app:nondim} and 
summarised in Tab.~\ref{tab:nondim}.

\section{Relativistic fluid dynamics}\label{sec:fdyn}

In this section, the common framework used in later sections for the analysis of 
the evolution of longitudinal waves is presented. 
Subsection~\ref{sec:fdyn:kinetic} introduces a brief review of the 
connection between the relativistic Boltzmann equation and the macroscopic hydrodynamics 
equations, which are written in linearised form in Subsec.~\ref{sec:fdyn:hydro}.
The equations which serve as the basis for the analysis of longitudinal waves are 
presented in Subsec.~\ref{sec:fdyn:long}.

\subsection{Relativistic kinetic theory}\label{sec:fdyn:kinetic}

This paper is focused on the relativistic Boltzmann equation 
for massless particles in the Anderson-Witting approximation for the collision 
term, which reads \cite{anderson74a}:
\begin{equation}
 p^\mu \partial_\mu f = \frac{p \cdot u_L}{\tau} (f - \feq_L),
 \label{eq:boltz}
\end{equation}
where it is assumed for simplicity that the relaxation time $\tau$ is constant.
The equilibrium distribution $\feq$ is taken to be the Maxwell-J\"uttner distribution function:
\begin{equation}
 \feq_L = \frac{n_L}{8\pi T_L^3} \exp\left(\frac{p\cdot u_L}{T_L}\right).
 \label{eq:feq}
\end{equation}
In the above, $n_L$ represents the particle number density, $u_L^\mu$ is the 
macroscopic four-velocity, $T_L$ is the local temperature and $p^\mu$ is the on-shell 
particle four-momentum. The quantities bearing the subscript $L$ are expressed in the Landau (energy)
frame \cite{anderson74a,landau87}.

The transition from the Boltzmann equation \eqref{eq:boltz} to relativistic hydrodynamics 
is done by considering the macrosopic four-flow vector $N^\mu$ and stress-energy tensor (SET) $T^{\mu\nu}$,
which are obtained by integrating the distribution function over the momentum space:
\begin{equation}
 N^\mu = \int \frac{d^3p}{p^0} f\,p^\mu, \qquad 
 T^{\mu\nu} = \int \frac{d^3p}{p^0} f\,p^\mu p^\nu.
 \label{eq:NT_def}
\end{equation}
Substituting $\feq_L$ \eqref{eq:feq} into Eq.~\eqref{eq:NT_def} gives the equilibrium 
four-flow vector $N^\mu_{(\mathrm{eq})}$ and SET $T^{\mu\nu}_{(\mathrm{eq})}$:
\begin{equation}
 N^\mu_{(\mathrm{eq})} = n_L u^\mu_L, \qquad 
 T^{\mu\nu}_{(\mathrm{eq})} = (E_L + P_L) u^\mu_L u^\nu_L + P_L \eta^{\mu\nu}.
\end{equation}
The Landau velocity $u_L^\mu$ is defined as the eigenvector of $T^{\mu\nu}$ corresponding to the 
Landau energy density $E_L$:
\begin{equation}
 T^\mu{}_{\nu} u^\nu_L = -E_L u^\mu_L.\label{eq:landau_def}
\end{equation}
For massless particles, $E_L = 3P_L$ and the Landau pressure $P_L = n_L T_L$ is used to define the 
Landau temperature $T_L$, while the Landau particle number density $n_L$ is obtained by contracting 
$N^\mu$ with $u^\mu_L$:
\begin{equation}
 n_L = -N_\mu u^\mu_L.
\end{equation}

Multiplying the Boltzmann equation \eqref{eq:boltz} by the collision invariants 
$\psi\in\{1,p^\mu\}$ and integrating with respect to the momentum space, the following 
conservation equations are obtained:
\begin{equation}
 \partial_\mu N^\mu = 0, \qquad \partial_\nu T^{\mu\nu} = 0.
 \label{eq:cons}
\end{equation}

Due to its simplicity and pedagogical value, the Eckart (particle)
frame will be employed in this paper, 
where the macroscopic velocity $u^\mu$ is defined as the unit 
vector parallel to $N^\mu$ \cite{eckart40,rezzolla13}:
\begin{equation}
 u^\mu = N^\mu / \sqrt{-N^2}.
\end{equation}
With respect to $u^\mu$, $N^\mu$ and the SET $T^{\mu\nu}$ can be decomposed as:
\begin{gather}
 N^\mu = nu^\mu, \nonumber\\
 T^{\mu\nu} = E u^\mu u^\nu + (P + \omegabar) \Delta^{\mu\nu} + 
 u^\mu q^\nu + q^\nu u^\mu + \Pi^{\mu\nu},
 \label{eq:macro}
\end{gather}
where $\Delta^{\mu\nu} = \eta^{\mu\nu} + u^\mu u^\nu$ is the projector on the hypersurface 
orthogonal to $u^\mu$.
The particle number density $n$, energy density $E$, isotropic pressure $P + \omegabar$, 
heat flux $q^\mu$ and shear stress tensor $\Pi^{\mu\nu}$ can be obtained as follows 
\cite{bouras10,rezzolla13}:
\begin{gather}
 n = -u_\mu N^\mu, \qquad E = u_\mu u_\nu T^{\mu\nu}, \qquad 
 P = \frac{1}{3} \Delta_{\mu\nu} T^{\mu\nu},\nonumber\\
 \Pi^{\mu\nu} = \left(\Delta^\mu{}_\lambda \Delta^\nu{}_\kappa - 
 \frac{1}{3} \Delta^{\mu\nu} \Delta_{\lambda\kappa} \right) T^{\lambda\kappa},\nonumber\\
 q^\mu = - \Delta^{\mu}{}_\nu u_\lambda T^{\nu\lambda},
\end{gather}
while the dynamic pressure $\omegabar = 0$ for massless particles, when $E = 3P$.
In general, the Eckart quantities introduced above are different from the corresponding 
quantities defined in the Landau frame (more details will be given in Sec.~\ref{sec:fdyn:hydro}).

The system \eqref{eq:cons} consisting of $5$ equations is not closed, 
since $q^\mu$ and $\Pi^{\mu\nu}$ are {\it a priori} unconstrained.
The constitutive relations which close this system 
corresponding to the first- and second-order relativistic 
hydrodynamics frameworks will be discussed in Secs.~\ref{sec:hydro1} and \ref{sec:hydro2}.

\subsection{Linearised relativistic hydrodynamics} \label{sec:fdyn:hydro}

Let us now consider a system which is homogeneous along the $x$ and $y$ directions. 
In this case, 
the AWB equation \eqref{eq:boltz} reduces to \cite{blaga17prc}:
\begin{gather}
 \partial_t f + \xi \partial_z f = -\frac{\gamma_L(1 - \beta_L \xi)}{\tau} 
 (f - \feq_L), \nonumber\\
 \feq_L = \frac{n_L}{8\pi T_L^3} \exp\left[-\frac{p \gamma_L}{T_L}(1 - \beta_L \xi)\right],
 \label{eq:boltz_z}
\end{gather}
where $\xi = p^z / p$ represents the particle velocity along the $z$ axis, 
taking values in $[-1, 1]$.
Taking into account the constraints $u_\mu q^\mu = 0$ and $u_\mu \Pi^{\mu\nu} = 0$, 
the variables $u^\mu$, $q^\mu$ and $\Pi^{\mu\nu}$ can be taken as follows 
\cite{blaga17prc,bouras10}:
\begin{gather}
 u^\mu \partial_\mu = \gamma(\partial_t + \beta \partial_z), \qquad 
 q^\mu \partial_\mu = q(\beta \partial_t + \partial_z), \nonumber\\
 \Pi^{\mu\nu} = \Pi
 \begin{pmatrix}
  \beta^2 \gamma^2  & 0 & 0 & \beta \gamma^2 \\
  0 & -\frac{1}{2} & 0 & 0\\
  0 & 0 & -\frac{1}{2} & 0\\
  \beta \gamma^2 & 0 & 0 & \gamma^2
 \end{pmatrix},\label{eq:macro1D}
\end{gather}
where $\gamma = (1 - \beta^2)^{-1/2}$ is the Lorentz factor corresponding to the velocity $\beta$.
The Landau frame can be constructed analytically by solving the eigenvalue equation 
\eqref{eq:landau_def} \cite{blaga17prc,bouras10}:
\begin{align}
 E_L =& \frac{1}{2}\left[T^{00} - T^{zz} + \sqrt{(T^{00} + T^{zz})^2 - 4(T^{0z})^2}\right],\nonumber\\
 \beta_L =& \frac{T^{0z}}{E_L + T^{zz}}.
 \label{eq:landau}
\end{align}

In order to arrive at the linearised form of Eqs.~\eqref{eq:cons}, 
$n$ and $P$ can be written as:
\begin{equation}
 n = n_0 + \delta n, \qquad P = P_0 + \delta P,
\end{equation}
where $\delta n / n_0$ and $\delta P / P_0 $ are quantities of order $O(\beta)$
and the limit $\beta \ll 1$ was considered.
Furthermore, $\Pi$ and $q$ are also of order $O(\beta)$,
since they represent non-equilibrium quantities.
Neglecting the terms of order $\beta^2$, $N^\mu$ and $T^{\mu\nu}$ 
\eqref{eq:macro} reduce to:
\begin{widetext}
\begin{equation}
 N^\mu \simeq (n_0 + \delta n, 0, 0, n_0 \beta)^T, \qquad
 T^{\mu\nu} \simeq 
 \begin{pmatrix}
  3(P_0 + \delta P) & 0 & 0 & 4\beta P_0 + q\\
  0 & P_0 + \delta P - \frac{\Pi}{2} & 0 & 0\\
  0 & 0 & P_0 + \delta P - \frac{\Pi}{2} & 0\\ 
  4\beta P_0 + q & 0 & 0 & P_0 + \delta P + \Pi
 \end{pmatrix},
\end{equation} 
\end{widetext}
while the Landau quantities $n_L$, $P_L$ and $\beta_L$ can be approximated through:
\begin{align}
 n_L \simeq& n_0 + \delta n,\nonumber\\
 P_L \simeq& P_0 + \delta P, \nonumber\\
 \beta_L \simeq& \beta + \frac{q}{4P_0}.
\end{align}

In the linearised approximation, the conservation equations \eqref{eq:cons} reduce to:
\begin{gather}
 \partial_t \delta n + n_0 \partial_z \beta = 0,\nonumber\\
 3\partial_t \delta P + 4P_0 \partial_z \beta + \partial_z q = 0,\nonumber\\
 4P_0 \partial_t \beta + \partial_t q + \partial_z \delta P + \partial_z \Pi = 0.
 \label{eq:cons_lin}
\end{gather}
Noting that $f - \feq_L$ is also of order $O(\beta)$, the Boltzmann equation \eqref{eq:boltz_z} 
can also be expressed in linearised form:
\begin{equation}
 \partial_t f + \xi \partial_z f \simeq -\frac{1}{\tau}(f - \feq_L),\label{eq:boltz_lin}
\end{equation}
where $\feq_L$ can be linearised as follows:
\begin{multline}
 \feq_L \simeq \frac{n_0}{8\pi T_0^3} e^{-p/T_0} 
 \left[1 + \frac{p \xi}{T_0} \left(\beta + \frac{q}{4P_0}\right) \right.\\
 \left.+ \frac{4 \delta n}{n_0} - \frac{3\delta P}{P_0} + \frac{p}{T_0} 
 \left(\frac{\delta P}{P_0} - \frac{\delta n}{n_0}\right)\right].
 \label{eq:feq_lin}
\end{multline}

\subsection{Longitudinal waves}\label{sec:fdyn:long}

Next, solutions of the following form are sought:
\begin{equation}
 \begin{pmatrix}
  \beta \\ q
 \end{pmatrix} = 
 \begin{pmatrix}
  \widetilde{\beta} \\ \widetilde{q}
 \end{pmatrix}
 \sin k z, \qquad 
 \begin{pmatrix}
  \delta n \\ \delta P \\ \Pi
 \end{pmatrix} = 
 \begin{pmatrix}
  \widetilde{\delta n} \\ \widetilde{\delta P_\alpha} \\ \widetilde{\Pi_\alpha}
 \end{pmatrix}  \cos k z,\label{eq:ansatz}
\end{equation}
where $k = 2\pi / L$ is the wave number and $L$ is the wavelength. The quantities with a tilde 
$\widetilde{M} \in \{\widetilde{\beta}, \widetilde{\delta n}, \widetilde{\delta P}, \widetilde{q}, 
\widetilde{\Pi}\}$ depend only on time $t$. Taking this dependence in the form:
\begin{equation}
 \widetilde{M} = \sum_{\alpha} M_{\alpha} e^{-\alpha t},\label{eq:alpha_def}
\end{equation}
Eq.~\eqref{eq:cons_lin} can be solved for each (constant) value of $\alpha$ independently, 
yielding a spectrum of linearly-independent modes satisfying:
\begin{subequations}\label{eq:cons_lin_alpha}
\begin{gather}
 \alpha \delta n_\alpha - k n_0 \beta_\alpha = 0,\label{eq:cons_lin_alpha:n}\\
 3\alpha\delta P_\alpha - 4k P_0 \beta_\alpha - k q_\alpha = 0, \label{eq:cons_lin_alpha:P}\\
 4\alpha P_0 \beta_\alpha + \alpha q_\alpha + k \delta P_\alpha + k \Pi_\alpha = 0.\label{eq:cons_lin_alpha:beta}
\end{gather}
\end{subequations}
The imaginary part of $\alpha$ represents the propagation angular frequency, 
while its real part causes the dissipative dampening of the wave. 
In order to solve the above set of equations, 
the constitutive equations for $q$ and $\Pi$ must be supplied separately.

The initial conditions for Eqs.~\eqref{eq:cons_lin_alpha} are given in the 
form 
\begin{align}
 \widetilde{\beta}(t = 0) =& \beta_0, \nonumber\\
 \widetilde{\delta n}(t = 0) =& \delta n_0, \nonumber\\
 \widetilde{\delta P}(t = 0) =& \delta P_0. \label{eq:init}
\end{align}
In this paper, the 
following sets of values for $\beta_0$, $\delta n_0$ and $\delta P_0$
will be considered:
\begin{itemize}
 \item {\it Case 1:} $\delta n_0 = \delta P_0 = 0$, $\beta_0 \neq 0$;
 \item {\it Case 2a:} $\delta n_0 = \beta_0 = 0$, $\delta P_0 \neq 0$;
 \item {\it Case 2b:} $\beta_0 = \delta P_0 = 0$, $\delta n_0  \neq 0$. 
\end{itemize}

In Secs.~\ref{sec:hydro1} and \ref{sec:hydro2}, the consitutive 
relations corresponding to the first- and second-order relativistic hydrodynamics
will be employed.
In Sec.~\ref{sec:mom}, a solution of Eqs.~\eqref{eq:cons_lin_alpha}
will be constructed starting from the Boltzmann equation \eqref{eq:boltz_lin} 
written in linearised form.

\section{First-order hydrodynamics}\label{sec:hydro1}

The equations of first-order relativistic hydrodynamics represent the analogue 
of the Navier-Stokes-Fourier equations of non-relativistic hydrodynamics.
In this formulation, the fields $n$, $u^\mu$ and $P$ are considered as fundamental 
variables. Since $u^\mu$ is normalised according to $u^2 = -1$, the 
theory contains five independent fields and is sometimes referred to as the 
five field theory \cite{cercignani02}. In this first-order framework, the constitutive equations for 
the heat flux $q^\mu$ and shear stress tensor $\Pi^{\mu\nu}$ represent algebraic 
relations linking them to the gradients of the fundamental fields via the transport 
coefficients $\lambda$ (heat conductivity) and $\eta$ (shear viscosity), respectively.
Since $q^\mu$ and $\Pi^{\mu\nu}$ respond instantaneously to changes in 
the fundamental fields, the ensuing system of equations is not hyperbolic \cite{rezzolla13},
rendering the theory non-causal. This issue can be remedied within the 
second-order relativistic hydrodynamics framework, as will be discussed in 
Sec.~\ref{sec:hydro2}. This section is focused on determining $\lambda$ and 
$\eta$ by comparing the analytical and numerical results for the 
attenuation process occuring in the longitudinal wave problem described in 
Sec.~\ref{sec:fdyn:long}.

\subsection{Constitutive relations}\label{sec:hydro1:const}

The constitutive equations for 
$q^\mu$ and $\Pi^{\mu\nu}$ can be written 
in the frame of the first-order relativistic hydrodynamics
as \cite{cercignani02,rezzolla13}:
\begin{align}
 q^\mu =& -\lambda \Delta^{\mu\nu}\left(\partial_\nu T - \frac{T}{E + P} \partial_\nu P\right),\nonumber\\
 \Pi^{\mu\nu} =& -2\eta \left[\frac{1}{2} \left(\Delta^{\mu\lambda} \Delta^{\nu\kappa} + \Delta^{\nu\lambda} 
 \Delta^{\mu\kappa}\right) - \frac{1}{3} \Delta^{\mu\nu} \Delta^{\lambda\kappa}\right]\nonumber\\ &\times \partial_\lambda u_\kappa,
 \label{eq:tcoeff}
\end{align}
where $\lambda$ and $\eta$ represent the coefficients of heat conductivity and shear 
viscosity $\eta$, respectively. At the level of the first-order hydrodynamics theory, 
it is not specified whether the macroscopic velocity $u^\mu$ appearing in the right hand side 
of the second line of Eq.~\eqref{eq:tcoeff} is defined in the Eckart or in the Landau frame. 
In this section, the Landau frame velocity will be considered, since this choice seems 
natural when the Anderson-Witting approximation is used for the collision term  
\cite{florkowski15,ryblewski15,denicol14,bouras10,denicol12,jaiswal13a,jaiswal13b,chattopadhyay15,jaiswal15}.
In Sec.~\ref{sec:mom}, it will be shown that this choice arises naturally when a moment-based
approach is used to solve the AWB equation \eqref{eq:boltz_lin}.

The connection between the Boltzmann equation \eqref{eq:boltz} and the constitutive 
equations given in Eq.~\eqref{eq:tcoeff} is commonly achieved via two paths: 
(a) the Chapman-Enskog expansion; and (b) Grad's 14 moments approximation. 
In the ultrarelativistic regime considered in this paper, the transport coefficients 
$\eta$ and $\lambda$ are given by:
\begin{equation}
 \eta = \eta_0 P \tau, \qquad \lambda = \lambda_0 n \tau,\label{eq:tcoeff_red}
\end{equation}
where the dimensionless constants $\eta_0$ and $\lambda_0$ are obtained 
using Grad's approximation and the Chapman-Enskog procedure as follows \cite{cercignani02}:
\begin{subequations}
\begin{align}
&\text{Grad method:} & \eta_{0,G} =& \frac{2}{3}, & 
\lambda_{0,G} =& \frac{4}{5},\label{eq:tcoeff_G}\\
&\text{Chapman-Enskog}:& \eta_{0,C-E} =& \frac{4}{5}, & 
\lambda_{0,C-E} =& \frac{4}{3}.\label{eq:tcoeff_CE}
\end{align}
\end{subequations}
The validity of the constitutive equations \eqref{eq:tcoeff} and of the above expressions 
for the transport coefficients is limited to the hydrodynamic regime, i.e. when the 
relaxation time $\tau$ is sufficiently small.

In the linearised approximation introduced in Sec.~\ref{sec:fdyn:hydro}, the constitutive equations 
\eqref{eq:tcoeff} reduce to:
\begin{align}
 q =& -\frac{\lambda P_0}{4n_0} \left(\frac{3 \partial_z \delta P}{P_0} - 
 \frac{4 \partial_z \delta n}{n_0} \right),\nonumber\\
 \Pi =& -\frac{4\eta}{3} \partial_z \left(\beta + \frac{q}{4P_0}\right), \label{eq:tcoeff_lin}
\end{align}
where $u_L^\mu \simeq (1, 0, 0, \beta + q / 4P_0)$ was used in the expression for $\Pi$.

\subsection{Longitudinal waves solution}\label{sec:hydro1:long}

The modes $q_\alpha$ and $\Pi_\alpha$ appearing in Eq.~\eqref{eq:cons_lin_alpha} can be found from 
Eq.~\eqref{eq:tcoeff_lin}:
\begin{align}
 q_\alpha =& \frac{k \lambda P_0}{4 n_0} \left(3\frac{\delta P_\alpha}{P_0} - 
 4\frac{\delta n_\alpha}{n_0}\right), \nonumber\\
 \Pi_\alpha =& -\frac{4 k \eta}{3} \left(\beta_\alpha + 
 \frac{q_\alpha}{4 P_0}\right).\label{eq:ansatz2}
\end{align}
Noting from Eq.~\eqref{eq:cons_lin_alpha:n} that 
\begin{equation}
 \frac{\delta n_\alpha}{n_0} = \frac{k}{\alpha} \beta_\alpha,
\end{equation}
Eq.~\eqref{eq:cons_lin_alpha:P} reduces to:
\begin{equation}
 P_0 \left(3\frac{\delta P_\alpha}{P_0} - 4\frac{\delta n_\alpha}{n_0}\right) 
 \left(\frac{\lambda k^2}{4n_0} - \alpha\right) = 0.
\end{equation}
According to Eq.~\eqref{eq:ansatz2}, the first parenthesis vanishes only when $q_\alpha = 0$.
Thus, the solution 
\begin{equation}
 \alpha_\lambda = \frac{k^2 \lambda}{4 n_0}\label{eq:alphal}
\end{equation}
corresponds to the only mode which dissipates heat.
In this case, Eq.~\eqref{eq:cons_lin_alpha} can be used to obtain:
\begin{align}
 \delta n_\lambda =& \frac{k n_0}{\alpha_\lambda} \beta_\lambda, &
 \delta P_\lambda =& 0, \nonumber\\
 q_\lambda =& -4P_0 \beta_\lambda, & \Pi_\lambda =& 0.\label{eq:delta_aux1}
\end{align}
It is remarkable that this mode induces no viscous dissipation.

Considering now that $q = 0$, Eq.~\eqref{eq:cons_lin_alpha:beta} reduces to:
\begin{equation}
 4P_0 \beta_\alpha \left(\alpha + \frac{k^2}{3\alpha} - \frac{k^2 \eta}{3 P_0} \right) = 0.
\end{equation}
The solution $\beta_\alpha = 0$ is trivial since in this case $\delta n_\alpha = \delta P_\alpha = 0$.
Setting the quantity inside the parenthesis equal to zero yields the following allowed values for 
$\alpha$:
\begin{equation}
 \alpha_{\pm} = \alpha_{d} \pm i \alpha_{o},\label{eq:alphapm}
\end{equation}
where the dampening ($\alpha_{d}$) and oscillatory ($\alpha_{o}$) parts of $\alpha_\pm$ read:
\begin{equation}
 \alpha_{d} = \frac{k^2 \eta}{6 P_0}, \qquad 
 \alpha_{o} = \frac{k}{\sqrt{3}} \sqrt{1 - \frac{3 \alpha_d^2}{k^2}},
 \label{eq:alphad_alphao}
\end{equation}
It is worth noting that the phase velocity $\alpha_o / k = c_s \sqrt{1 - \frac{3 \alpha_d^2}{k^2}}$
predicted in the first-order theory
is smaller than the sound speed $c_s = 1/\sqrt{3}$.
The amplitudes of the density and pressure perturbations $\delta n_\pm$ and $\delta P_\pm$ 
are given in terms of the velocity amplitudes $\beta_\pm$ as follows:
\begin{gather}
 \delta n_{\pm} = \frac{k n_0}{\alpha_{\pm}} \beta_{\pm},\qquad
 \delta P_{\pm} = \frac{4k P_0}{3\alpha_{\pm}} \beta_{\pm}.\label{eq:delta_aux2}
\end{gather}

Taking into account the above allowed values for $\alpha$, the general 
solution \eqref{eq:alpha_def} reads:
\begin{widetext}
\begin{equation}
\begin{pmatrix}
 \widetilde{\beta} \\
 \widetilde{\delta n} \\
 \widetilde{\delta P} \\
 \widetilde{q} \\
 \widetilde{\Pi} 
\end{pmatrix} =
\begin{pmatrix}
 \beta_\lambda \\
 \delta n_\lambda \\
 0 \\
 q_\lambda \\
 0
\end{pmatrix}
e^{-\alpha_\lambda t} 
+ \left[
\begin{pmatrix}
 \beta_{c} \\
 \delta n_{c} \\
 \delta P_{c} \\
 0 \\
 \Pi_c
\end{pmatrix}
\cos \alpha_{o} t +
\begin{pmatrix}
 \beta_{s} \\
 \delta n_{s} \\
 \delta P_{s} \\
 0 \\
 \Pi_s
\end{pmatrix}
\sin \alpha_{o} t\right] e^{-\alpha_d t}.
\label{eq:h1_sol}
\end{equation}
In the above, $\beta_\lambda$, $\beta_c = \beta_+ + \beta_-$ and $\beta_s = -i(\beta_+ - \beta_-)$ are
independent integration constants with respect to which the following definitions were made:
\begin{equation}
 \begin{pmatrix}
  \delta n_c\\
  \delta P_c
 \end{pmatrix} =
 \begin{pmatrix} 
  k n_0  \\ 
  4k P_0 / 3
 \end{pmatrix}
 \frac{\alpha_d \beta_c + \alpha_o \beta_s}{\alpha_d^2 + \alpha_o^2}, \qquad
 \begin{pmatrix}
  \delta n_s\\
  \delta P_s
 \end{pmatrix} =
 \begin{pmatrix} 
  kn_0 \\ 
  4kP_0 / 3
 \end{pmatrix}
 \frac{\alpha_d \beta_s - \alpha_o \beta_c}{\alpha_d^2 + \alpha_o^2}, \qquad
 \begin{pmatrix}
  \Pi_c \\
  \Pi_s
 \end{pmatrix}
 = -\frac{8 \alpha_d P_0}{k} 
 \begin{pmatrix}
  \beta_c \\
  \beta_s
 \end{pmatrix},
 \label{eq:delta_aux3}
\end{equation}
\end{widetext}
while $q_c = q_s = 0$. The other constants 
$\delta n_\lambda$, $\delta P_\lambda$, $\Pi_\lambda$ and $q_\lambda$ were already defined in 
Eq.~\eqref{eq:delta_aux1}.

The constants $\beta_\lambda$, $\beta_c$ and $\beta_s$ can be obtained by 
substituting the solution \eqref{eq:h1_sol} into the initial conditions 
\eqref{eq:init} yielding:
\begin{gather}
 \beta_\lambda + \beta_c = \beta_0, \qquad 
 \delta n_\lambda + \delta n_c = \delta n_0, \nonumber\\
 \delta P_c = \delta P_0.\label{eq:h1_init}
\end{gather}
The solution of Eq.~\eqref{eq:h1_init} can be written as:
\begin{align}
 \beta_\lambda =& \frac{\alpha_\lambda}{4k} \left(\frac{4 \delta n_0}{n_0} - \frac{3\delta P_0}{P_0}\right),\nonumber\\
 \beta_c =& \beta_0 - \frac{\alpha_\lambda}{4k} \left(\frac{4 \delta n_0}{n_0} - \frac{3\delta P_0}{P_0}\right),\nonumber\\
 \beta_s =& -\frac{\alpha_d}{\alpha_o} \beta_0 + 
 \frac{\alpha_\lambda \alpha_d}{k \alpha_o} \frac{\delta n_0}{n_0} \nonumber\\
 &+ \frac{3 \delta P_0}{4k P_0 \alpha_o} (\alpha_d^2 + \alpha_o^2 - \alpha_\lambda \alpha_d).
 \label{eq:init_sol}
\end{align}

The analytic solution presented in this section facilitates the study of the transport 
coefficients corresponding to a relativistic gas. Using the numerical method described in 
Appendix~\ref{app:num}, this system will be considered in the following subsections 
for the study of the ultrarelativistic 
limits of the shear viscosity $\eta$ and heat conductivity $\lambda$ arising from the 
AWB equation \eqref{eq:boltz}.

\subsection{Case 1: Adiabatic flow} \label{sec:hydro1:case1}

\begin{figure*}
\begin{center}
\begin{tabular}{cc}
\includegraphics[angle=270,width=0.48\linewidth]{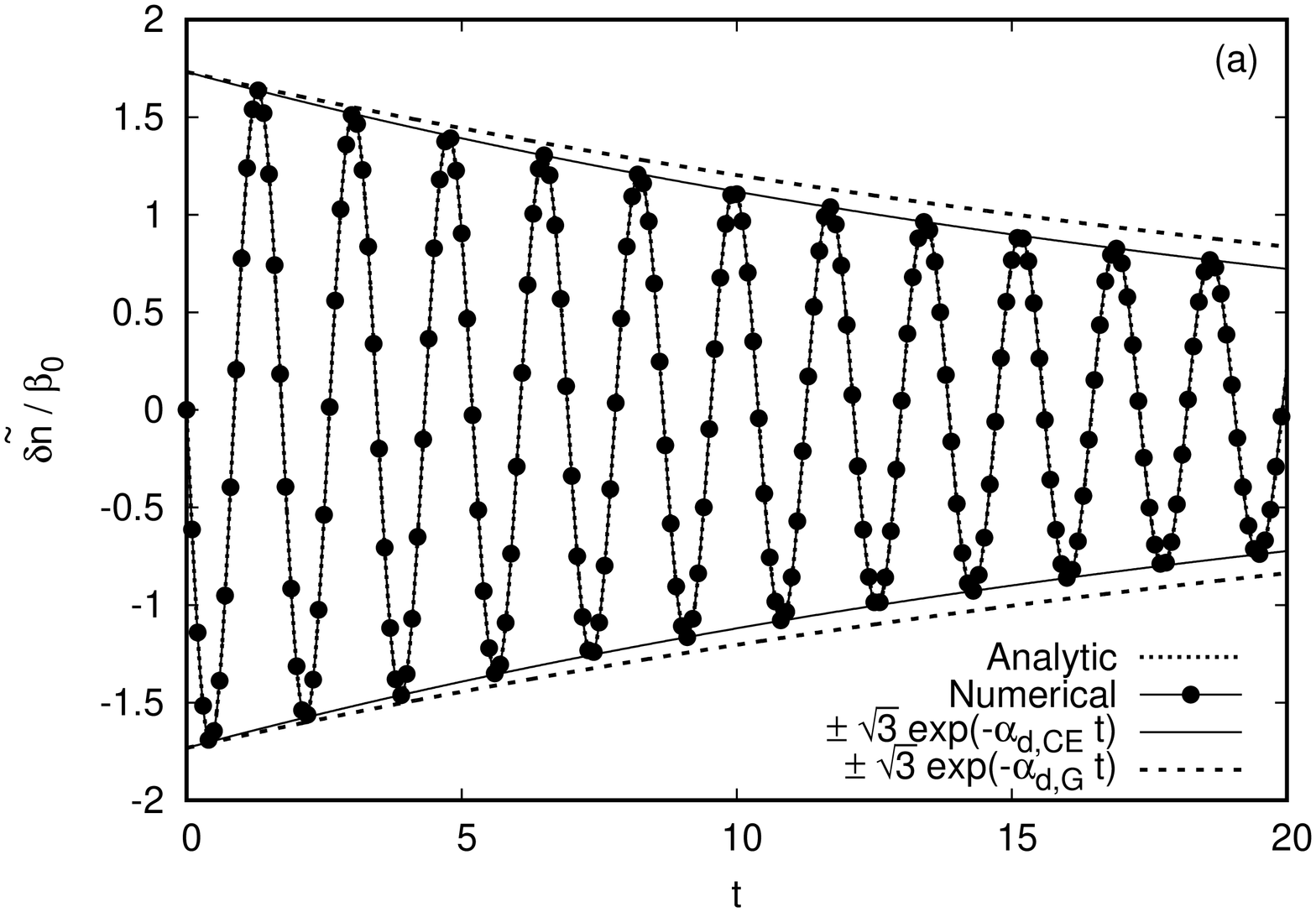} &
\includegraphics[angle=270,width=0.48\linewidth]{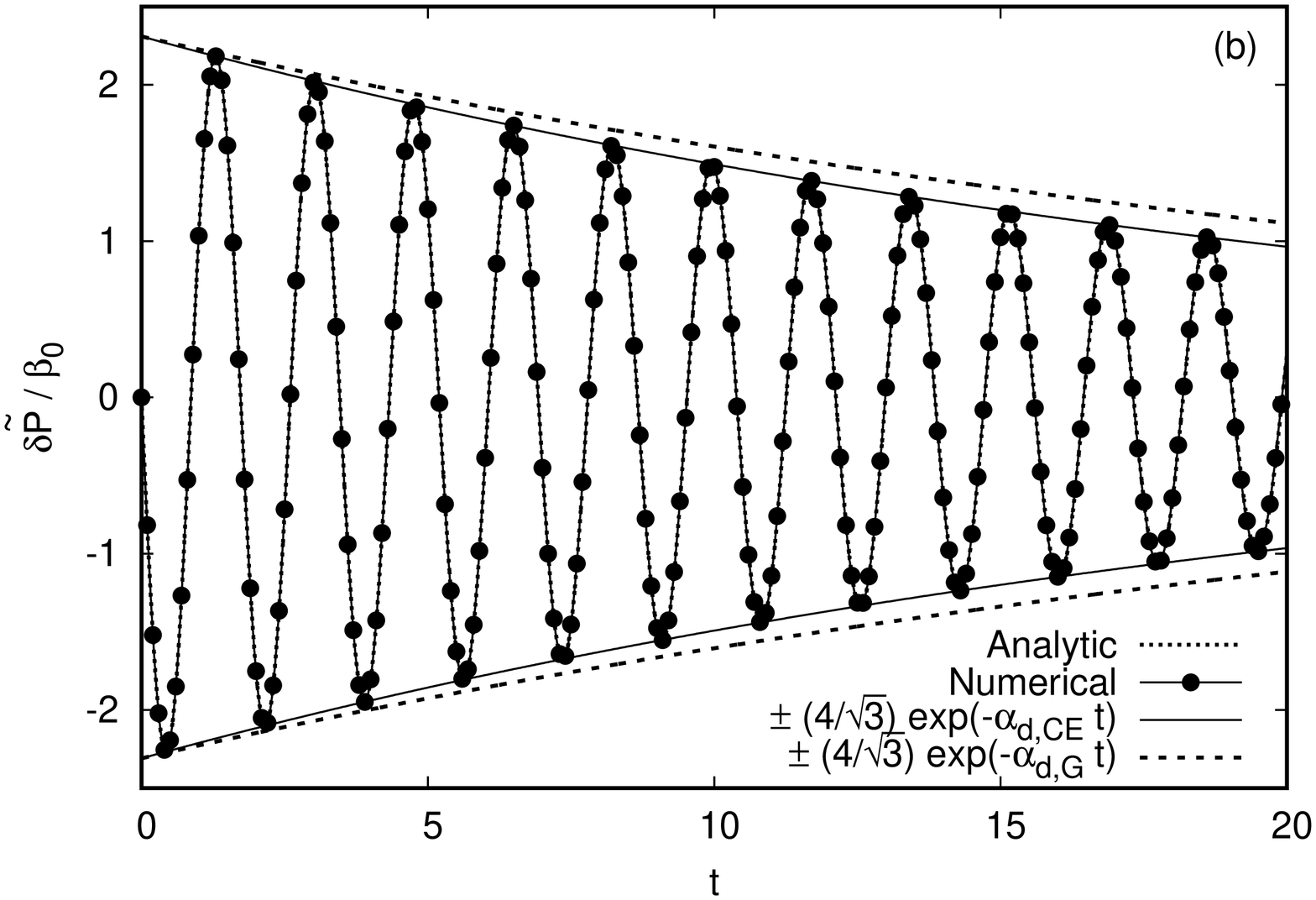} \\
\includegraphics[angle=270,width=0.48\linewidth]{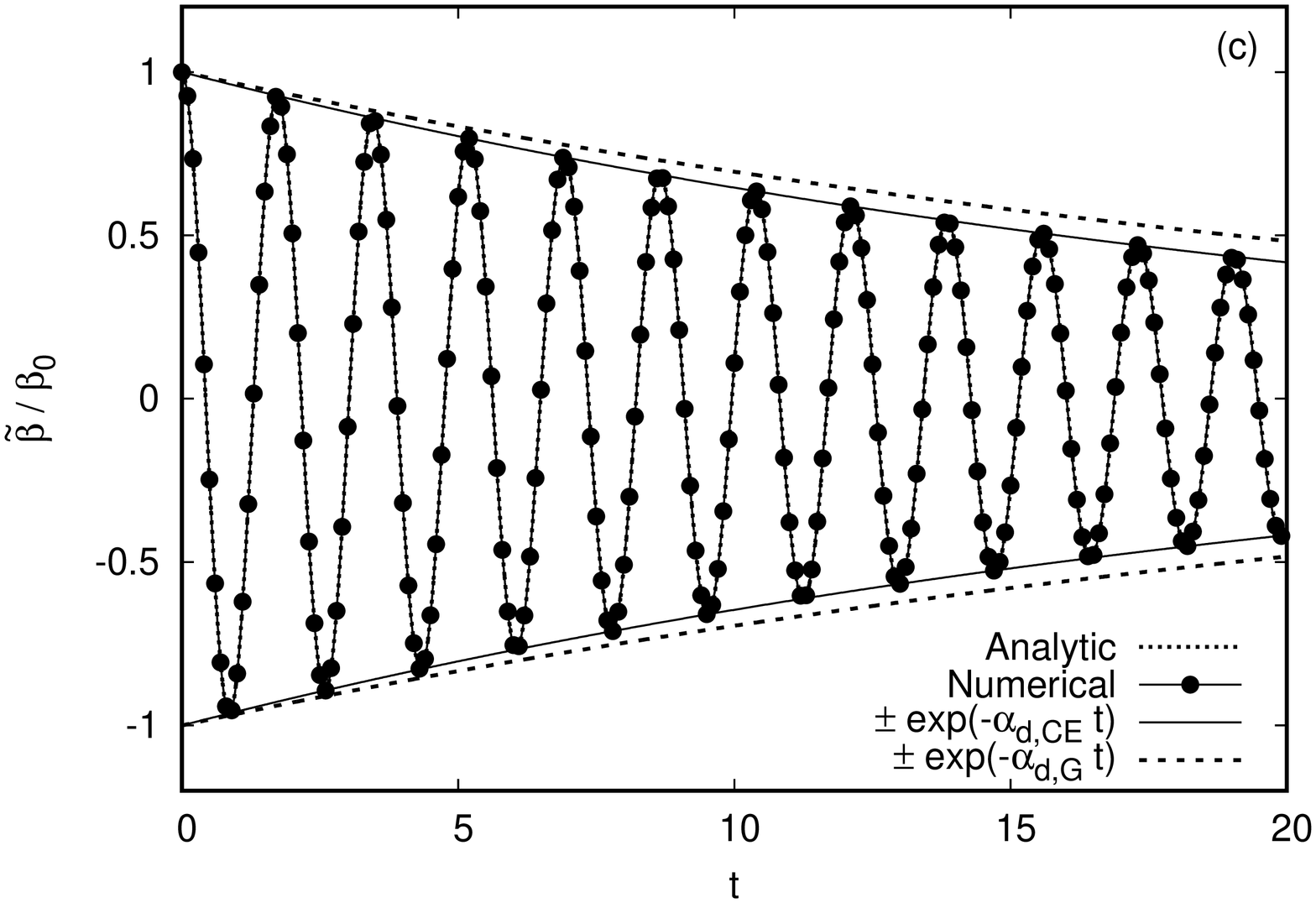} &
\includegraphics[angle=270,width=0.48\linewidth]{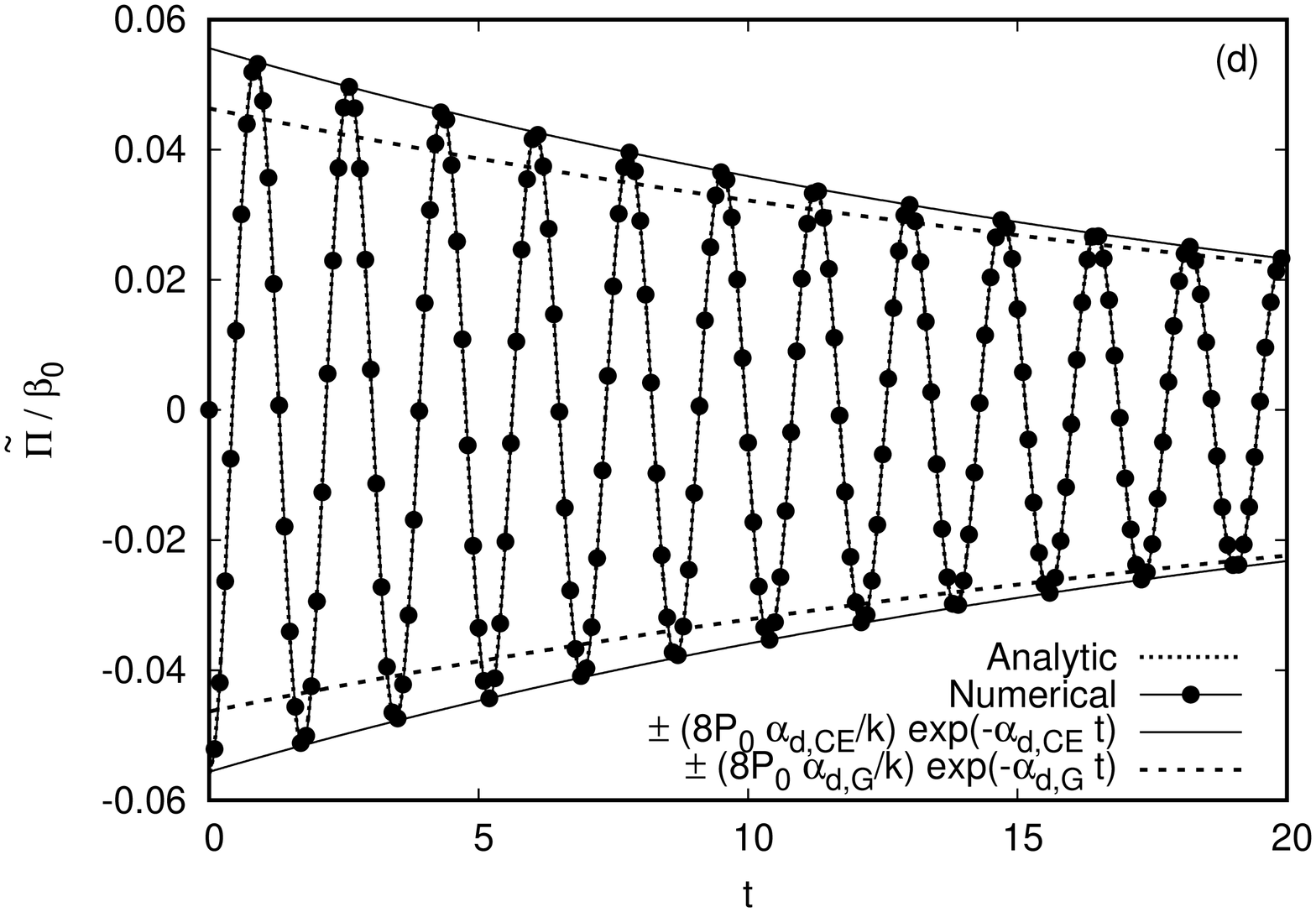}
\end{tabular}
\end{center}
\caption{
Comparison between the analytic solutions in Eq.~\eqref{eq:case1} 
corresponding to the Chapman-Enskog value for $\eta$ (continuous lines) and 
numerical results (dotted lines and points) for $\tau = 8.3 \times 10^{-3}$.
The vertical axes represent the values of (a) $\widetilde{\delta n}$, (b) $\widetilde{\delta P}$,
(c) $\widetilde{\beta}$ and (d) $\widetilde{\Pi}$, divided by $\beta_0$. 
The coefficients $\alpha_{d,{\rm CE}}$ and $\alpha_{d,{\rm G}}$ in the asymptotic dampening lines 
are obtained from Eq.~\eqref{eq:alphad_alphao} by substituting the Chapman-Enskog 
\eqref{eq:tcoeff_CE} and Grad \eqref{eq:tcoeff_G} expressions for $\eta$.
The system was initialised according to {\em Case 1},
i.e. $\delta n_0 = \delta P_0 = 0$ and $\beta_0 = 10^{-3}$.
}
\label{fig:case1-profiles}
\end{figure*}

First, an adiabatic flow is considered (i.e. $q = 0$) 
such that the shear viscosity $\eta$ can be isolated from the 
heat conductivity $\lambda$. This can be achieved when $3 n_0 \delta P_0 = 4 P_0 \delta n_0$.
This condition is equivalent to the requirement that the fugacity 
$\lambda_{\rm fug} = n^4 / P^3$ is constant in the initial state. Indeed, 
combining the first two relations in Eq.~\eqref{eq:cons_lin} gives:
\begin{equation}
 \partial_t \delta \lambda_{\rm fug} + \partial_z (q / P_0) = 0.
\end{equation}
The above equation (valid in the linearised regime) shows that if there is 
no heat flux present, the fugacity remains constant in time.

Furthermore, Eq.~\eqref{eq:init_sol} indicates that $\beta_\lambda = 0$ when 
$3 n_0 \delta P_0 = 4 P_0 \delta n_0$, while Eq.~\eqref{eq:delta_aux1}
implies that $\delta n_\lambda$, $\delta P_\lambda$, $q_\lambda$ and $\Pi_\lambda$ cancel.
Thus, the evolution of the fluid is completely independent of $\alpha_\lambda$,
enabling $\eta$ to be determined independently. For simplicity, the 
initialisation corresponding to {\it Case 1} in Sec.~\ref{sec:fdyn:long} will 
be considered (i.e., $\delta n_0 = \delta P_0 = 0$). 
In this limit, Eq.~\eqref{eq:init_sol} reduces to: 
\begin{equation}
 \beta_c = \beta_0, \qquad \beta_s = -\frac{\alpha_d}{\alpha_o} \beta_0, \qquad \beta_\lambda = 0,
\end{equation}
such that the exact solution \eqref{eq:h1_sol} reads:
\begin{gather}
 \begin{pmatrix}
  \widetilde{\beta}\\
  \widetilde{\Pi} 
 \end{pmatrix} = \beta_0 
 \begin{pmatrix}
  1 \\
  -8 \alpha_d P_0 / k
 \end{pmatrix}
 \left(\cos\alpha_o t - \frac{\alpha_d}{\alpha_o} \sin\alpha_o t\right) e^{-\alpha_d t},\nonumber\\
 \widetilde{\delta n} = -\frac{k n_0 \beta_0}{\alpha_o} e^{-\alpha_d t} \sin\alpha_o t, \nonumber\\
 \widetilde{\delta P} = -\frac{4k P_0 \beta_0}{3 \alpha_o} e^{-\alpha_d t} \sin\alpha_o t,
 \label{eq:case1}
\end{gather}
where $\alpha_d$ and $\alpha_o$ are given in terms of $\eta$ in Eq.~\eqref{eq:alphad_alphao}.

The analytic results in Eq.~\eqref{eq:case1} are represented in 
Fig.~\ref{fig:case1-profiles} for the 
initial conditions $\beta_0 = 10^{-3}$ and $\delta n_0 = \delta P_0 = 0$
alongside the corresponding numerical results obtained using the method described 
in Appendix~\ref{app:num}. The relaxation time was taken to be $\tau = 0.0083$, such that 
both the dampening and the oscillatory characteristics of the solutions can be highlighted 
on the same timescale. The first entry in the legend (fine dotted lines) 
corresponds to the analytic expressions in Eqs.~\eqref{eq:case1}, where $\alpha_d$
and $\alpha_o$ are computed using the Chapman-Enkog value for $\eta$. The numerical 
results are indistinguishable from the analytic predictions. 

Also in the plots in Fig.~\ref{fig:case1-profiles}, the dampening caused by the 
$\exp(-\alpha_d t)$ factor in Eqs.~\eqref{eq:case1} is represented when $\alpha_d$ 
is calculated using the Chapman-Enskog and Grad expressions for $\eta$. In the amplitude of 
the dampening terms in $\widetilde{\beta} / \beta_0$ and $\widetilde{\Pi} / \beta_0$, 
the approximation $\sqrt{1 + \alpha_d^2 / \alpha_o^2} \simeq 1$ was used. 
It can be seen that the dampening predicted by the 
analytic solution when the Grad expression for $\eta$ is used does not match the 
numerical results.

\subsection{Cases 2a and 2b: Non-adiabatic flow}\label{sec:hydro1:case2}

\begin{figure}
\begin{center}
\begin{tabular}{c}
\includegraphics[angle=270,width=0.98\linewidth]{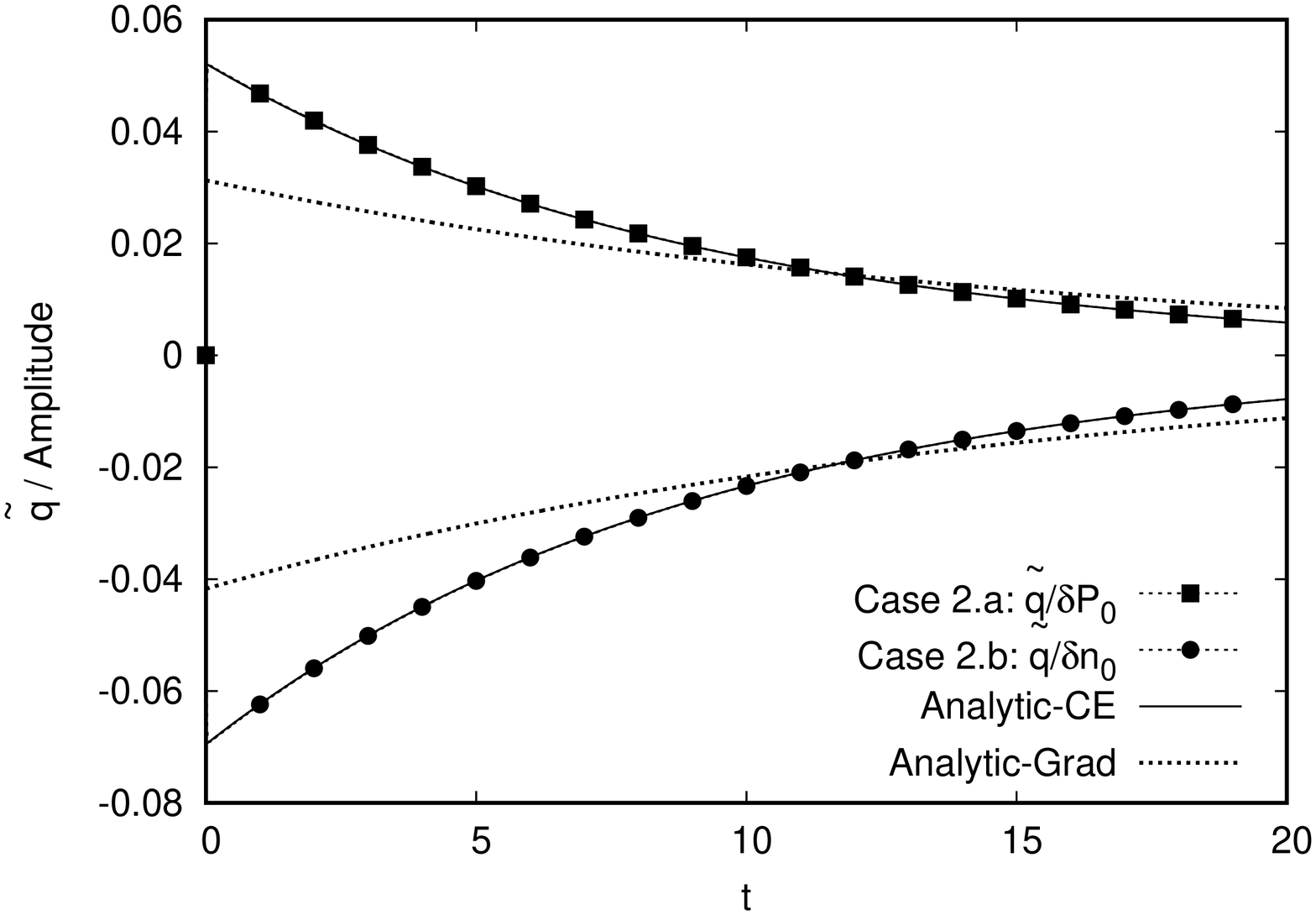}
\end{tabular}
\end{center}
\caption{
The time evolution of $\widetilde{q}$ for the initialisations corresponding to 
{\em Case 2 (a)} (i.e.~$\delta n_0 = \beta_0 = 0$ and $\delta P_0 = 10^{-3}$) 
and {\em Case 2 (b)} (i.e.~$\delta P_0 = \beta_0 = 0$ and $\delta n_0 = 10^{-3}$).
The dotted lines with points represent the numerical results. 
The analytic solution \eqref{eq:case2} is represented for the 
two cases using solid lines when $\alpha_\lambda$ is computed using the Chapman-Enskog 
expression for $\lambda$ \eqref{eq:tcoeff_CE}, while the dotted lines correspond to the case 
when the Grad expression \eqref{eq:tcoeff_G} is used.
The relaxation time was set to $\tau = 0.0083$.
}
\label{fig:case2-profiles}
\end{figure}

The coefficient $\alpha_\lambda$ can be investigated most easily by considering the decay of the 
amplitude $\widetilde{q}$ of the heat flux. The system will thus be initialised 
accordint to {\it Case 2a} ($\delta n_0 = \beta_0 = 0$ and $\delta P_0 = 10^{-3}$) and 
{\it Case 2b} ($\delta P_0 = \beta_0 = 0$ and $\delta n_0 = 10^{-3}$) 
described in Sec.~\ref{sec:fdyn:long}, while $\tau = 0.0083$. 
According to Eqs.~\eqref{eq:delta_aux1} and \eqref{eq:init_sol},
$\widetilde{q}$ takes the following form:
\begin{equation}
 \widetilde{q} = \frac{\alpha_\lambda P_0}{k} \left(\frac{3\delta P_0}{P_0} - \frac{4 \delta n_0}{n_0}\right) 
 e^{-\alpha_\lambda t}.\label{eq:case2}
\end{equation}
The above analytic result is compared in Fig.~\ref{fig:case2-profiles} to the numerical 
results obtained using the method described in Appendix~\ref{app:num}. 
For each of the two cases mentioned above, three curves are represented. 
The numerical results (dashed lines and points) are 
overlapped with the analytic prediction \eqref{eq:case2} when $\alpha_\lambda$ is calculated 
using the Chapman-Enskog expression for $\lambda$ (continuous line). The 
analytic prediction \eqref{eq:case2} corresponding to the case when $\alpha_\lambda$ is obtained 
using the Grad expression for $\lambda$ (dashed line) 
is clearly not consistent with the numerical results. 
The points at $t = 0$ indicate that in the numerical simulations, the system was 
initialised using an equilibrium state, in which the heat flux vanishes. 
In the five-field theory, the initial value of $\widetilde{q}$ does not represent a 
free parameter, as can be seen from the solution in Eq.~\eqref{eq:case2}.
However, since $\tau$ is small, the system quickly relaxes towards the 
five-field theory prediction \eqref{eq:case2}.
This relaxation process will be further considered in Sec.~\ref{sec:hydro2}.

\subsection{Limits of the linearised hydrodynamics equations}\label{sec:hydro1:lin}

\begin{figure}
\begin{center}
\begin{tabular}{c}
\includegraphics[angle=270,width=0.98\linewidth]{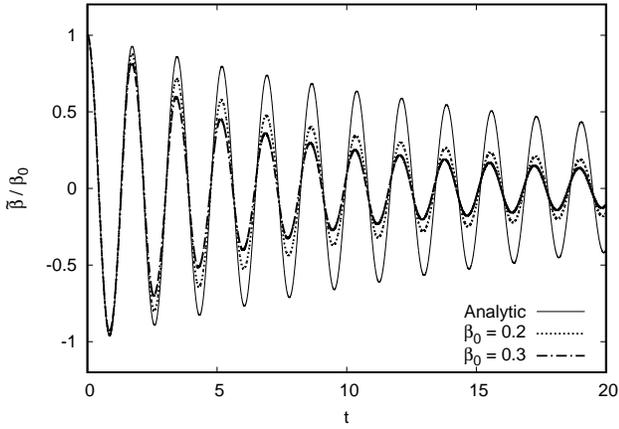} \\
\end{tabular}
\end{center}
\caption{
The ratio $\overline{\beta} / \beta_0$ for various values of $\beta_0$
corresponding to {\em Case 1} presented in Sec.~\ref{sec:fdyn:long}.
As $\beta_0$ increases, the time evolution of $\widetilde{\beta}/\beta_0$ 
departs from the solution \eqref{eq:case1}, indicating that nonlinear effects become important.
}
\label{fig:vel-betas}
\end{figure}

\begin{figure}
\begin{center}
\begin{tabular}{c}
\includegraphics[angle=270,width=0.98\linewidth]{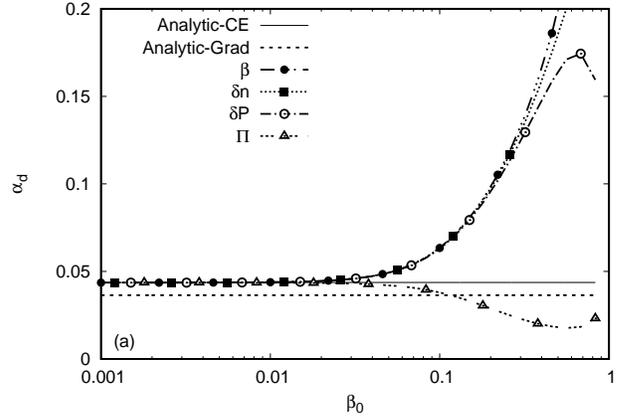} \\
\includegraphics[angle=270,width=0.98\linewidth]{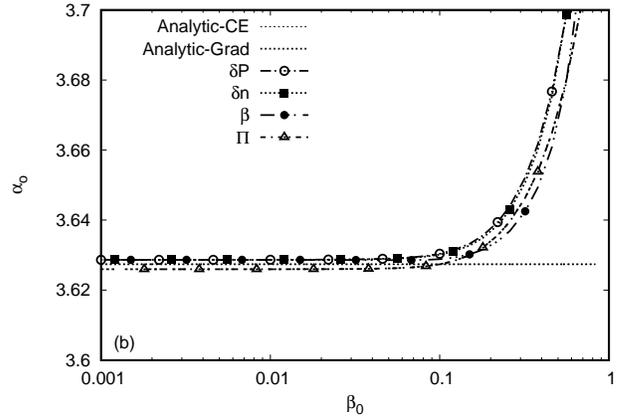} \\
\includegraphics[angle=270,width=0.98\linewidth]{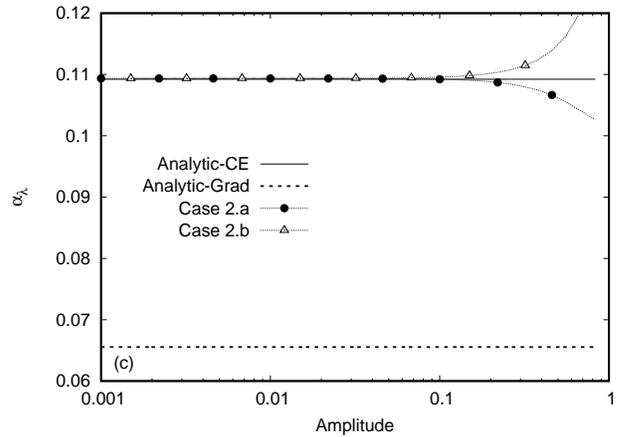}
\end{tabular}
\end{center}
\caption{
The dependence of (a) $\alpha_d$ and (b) $\alpha_o$ on $\beta_0$; the four
curves correspond to the two-parameter nonlinear fits of $\widetilde{\delta n}$, 
$\widetilde{\delta P}$, $\widetilde{\beta}$ and $\widetilde{\Pi}$ in 
Eq.~\eqref{eq:case1} to the corresponding numerical data, as described in 
Subsec.~\ref{sec:hydro1:lin}.
(c) The dependence of $\alpha_\lambda$, obtained using a nonlinear fit of 
Eq.~\eqref{eq:case2} to the numerical data, on the amplitudes 
$\delta P_0$ ({\em Case 2a}) and $\delta n_0$ ({\em Case 2b})
of the initial perturbation.
The relaxation time was always kept at $\tau = 0.0083$.
}
\label{fig:reg-betas}
\end{figure}

Next, an assessment of the limits within which the solution of the linearised 
equations \eqref{eq:cons_lin} is applicable is performed. 
In order to reduce the rarefaction effects, $\tau$ is fixed at $\tau = 0.0083$ 
throughout this subsection.

The solution \eqref{eq:case1} predicts that, for {\it Case 1}, the time evolution 
of $\widetilde{\beta}$ is damped according to the factor $\exp(-\alpha_d t)$, with 
$\alpha_d$ \eqref{eq:alphad_alphao} being independent of the magnitude $\beta_0$ 
of the perturbation. Figure~\ref{fig:vel-betas} shows that this is not the case: 
while at small values of $\beta_0$, $\widetilde{\beta}$ follows closely the analytic 
prediction [as confirmed in Fig.~\ref{fig:case1-profiles}(c)], at larger values of $\beta$, 
the dampening is enhanced compared to the linearised limit \eqref{eq:alphad_alphao}. 

In order to test the versatility of the functional form of the solution 
corresponding to the linearised regime, the parameters $\alpha_d$, $\alpha_o$ 
and $\alpha_\lambda$ are determined using nonlinear fits of the 
analytic solutions \eqref{eq:case1} and \eqref{eq:case2} to the 
corresponding numerical data. 
The coefficients $\alpha_d$ and $\alpha_o$ 
are obtained by treating them as free parameters while performing 
a two-parameter nonlinear fit of 
$\widetilde{\delta n}$, $\widetilde{\beta}$, $\widetilde{\delta P}$ 
and $\widetilde{\Pi}$ given in Eq.~\eqref{eq:case1} 
for the initial conditions corresponding to {\em Case 1}
(i.e. $\delta n_0 = \delta P_0 = 0$ and various values of $\beta_0$)
to the corresponding numerical results.
The coefficient $\alpha_\lambda$ is obtained by performing a 
one-parameter nonlinear fit of 
$\widetilde{q}$ with respect to the numerical results 
with the initial conditions described in 
{\em Case 2a} ($\beta_0 = 0$, $\delta n_0 = 0$ and various values for $\delta P_0$) and 
{\em Case 2b} ($\beta_0 = 0$, $\delta P_0 = 0$ and various values for $\delta n_0$).

The dependence of $\alpha_d$, $\alpha_o$ and $\alpha_\lambda$ on the amplitude of 
the perturbations 
is presented in the plots (a), (b) and (c) of Fig.~\ref{fig:reg-betas}, respectively.
The horizontal axis in Fig.~\ref{fig:reg-betas}(c) 
represents the amplitude of the initial perturbation, i.e. $\delta P_0$ for {\em Case 2a} and 
$\delta n_0$ for {\em Case 2b}. All of the above plots show the analytic predictions \eqref{eq:alphad_alphao}
and \eqref{eq:alphal} for $\alpha_d$, $\alpha_o$ and $\alpha_\lambda$, specialised to the cases when the 
transport coefficients $\eta$ and $\lambda$ are computed using the Chapman-Enskog \eqref{eq:tcoeff_CE} and 
the Grad \eqref{eq:tcoeff_G} expressions. The results clearly favor the Chapman-Enskog expressions. These plots also indicate that the analytic analysis performed in Sec.~\ref{sec:hydro1:long} in the context of the 
linearised hydrodynamic equations loses applicability when the perturbation amplitudes 
$\beta_0$, $\delta n_0 / n_0$ or $\delta P_0 / P_0$ are larger than $\sim 0.05$.

\subsection{Limits of the first-order hydrodynamics regime}\label{sec:hydro1:hydro}

\begin{figure*}
\begin{center}
\begin{tabular}{cc}
\includegraphics[angle=270,width=0.48\linewidth]{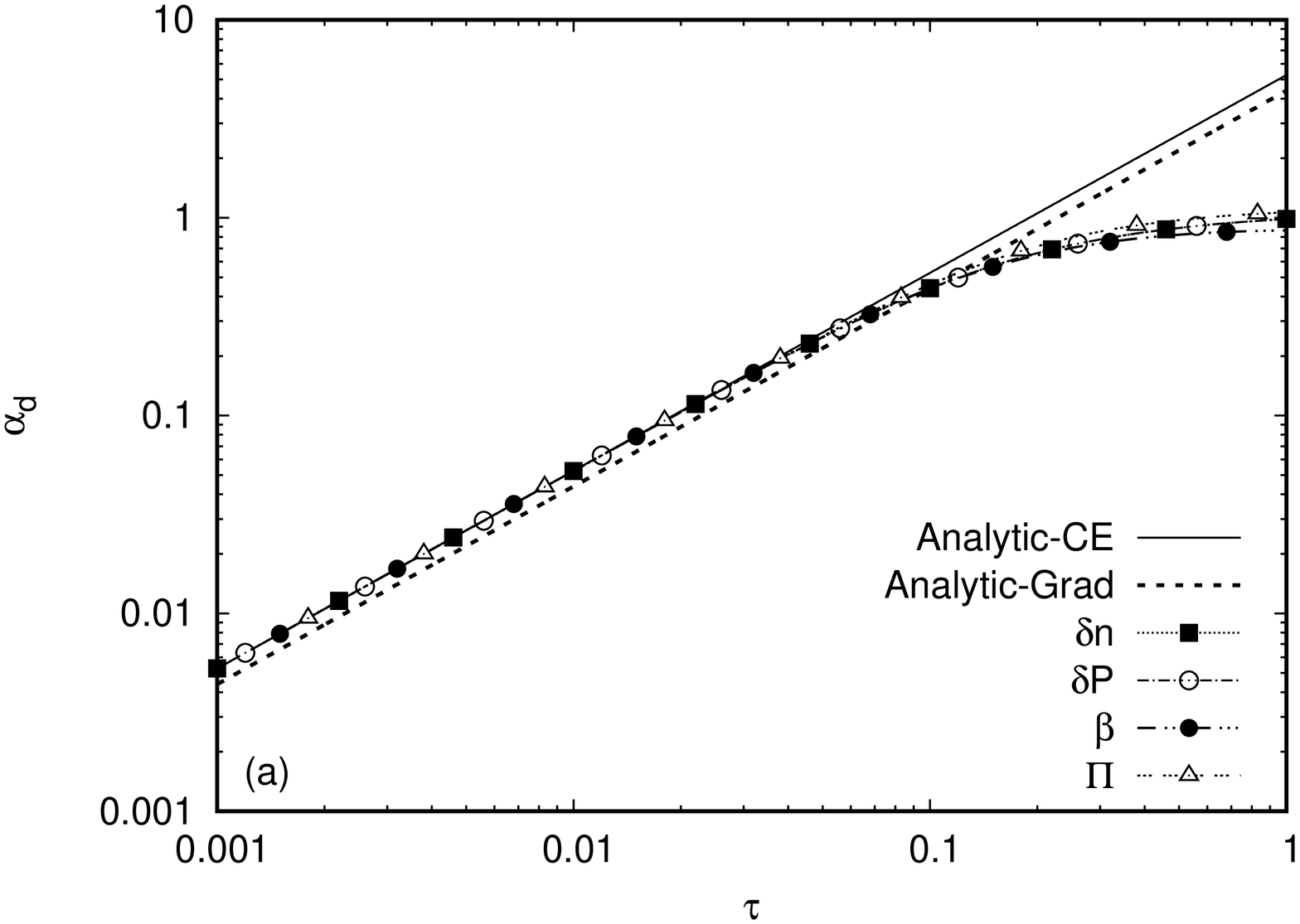} &
\includegraphics[angle=270,width=0.48\linewidth]{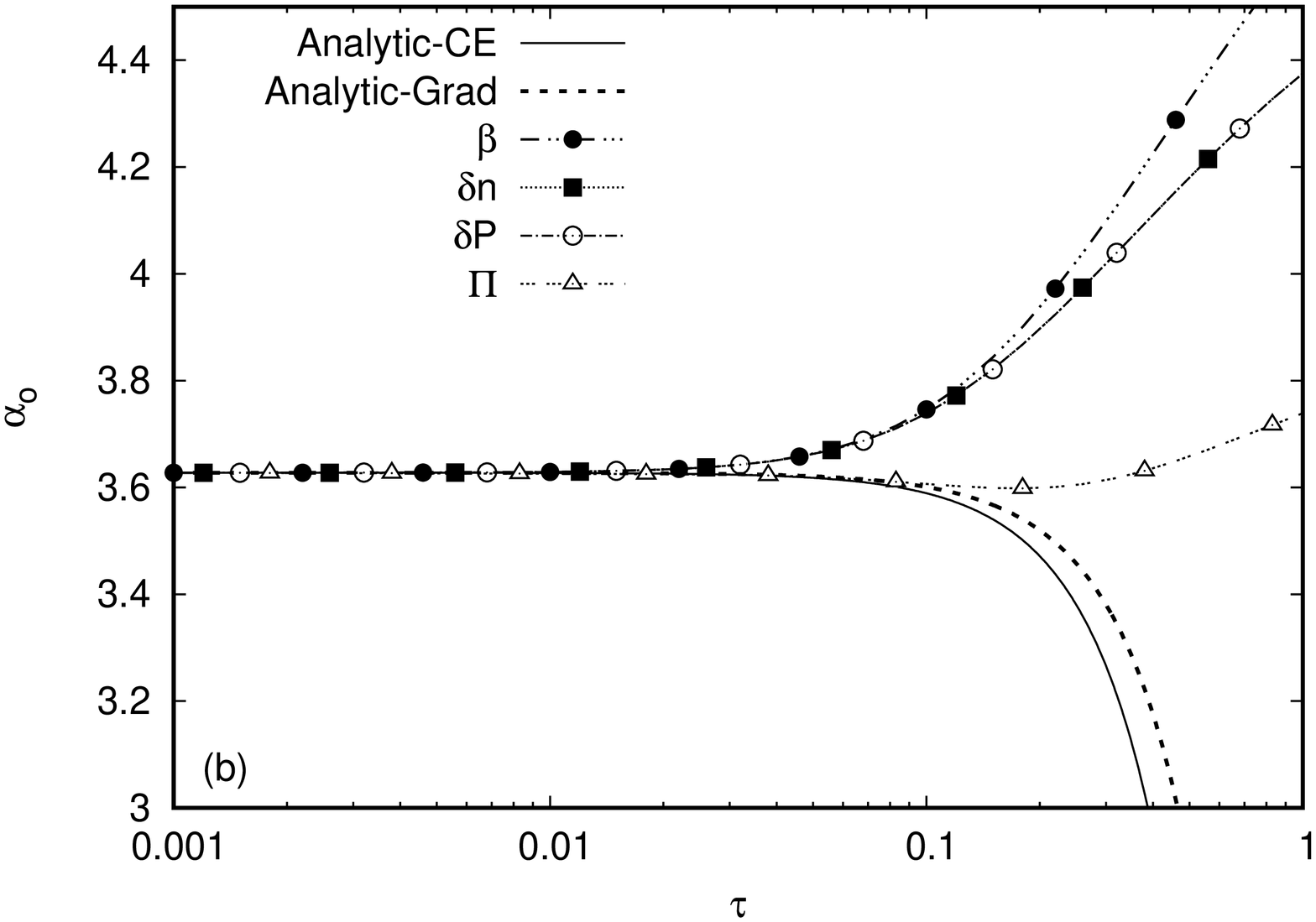} \\
\includegraphics[angle=270,width=0.48\linewidth]{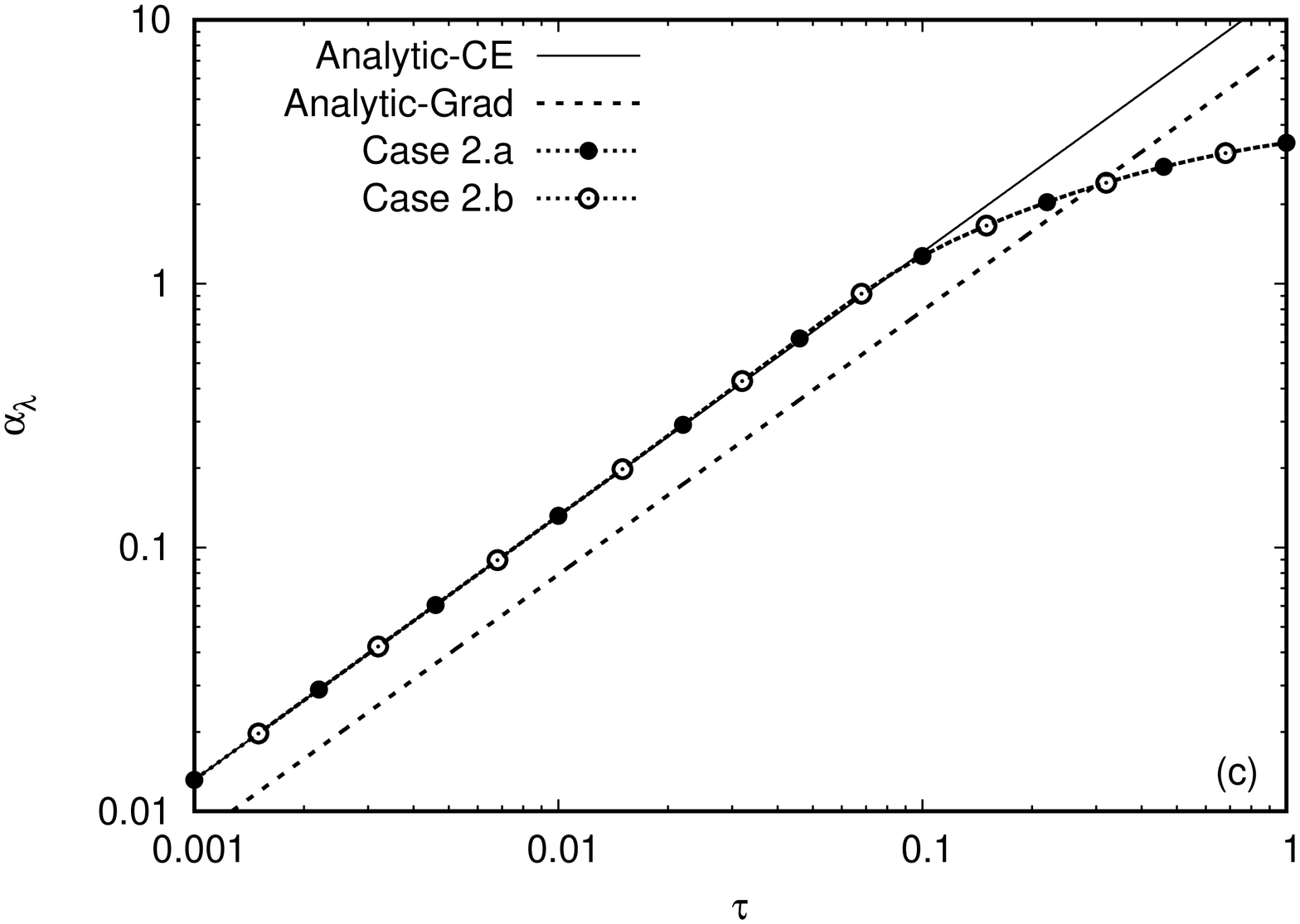} &
\includegraphics[angle=270,width=0.48\linewidth]{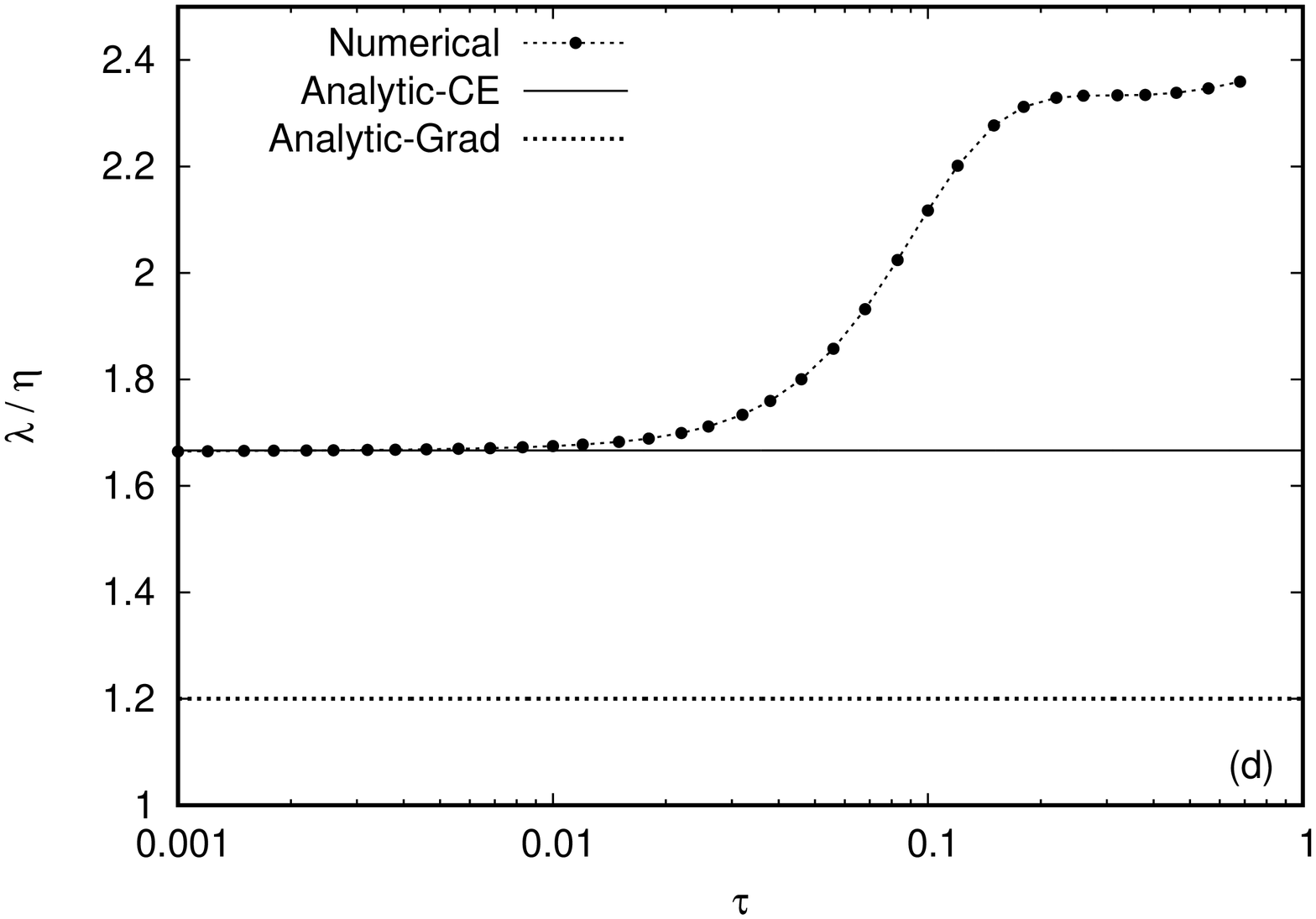} 
\end{tabular}
\end{center}
\caption{
Graphical representation of the dependence on $\tau$ of (a) $\alpha_d$, (b) $\alpha_o$ 
and (c) $\alpha_\lambda$ obtained using the nonlinear fitting procedure described 
in Subsec.~\ref{sec:hydro1:lin}. 
In (d), the ratio $\alpha_\lambda / \alpha_d$ is represented with respect to $\tau$,
where $\alpha_d$ is obtained using the two-parameter nonlinear fit of $\widetilde{\beta}$
\eqref{eq:case1} on the numerical data, while $\alpha_\lambda$ is obtained 
using a one-parameter nonlinear fit of $\widetilde{q}$ \eqref{eq:case2} for 
{\em Case 2a}. The perturbation amplitude is in all cases set to $10^{-3}$.
On each plot, the analytic curves corresponding to the Chapman-Enskog 
and Grad methods are displayed.
}
\label{fig:reg-tau}
\end{figure*}

It is known that the constitutive equations \eqref{eq:cons} are valid only 
when $\tau$ is small \cite{cercignani02}. This subsection is focused
on investigating the validity of the analysis presented in Sec.~\ref{sec:hydro1:long} 
as the relaxation time $\tau$ is increased. 

In order to test the effect of increasing $\tau$, the perturbations are kept at 
a small value, i.e. 
$\beta_0 = 10^{-3}$ (for {\it Case 1}),
$\delta P_0 = 10^{-3}$ (for {\it Case 2a}), or $\delta n_0 = 10^{-3}$ (for {\it Case 2b}).
The plots in Fig.~\ref{fig:reg-tau} show the dependence of (a) $\alpha_d$, (b) $\alpha_o$,
(c) $\alpha_\lambda$ and (d) the ratio $\lambda / \eta = 2\alpha_\lambda n_0 / 3\alpha_d P_0$ 
on the value of $\tau$. 
As before, the analytic predictions for the 
dependence of these coefficients on $\tau$ is also shown for the cases when the transport 
coefficients $\eta$ and $\lambda$ are obtained using the Chapman-Enskog \eqref{eq:tcoeff_CE} 
and Grad \eqref{eq:tcoeff_G} expressions. For $\tau < 0.1$, the numerical results clearly 
favor the Chapman-Enskog expression. Plot (d) confirms that for small values of $\tau$, 
the ratio $\lambda / \eta$ is equal to 
$5/3$, as predicted in the Chapman-Enskog theory \eqref{eq:tcoeff_CE}. 
This value is in agreement with the high chemical potential limit of Fig.~2 
in Ref.~\cite{jaiswal15}.

While for $\tau \gtrsim 0.1$, the constitutive equations \eqref{eq:cons} no longer hold, our nonlinear 
fit analysis seems to indicate that the dampening coefficients $\alpha_d$ and $\alpha_\lambda$
plateau at large $\tau$. This conclusion is not necessarily meaningful, since the ansatz \eqref{eq:ansatz}
that the time dependence of $\delta n$, $\delta P$ and $\beta$ is of the form $e^{-\alpha t}$ with 
constant $\alpha$ is not guaranteed to be valid in the transition regime. It is certain that the time 
dependence of the above quantities is more complex at large values of $\tau$,
since the dissipative exponential attenuation is replaced by a polynomial dispersive attenuation 
in the ballistic regime, as will be shown in Sec.~\ref{sec:bal}.

\subsection{Summary} \label{sec:hydro1:summary}

In this section, the attenuation of a longitudinal wave was analysed using the 
equations of first-order hydrodynamics. The results obtained by numerically solving 
the AWB equation at small values of the relaxation time $\tau$ showed an excellent 
agreement with the analytic results when the transport coefficinets $\lambda$ and 
$\eta$ were obtained using the Chapman-Enskog method. The analytic results 
corresponding to the transport coefficients obtained using the Grad method 
exhibited a clear discrepancy compared to the numerical results.

The validity of the linearised form of the AWB equation \eqref{eq:boltz_lin} 
was further tested by comparing the numerical and analytic results at increasing 
values of the wave amplitudes, while keeping $\tau = 0.0083$. 
A visible discrepancy can be seen at wave amplitudes 
of $\beta_0, \delta n_0 / n_0, \delta P_0 / P_0 \gtrsim 0.1$. The 
applicability of the functional form of the solution of the first-order hydrodynamics 
equations was further considered 
by numerically fitting the attenuation coefficients $\alpha_\lambda$ and 
$\alpha_\eta$, as well as the oscillation frequency $\alpha_o$, to the numerical results.
The above analysis shows that the best fit parameters differ significantly from the analytic 
prediction when the wave amplitude is $\gtrsim 0.05$. At smaller values of the wave 
amplitudes, the numerical results clearly favored the Chapman-Enskog predictions
for the transport coefficients, while the Grad predictions showed a visible 
discrepancy to the numerical results.

Finally, the validity of the first-order hydrodynamics equations was investigated
at increasing values of $\tau$. The dampening coefficients $\alpha_d$ and 
$\alpha_\lambda$ were found to be directly 
proportional to $\tau$ only for $\tau \lesssim 0.1$,
confirming the Chapman-Enskog prediction.
At larger values of $\tau$, they increase at a significantly slower rate, signaling the 
breakdown of the first-order hydrodynamics formulation when $\tau \gtrsim 0.1$.
The Grad prediction was in clear disagreement with the numerical results for all 
values of $\tau$. 

\section{Second-order hydrodynamics}\label{sec:hydro2}

The five-field theory provides the constitutive equations for $q^\mu$ and $\Pi^{\mu\nu}$ in 
the form given in Eq.~\eqref{eq:tcoeff}. Since these constitutive equations do not represent 
evolution equations, $q^\mu$ and $\Pi^{\mu\nu}$ are fully determined by the spatial and temporal 
gradients of $n$, $u^\mu$ and $P$. In particular, their initial values at $t = 0$ cannot 
be set to arbitrary values. In this section, the second-order extension 
of the five-field theory will be employed in order to study the 
relaxation process of the heat flux and shear stress from their initial vanishing value 
to the value required through the constitutive equations of the five-field theory.

\subsection{Constitutive relations}\label{sec:hydro2:const}

There are many variations of the form in which the 
equations of second-order hydrodynamics (also known as extended irreversible 
thermodynamics \cite{rezzolla13}) are presented, essentially due to the route adopted 
in deriving them \cite{israel79,hiscock83,el10,denicol10,jaiswal13b,chattopadhyay15}.
Only the form introduced in Refs.~\cite{hiscock83,rezzolla13} will be further 
considered, according to which $q^\mu$ and $T^{\mu\nu}$ satisfy the following equations:
\begin{widetext}
\begin{align}
 q^\mu =& -\lambda T \Delta^{\mu\nu}\Bigg[ \partial_\nu \ln T + 
 u^\rho \partial_\rho u_\nu + \beta_1 u^\rho \partial_\rho q_\nu - 
 \alpha_1 \partial_\rho \Pi^\rho{}_\nu 
 - (1 - \gamma_1) T \Pi^\rho{}_\nu \partial_\rho 
 \left(\frac{\alpha_1}{T}\right) + 
 \frac{1}{2} T q_\nu \partial_\rho \left(\frac{\beta_1 u^\rho}{T}\right)\Bigg],\nonumber\\
 \Pi^{\mu\nu} =&-2\eta \left[\frac{1}{2}\left(\Delta^{\mu\lambda} \Delta^{\nu\kappa} + \Delta^{\nu\lambda} \Delta^{\mu\kappa}\right) -
 \frac{1}{3} \Delta^{\mu\nu} \Delta^{\lambda\kappa}\right] \nonumber\\
 &\times \left[\partial_\lambda u_\kappa + \beta_2 u^\rho \partial_\rho \Pi_{\lambda\kappa} - 
 \alpha_1 \partial_\lambda q_\kappa - \gamma_1 T q_\lambda \partial_\kappa \left(\frac{\alpha_1}{T}\right) +
 \frac{1}{2} T \Pi_{\lambda\kappa} \partial_\rho \left(\frac{\beta_2}{T} u^\rho\right)\right],
 \label{eq:h2_tcoeff}
\end{align}
where the thermodynamic coefficients $\beta_1$, $\beta_2$ and $\gamma_1$ and 
the viscous-heat flux coupling coefficient $\alpha_1$ are {\it a priori} not 
known. 
After performing the linearisation described in Sec.~\ref{sec:fdyn:hydro},
Eq.~\eqref{eq:h2_tcoeff} becomes:
\begin{subequations}\label{eq:h2_tcoeff_lin_alpha}
\begin{align}
 \tau_q \partial_t q + q =& -\frac{\lambda P_0}{4n_0}\left(\frac{3\partial_z \delta P}{P_0} - 
 \frac{4\partial_z \delta n}{n_0}\right)
 +\frac{\lambda}{4n_0} (1 + 4\alpha_1 P_0) \partial_z \Pi,
 \label{eq:h2_tcoeff_lin_alpha:q}\\
 \tau_\Pi \partial_t \Pi + \Pi =& -\frac{4\eta}{3} \partial_z \left(\beta - \alpha_1 q\right),
 \label{eq:h2_tcoeff_lin_alpha:Pi}
\end{align}
\end{subequations}
\end{widetext}
where Eq.~\eqref{eq:cons_lin} was used to eliminate the time derivative of 
$\beta$ in Eq.~\eqref{eq:h2_tcoeff_lin_alpha:q}, while the relaxation times were defined as
$\tau_q = \lambda(T \beta_1 - 1/ 4n_0)$ and $\tau_\Pi = 2\eta \beta_2$.
In order for the constitutive equation for $q$ and $\Pi$ to match 
the first-order versions \eqref{eq:tcoeff_lin} in the limit $\tau_q, \tau_\Pi \rightarrow 0$,
we set $\alpha_1 = -1/4P_0$ and Eqs.~\eqref{eq:h2_tcoeff_lin_alpha} reduce to:
\begin{subequations}\label{eq:h2_tcoeff_lin}
\begin{align}
 \tau_q \partial_t q + q =& -\frac{\lambda P_0}{4n_0}\left(\frac{3\partial_z \delta P}{P_0} - 
 \frac{4\partial_z \delta n}{n_0}\right),\label{eq:h2_tcoeff_lin:q}\\
 \tau_\Pi \partial_t \Pi + \Pi =& -\frac{4\eta}{3} \partial_z \left(\beta + \frac{q}{4P_0}\right),
 \label{eq:h2_tcoeff_lin:Pi}
\end{align}
\end{subequations}
By analogy to Eq.~\eqref{eq:tcoeff_red}, the reduced relaxation times 
$\tau_{q,0}$ and $\tau_{\Pi, 0}$ can be introduced through:
\begin{equation}
 \tau_q = \tau_{q,0} \tau, \qquad 
 \tau_\Pi = \tau_{\Pi,0} \tau.\label{eq:h2_tcoeff_red}
\end{equation}
In general, the values of $\tau_{q,0}$ and $\tau_{\Pi, 0}$ are determined by 
the properties of the collision term in the Boltzmann equation. When the 
Anderson-Witting single relaxation time approximation is used, the following 
values for $\tau_{q,0}$ and $\tau_{\Pi, 0}$ are commonly employed 
within both the Chapman-Enskog- and moment-like methods 
\cite{denicol12,jaiswal13a,bhalerao14,chattopadhyay15,tinti17}:
\begin{equation}
 \tau_{\Pi, 0} = 1, \qquad \tau_{q,0} = 1.\label{eq:h2_tcoeff_CE}
\end{equation}

It is interesting to test the hyperbolicity of the resulting set of equations. 
Equations \eqref{eq:cons_lin} and \eqref{eq:h2_tcoeff_lin} can be written 
in the following form:
\begin{equation}
 \partial_t \mathbb{U} + \partial_z (\mathbb{A} \mathbb{U}) = \mathbb{S},
 \label{eq:hydro2_sys}
\end{equation}
where 
\begin{align}
 \mathbb{U} =& \left(\frac{\delta n}{n_0}, \frac{\delta P}{P_0}, \beta, \frac{q}{P_0}, \frac{\Pi}{P_0}\right)^T, \nonumber\\
 \mathbb{S} =& \left(0, 0, \frac{q}{4 \tau_q P_0},
 -\frac{q}{\tau_q P_0}, -\frac{\Pi}{\tau_{\Pi} P_0}\right)^T,
\end{align}
while $\mathbb{A}$ is given by:
\begin{equation}
 \mathbb{A} =
 \begin{pmatrix}
  0 & 0 & 1 & 0 & 0\\
  0 & 0 & \frac{4}{3} & \frac{1}{3} & 0\\
  \frac{\lambda_0}{4 \tau_{q,0}} & \frac{1}{4} - \frac{3\lambda_0}{16\tau_{q,0}} &
  0 & 0& \frac{1}{4} \\
  -\frac{\lambda_0}{\tau_{q,0}} & \frac{3\lambda_0}{4\tau_{q,0}} & 0 & 0 & 0\\
  0 & 0 & \frac{4\eta_0}{3\tau_{\Pi, 0}} & \frac{\eta_0}{3 \tau_{\Pi,0}} & 0
 \end{pmatrix}.
\end{equation}
The five eigenvalues of $\mathbb{A}$ can be found analytically and are given by
$a_0 = 0$ and
\begin{equation}
 a_{\lambda,\pm} = \pm\frac{1}{2}\sqrt{\frac{\lambda_0}{\tau_{q,0}}}, \qquad 
 a_{\eta,\pm} = \pm \frac{1}{\sqrt{3}} 
 \sqrt{1 + \frac{\eta_0}{\tau_{\Pi,0}}}.
\end{equation}
Since the eigenvalues of $\mathbb{A}$ are real, 
the system of equations \eqref{eq:hydro2_sys} is hyperbolic \cite{rezzolla13,toro09}.

\subsection{Longitudinal waves: modes}\label{sec:hydro2:longalpha}

Employing the ans\"atze \eqref{eq:ansatz} and \eqref{eq:alpha_def} allows 
$\Pi_\alpha$ to be expressed from Eq.~\eqref{eq:h2_tcoeff_lin:Pi} as follows:
\begin{equation}
 \Pi_\alpha = -\frac{4k \eta}{3(1 - \alpha \tau_\Pi)} \left(\beta_\alpha + \frac{q_\alpha}{4P_0}\right).
 \label{eq:h2_Pi_alpha}
\end{equation}
Furthermore, Eq.~\eqref{eq:cons_lin_alpha:P} allows $\delta P_\alpha$ to be written as:
\begin{equation}
 \delta P_\alpha = \frac{4 k P_0}{3\alpha} \left(\beta_\alpha + \frac{q_\alpha}{4P_0}\right).
 \label{eq:h2_dP_alpha}
\end{equation}
Inserting Eq.~\eqref{eq:h2_Pi_alpha} and \eqref{eq:h2_dP_alpha} into Eq.~\eqref{eq:cons_lin_alpha:beta} yields:
\begin{equation}
 \left[\alpha + \frac{k^2}{3\alpha} - \frac{\eta k^2}{3P_0(1 - \alpha \tau_\Pi)}\right] 
 \left(\beta_\alpha + \frac{q_\alpha}{4P_0}\right) = 0.\label{eq:h2_eig}
\end{equation}
The above equation is satisfied when either of the two factors cancel. These two cases 
are discussed separately below.

\paragraph{$q_\alpha = -4P_0 \beta_\alpha$.}
This case is also recovered for $q_\lambda$ in the first-order 
approximation \eqref{eq:delta_aux1}. In this case, Eqs.~\eqref{eq:h2_Pi_alpha} and 
\eqref{eq:h2_dP_alpha} show that $\Pi_\alpha = \delta P_\alpha = 0$. The values of 
$\alpha$ corresponding to this regime can be found from 
Eq.~\eqref{eq:h2_tcoeff_lin:q}, which reduces to:
\begin{equation}
 \left[\frac{\lambda k^2}{\alpha n_0} - 4(1 - \alpha\tau_q)\right] \beta_\alpha = 0.
 \label{eq:h2_alphal_eq}
\end{equation}
The case $\beta_\alpha = 0$ trivially corresponds to a vanishing perturbation (i.e. $\delta n_\alpha = \delta P_\alpha = 0$).
Setting the term inside the square bracket to $0$ yields the following values for $\alpha$:
\begin{align}
 \alpha_{\lambda,+} =& \frac{1}{2\tau_q} \left(1 + \sqrt{1 - \frac{k^2 \lambda \tau_q}{n_0}}\right)
 \simeq \frac{1}{\tau_q} - \alpha_\lambda + O(\tau^3),\nonumber\\
 \alpha_{\lambda,-} =& \frac{1}{2\tau_q} \left(1 - \sqrt{1 - \frac{k^2 \lambda \tau_q}{n_0}}\right)
 \simeq \alpha_\lambda + O(\tau^3).\label{eq:h2_al}
\end{align}
It can be seen that, in the small $\tau$ limit, 
$\alpha_{\lambda,-}$ corresponds to the dampening coefficient 
$\alpha_\lambda = k^2 \lambda / 4n_0$ identified in the first-order theory \eqref{eq:alphal}.
The term $\alpha_{\lambda,+}$ allows $\widetilde{q}$ to relax from an arbitrary 
initial condition at $t = 0$ to the first-order expression \eqref{eq:tcoeff_lin}
on a timescale of order $\tau_q$.
The modes corresponding to the above values of $\alpha_{\lambda,\pm}$ can be written in 
terms of the amplitudes $\beta_{\lambda,\pm}$ as follows:
\begin{gather}
 \delta n_{\lambda, \pm} = \frac{k n_0}{\alpha_{\lambda,\pm}} \beta_{\lambda,\pm}, \qquad 
 q_{\lambda, \pm} = -4P_0 \beta_{\lambda,\pm},\nonumber\\
 \delta P_{\lambda,\pm} = \Pi_{\lambda, \pm} = 0.
 \label{eq:h2_solpm}
\end{gather}

The square root in the definition of $\alpha_{\lambda,\pm}$ \eqref{eq:h2_al} becomes imaginary when 
$\tau > \tau_{\lambda,{\rm lim}}$, where
\begin{equation}
 \tau_{\lambda,{\rm lim}} = \frac{1}{k \sqrt{\tau_{q,0} \lambda_0}}.
 \label{eq:h2_taullim}
\end{equation}
When $0 < \tau < \tau_{\lambda,{\rm lim}}$, the modes corresponding to 
$\alpha_{\lambda,\pm}$ are overdamped (evanescent),
while when $\tau > \tau_{\lambda, {\rm lim}}$, the underdamped regime settles in.
In order to treat both regimes within a unitary framework, 
it is convenient to cast Eq.~\eqref{eq:h2_al} in the form $\alpha_{\lambda,\pm} = 
\alpha_{\lambda,d} \pm \alpha_{\lambda,o} = \alpha_{\lambda,d} \pm i \overline{\alpha}_{\lambda,o}$, 
where $\alpha_{\lambda, d} = 1 / 2\tau_q$ and 
\begin{align}
 \alpha_{\lambda,o} =& \alpha_{\lambda, d} \sqrt{1 - \frac{k^2 \lambda \tau_q}{n_0}},
 \nonumber\\
 \overline{\alpha}_{\lambda,o} =& \alpha_{\lambda, d} 
 \sqrt{\frac{k^2 \lambda \tau_q}{n_0} - 1}.
 \label{eq:h2_ald_alo}
\end{align}

\paragraph{$q_\alpha = 0$.}
Next, the case when the first bracket in Eq.~\eqref{eq:h2_eig} vanishes is considered.
Substituting Eqs.~\eqref{eq:cons_lin_alpha:n} and \eqref{eq:h2_dP_alpha} in
Eq.~\eqref{eq:h2_tcoeff_lin:q} yields:
\begin{equation}
 \left[\frac{\lambda k^2}{\alpha n_0} - 4(1 - \alpha \tau_q)\right] q_\alpha = 0.
 \label{eq:h2_q0}
\end{equation}
The term inside the square bracket is identical to the one in Eq.~\eqref{eq:h2_alphal_eq}. 
Setting this term to $0$ yields the same values for $\alpha$ as considered 
previously, when $q_\alpha = -4 P_0 \beta_\alpha$. To obtain a different 
set of values for $\alpha$, Eq.~\eqref{eq:h2_q0} is now solved by setting 
$q_\alpha = 0$.

The allowed values for $\alpha$ can be found by solving the following 
cubic equation:
\begin{equation}
 \left(\frac{3\alpha^2}{k^2} + 1\right) (1 - \alpha \tau_\Pi)- \frac{\eta \alpha}{P_0} = 0.
 \label{eq:h2_ae_eq}
\end{equation}
Equation \eqref{eq:h2_ae_eq} has the roots $\alpha\in \{\alpha_{\eta, r}, \alpha_{\eta,\pm}\}$, 
where the notation $\alpha_{\eta,\pm} = \alpha_{\eta, d} \pm i \alpha_{\eta, o}$ 
is introduced by analogy to the first-order case \eqref{eq:alphapm}. 
The exact expressions for the coefficients $\alpha_{\eta, r}$, 
$\alpha_{\eta, d}$ and $\alpha_{\eta, o}$ are:
\begin{widetext}
\begin{align}
 \alpha_{\eta,r} =& \frac{1}{3\tau_\Pi}\left\{1 + 
 \frac{1}{R_\eta}\left[1 - k^2 \tau_{\Pi}^2\left(1+\frac{\eta}{\tau_\Pi P_0}\right)\right] + 
 R_\eta\right\} \simeq \frac{1}{\tau_\Pi} - 2 \alpha_d + O(\tau^3),\nonumber\\
 \alpha_{\eta,d} =& \frac{1}{3\tau_\Pi}\left\{1 - 
 \frac{1}{2R_\eta}\left[1 - k^2 \tau_{\Pi}^2\left(1+\frac{\eta}{\tau_\Pi P_0}\right)\right] - 
 \frac{R_\eta}{2}\right\} \simeq \alpha_d + O(\tau^3),\nonumber\\
 \alpha_{\eta,o} =& \frac{\sqrt{3}}{6\tau_\Pi} \left\{ 
 \frac{1}{R_\eta} \left[1 - k^2 \tau_{\Pi}^2\left(1+\frac{\eta}{\tau_\Pi P_0}\right)\right] - R_\eta\right\} \simeq 
 \alpha_o + O(\tau^2),\label{eq:h2_ae}
\end{align}
where 
\begin{align}
 R_\eta =& 
 \begin{cases}
  \left[1 - 3k \tau_\Pi \sqrt{R_{\eta,{\rm aux}}} + 3k^2 \tau_\Pi^2\left(1 - \frac{\eta}{2P_0 \tau_\Pi}\right)\right]^{1/3}, &
  0 < \tau < \tau_{\eta, {\rm lim}}, \\
  -\left[-1 + 3k \tau_\Pi \sqrt{R_{\eta, {\rm aux}}} - 3k^2 \tau_\Pi^2\left(1 - \frac{\eta}{2P_0 \tau_\Pi}\right)\right]^{1/3}, &
  \tau > \tau_{\eta, {\rm lim}},
 \end{cases}\label{eq:h2_Retadef}\\
 R_{\eta, {\rm aux}} =& 1 + 
 \frac{2}{3} k^2 \tau_\Pi^2 \left(1 - \frac{5\eta}{2P_0 \tau_\Pi} - \frac{\eta^2}{8P_0^2 \tau_\Pi^2}\right) + 
 \frac{k^4 \tau_\Pi^4}{9} \left(1 + \frac{\eta}{P_0\tau_\Pi}\right)^3.
 \label{eq:h2_Reaux}
\end{align}
\end{widetext}
The parameter $\tau_{\eta, \rm lim}$ is defined as the value of $\tau$ at which the expression under the 
cubic root in Eq.~\eqref{eq:h2_Retadef} vanishes. It is given by:
\begin{equation}
 \tau_{\eta, {\rm lim}} = \frac{1}{k\tau_{\Pi,0} \sqrt{1 + \eta_0 / \tau_{\Pi,0}}}.
 \label{eq:h2_tauelim}
\end{equation}
The definition \eqref{eq:h2_Retadef} of $R_\eta$ ensures that the coefficients $\alpha_{\eta, *}$ ($* \in \{r, d, o\}$), 
defined in Eq.~\eqref{eq:h2_ae}, are real for all positive values of $\tau$, provided that $R_{\eta, {\rm aux}} > 0$. 
In order to investigate the latter inequality, it is convenient to consider $\eta / \tau_\Pi P_0$ as an independent
parameter. In this case, the roots of $R_{\eta, {\rm aux}}$ \eqref{eq:h2_Reaux} are:
\begin{multline}
 (k^2 \tau_\Pi^2)_\pm = \frac{3}{8 (1 + \eta / \tau_\Pi P_0)^3}
 \Bigg\{\frac{20 \eta}{\tau_\Pi P_0} - 8 \\
 + \left(\frac{\eta}{\tau_\Pi P_0}\right)^2
 \left[1 \pm \left(1 - \frac{8\tau_\Pi P_0}{\eta}\right)\right]^{3/2}\Bigg\}.
\end{multline}
It can be seen that $(k^2 \tau_\Pi^2)_\pm$ is real only when 
$\eta / \tau_\Pi P_0 \ge 8$. In the hydrodynamic 
regime, the Chapman-Enskog expansion \eqref{eq:tcoeff_CE} together with Eq.~\eqref{eq:h2_tcoeff_CE} predict that 
$\eta / \tau_\Pi P_0 = 4/5$, which is much smaller than $8$. For the sake of simplicity, 
the case when $(k^2 \tau_\Pi^2)_\pm$ are real will not be consider in this paper. 
Instead, the coefficients $\alpha_{\eta, *}$ \eqref{eq:h2_ae} will be assumed to 
be real for all values of $\tau$ considered in this section. It is worth mentioning that 
setting $\eta_0 = 4/5$, $\tau_{\Pi,0} = 1$ and $k = 2\pi$ in Eq.~\eqref{eq:h2_tauelim} yields 
$\tau_{\eta, {\rm lim}} \simeq 0.12$, which is within the range of values of $\tau$ considered 
in Subsec.~\ref{sec:hydro2:case1}.

The small $\tau$ expansion of Eq.~\eqref{eq:h2_ae} reveals the first-order coefficients 
$\alpha_d = k^2\eta / 6P_0$ and $\alpha_o = k/\sqrt{3} + O(\tau^2)$ 
reported in Eq.~\eqref{eq:alphad_alphao}.
The coefficient $\alpha_{\eta,r}$ was not present in the first-order theory and 
in this case it corresponds to the mode that ensures the relaxation of $\widetilde{\Pi}$ 
from $\widetilde{\Pi}(t= 0) = 0$ to the value predicted through the constitutive equation 
of the first-order theory.

\subsection{Longitudinal waves: solution}\label{sec:hydro2:longsol}

The full solution of the longitudinal wave problem can be written in general in the following form:
\begin{equation}
 \widetilde{M} = \widetilde{M}_\lambda + \widetilde{M}_\eta,\label{eq:h2_solM}
\end{equation}
where $\widetilde{M} \in \{ \widetilde{\beta}, \widetilde{\delta n}, \widetilde{\delta P}, 
\widetilde{q}, \widetilde{\Pi}\}$.
The subscripts $\lambda$ and $\eta$ refer to the parts of the solution corresponding to $\alpha_{\lambda, \pm}$ 
and $\alpha_{\eta, *}$ ($* \in \{r, d, o\}$), respectively.

\begin{widetext}
According to Eq.~\eqref{eq:h2_solpm}, $\widetilde{\delta P}_\lambda = \widetilde{\Pi}_\lambda = 0$,
while $\widetilde{\beta}_\lambda$, $\widetilde{\delta n}_\lambda$ and $\widetilde{q}_\lambda$ 
can be written for $\tau < \tau_{\lambda,{\rm lim}}$ \eqref{eq:h2_taullim} as:
\begin{equation}
 \begin{pmatrix}
  \widetilde{\beta}_\lambda \\
  \widetilde{\delta n}_\lambda \\
  \widetilde{q}_\lambda
 \end{pmatrix}_{\tau_q < \tau_{\lambda,{\rm lim}}}
 = e^{-\alpha_{\lambda,d} t}  
 \left[ 
 \begin{pmatrix}
  \beta_{\lambda,c} \\
  \delta n_{\lambda,c}\\
  q_{\lambda, c} 
 \end{pmatrix}
 \cosh \alpha_{\lambda,o} t + 
 \begin{pmatrix}
  \beta_{\lambda,s} \\
  \delta n_{\lambda,s}\\
  q_{\lambda, s}
 \end{pmatrix}
 \sinh \alpha_{\lambda, o}t \right].
 \label{eq:h2_solld}
\end{equation}
When $\tau > \tau_{\lambda,{\rm lim}}$, the hyperbolic functions in the above expression become trigonometric functions:
\begin{equation}
 \begin{pmatrix}
  \widetilde{\beta}_\lambda \\
  \widetilde{\delta n}_\lambda \\
  \widetilde{q}_\lambda
 \end{pmatrix}_{\tau_q > \tau_{\lambda,{\rm lim}}}
 = e^{-\overline{\alpha}_{\lambda,d} t} 
 \left[ 
 \begin{pmatrix}
  \overline{\beta}_{\lambda,c} \\
  \overline{\delta n}_{\lambda,c}\\
  \overline{q}_{\lambda, c} 
 \end{pmatrix}
 \cos \overline{\alpha}_{\lambda,o} + 
 \begin{pmatrix}
  \overline{\beta}_{\lambda,s} \\
  \overline{\delta n}_{\lambda,s}\\
  \overline{q}_{\lambda, s}
 \end{pmatrix}
 \sin \overline{\alpha}_{\lambda, o}\right].
 \label{eq:h2_sollo}
\end{equation}
The constants $\beta_{\lambda, *}$ and $\overline{\beta}_{\lambda, *}$ (where $* \in \{c, s\}$) 
are fixed by the initial conditions, while:
\begin{align}
 q_{\lambda, *} =& -4P_0 \beta_{\lambda, *}, &
 \overline{q}_{\lambda, *} =& -4P_0 \overline{\beta}_{\lambda, *}, \nonumber\\
 \delta n_{\lambda, c} =& 
 kn_0 \frac{\alpha_{\lambda, d} \beta_{\lambda, c} + \alpha_{\lambda, o} \beta_{\lambda, s}}
 {\alpha_{\lambda,d}^2 - \alpha_{\lambda,o}^2}, &
 \overline{\delta n}_{\lambda,c} =& kn_0 \frac{\overline{\alpha}_{\lambda, d} \overline{\beta}_{\lambda, c} + 
 \overline{\alpha}_{\lambda, o} \overline{\beta}_{\lambda, s}}
 {\overline{\alpha}_{\lambda,d}^2 + \overline{\alpha}_{\lambda,o}^2}, \nonumber\\
 \delta n_{\lambda, s} =& 
 kn_0 \frac{\alpha_{\lambda, d} \beta_{\lambda, s} + \alpha_{\lambda, o} \beta_{\lambda, c}}
 {\alpha_{\lambda,d}^2 - \alpha_{\lambda,o}^2}, &
 \overline{\delta n}_{\lambda,s} =& kn_0 \frac{\overline{\alpha}_{\lambda, d} \overline{\beta}_{\lambda, s} - 
 \overline{\alpha}_{\lambda, o} \overline{\beta}_{\lambda, c}}
 {\overline{\alpha}_{\lambda,d}^2 + \overline{\alpha}_{\lambda,o}^2}.
\end{align}

The part of the solution \eqref{eq:h2_solM}  corresponding to $\alpha_{\eta, *}$ 
($* \in \{r, d, o\}$) can be written as:
\begin{align}
\begin{pmatrix}
 \widetilde{\beta}_\eta \\
 \widetilde{\delta n}_\eta \\
 \widetilde{\delta P}_\eta \\
 \widetilde{\Pi}_\eta 
\end{pmatrix} =
\begin{pmatrix}
 \beta_{\eta, r} \\
 \delta n_{\eta, r}\\
 \delta P_{\eta, r}\\
 \Pi_{\eta, r}
\end{pmatrix} e^{-\alpha_{\eta, r}} + \left[
\begin{pmatrix}
 \beta_{\eta, c} \\
 \delta n_{\eta, c}\\
 \delta P_{\eta, c}\\
 \Pi_{\eta, c}
\end{pmatrix} \cos \alpha_{\eta, o} + 
\begin{pmatrix}
 \beta_{\eta, s} \\
 \delta n_{\eta, s}\\
 \delta P_{\eta, s}\\
 \Pi_{\eta, s}
\end{pmatrix} \sin \alpha_{\eta, o}\right]
e^{-\alpha_{\eta, d}},
\label{eq:hydro2:sol}
\end{align}
while $\widetilde{q}_\eta = 0$.
The integration constants $\beta_{\eta, *}$ are fixed by the initial conditions, while 
\begin{gather}
 \begin{pmatrix}
  \delta n_{\eta, r} \\
  \delta P_{\eta, r} \\
  \Pi_{\eta, r} 
 \end{pmatrix} = 
 \begin{pmatrix}
  k n_0 / \alpha_{\eta,r}\\
  4 P_0 k / 3 \alpha_{\eta,r}\\
  -4k \eta / 3 (1 - \alpha_{\eta, r} \tau_\Pi)
 \end{pmatrix} \beta_{\eta,r}, \nonumber\\
 \begin{pmatrix}
  \delta n_{\eta, c} \\
  \delta P_{\eta, c} \\
 \end{pmatrix} = 
 \begin{pmatrix}
  k n_0 \\
  4 P_0 k / 3 
 \end{pmatrix} 
 \frac{\alpha_{\eta, d} \beta_{\eta,c} + \alpha_{\eta, o} \beta_{\eta, s}}{\alpha_{\eta,d}^2 + \alpha_{\eta,o}^2}, \qquad
 \begin{pmatrix}
  \delta n_{\eta, s} \\
  \delta P_{\eta, s} \\
 \end{pmatrix} = 
 \begin{pmatrix}
  k n_0 \\
  4 P_0 k / 3 
 \end{pmatrix} 
 \frac{\alpha_{\eta, d} \beta_{\eta,s} - \alpha_{\eta, o} \beta_{\eta, c}}{\alpha_{\eta,d}^2 + \alpha_{\eta,o}^2}, \nonumber\\
 \Pi_{\eta, c} = -\frac{4k \eta}{3} \frac{(1 - \alpha_{\eta, d} \tau_\Pi) \beta_{\eta,c} - \alpha_{\eta, o} \tau_\Pi \beta_{\eta, s}}
 {(1 - \alpha_{\eta,d} \tau_\Pi)^2 + (\alpha_{\eta,o} \tau_\Pi)^2}, \qquad 
 \Pi_{\eta, s} = -\frac{4k \eta}{3} \frac{(1 - \alpha_{\eta, d} \tau_\Pi) \beta_{\eta,s} + \alpha_{\eta, d} \tau_\Pi \beta_{\eta, c}}
 {(1 - \alpha_{\eta,d} \tau_\Pi)^2 + (\alpha_{\eta,o} \tau_\Pi)^2}.
 \label{eq:h2_consteta}
\end{gather}
The initial conditions \eqref{eq:init} for $\widetilde{\beta}$, $\widetilde{\delta n}$ and  $\widetilde{\delta P}$ are supplemented by 
\begin{equation}
 \widetilde{q}(t = 0) = 0, \qquad \widetilde{\Pi}(t = 0) = 0.
 \label{eq:init2}
\end{equation}
Substituting the solution \eqref{eq:h2_solM} into the initial conditions equations 
\eqref{eq:init} and \eqref{eq:init2} yields:
\begin{gather}
 \beta_{\lambda,c} + \beta_{\eta, r} + \beta_{\eta, c} = \beta_0, \qquad 
 \delta n_{\lambda,c} + \delta n_{\eta, r} + \delta n_{\eta, c} = \delta n_0, \nonumber\\
 \delta P_{\eta, r} + \delta P_{\eta, c} = \delta P_0, \qquad
 q_{\lambda,c} = 0, \qquad 
 \delta \Pi_{\eta, r} + \delta \Pi_{\eta, c} = 0.\label{eq:h2_init}
\end{gather}
The equation $q_{\lambda,c} = 0$ implies that $\beta_{\lambda, c} = 0$ (this is also true when 
$\tau_q > \tau_{\lambda, {\rm lim}}$, i.e. 
$\overline{q}_{\lambda, c} = \overline{\beta}_{\lambda, c} = 0$). 
Furthermore, noting that $\delta P_{\eta, *} = 4 P_0 \delta n_{\eta, *}/ 3n_0$, the second equality 
in Eq.~\eqref{eq:h2_init} implies:
\begin{subequations}\label{eq:h2_init_sol}
\begin{equation}
 \begin{pmatrix}
  \beta_{\lambda, s}\\
  \overline{\beta}_{\lambda, s}
 \end{pmatrix} = \frac{1}{4k}
 \begin{pmatrix}
  (\alpha_{\lambda,d}^2 - \alpha_{\lambda,o}^2) / \alpha_{\lambda, o}\\
  (\overline{\alpha}_{\lambda,d}^2 + \overline{\alpha}_{\lambda,o}^2) / \overline{\alpha}_{\lambda, o}
 \end{pmatrix}
 \left(\frac{4\delta n_0}{n_0} - \frac{3\delta P_0}{P_0}\right).
 \label{eq:h2_init_soll}
\end{equation}
Next, $\beta_{\eta, r}$ and $\beta_{\eta, s}$ can be written as:
\begin{align}
 \beta_{\eta,r} =& -\frac{\alpha_{\eta,r}(1 - \alpha_{\eta,r} \tau_\Pi)}
 {\tau_\Pi [\alpha_{\eta,o}^2+(\alpha_{\eta,d}-\alpha_{\eta,r})^2]} 
 \left[\beta_0 - \frac{3\tau_\Pi \delta P_0}{4kP_0} (\alpha_{\eta,d}^2 + \alpha_{\eta,o}^2)]\right],\nonumber\\
 \beta_{\eta, s} =& \frac{\alpha_{\eta, d}^2 + \alpha_{\eta, o}^2}
 {4k \alpha_{\eta, o}} \frac{3\delta P_0}{P_0}
 - \frac{\alpha_{\eta, d}}{\alpha_{\eta, o}} \beta_0 - 
 \frac{\alpha_{\eta, d}^2 + \alpha_{\eta, o}^2 - 
 \alpha_{\eta, d} \alpha_{\eta, r}}{\alpha_{\eta, r} \alpha_{\eta, o}} 
 \beta_{\eta, r},
 \label{eq:h2_init_sole}
\end{align}
while $\beta_{\eta,c} = \beta_0 - \beta_{\eta, r}$.
\end{subequations}
\end{widetext}

The above analytic results will now be employed to study the 
relaxation process for $\widetilde{q}$ 
and $\widetilde{\Pi}$ in the context of {\it Case 1} and {\it Case 2b} 
introduced in Sec.~\ref{sec:fdyn:long}.

\subsection{Case 1: Adiabatic flow}\label{sec:hydro2:case1}

\begin{figure*}
\begin{center}
\begin{tabular}{cc}
\includegraphics[angle=270,width=0.48\linewidth]{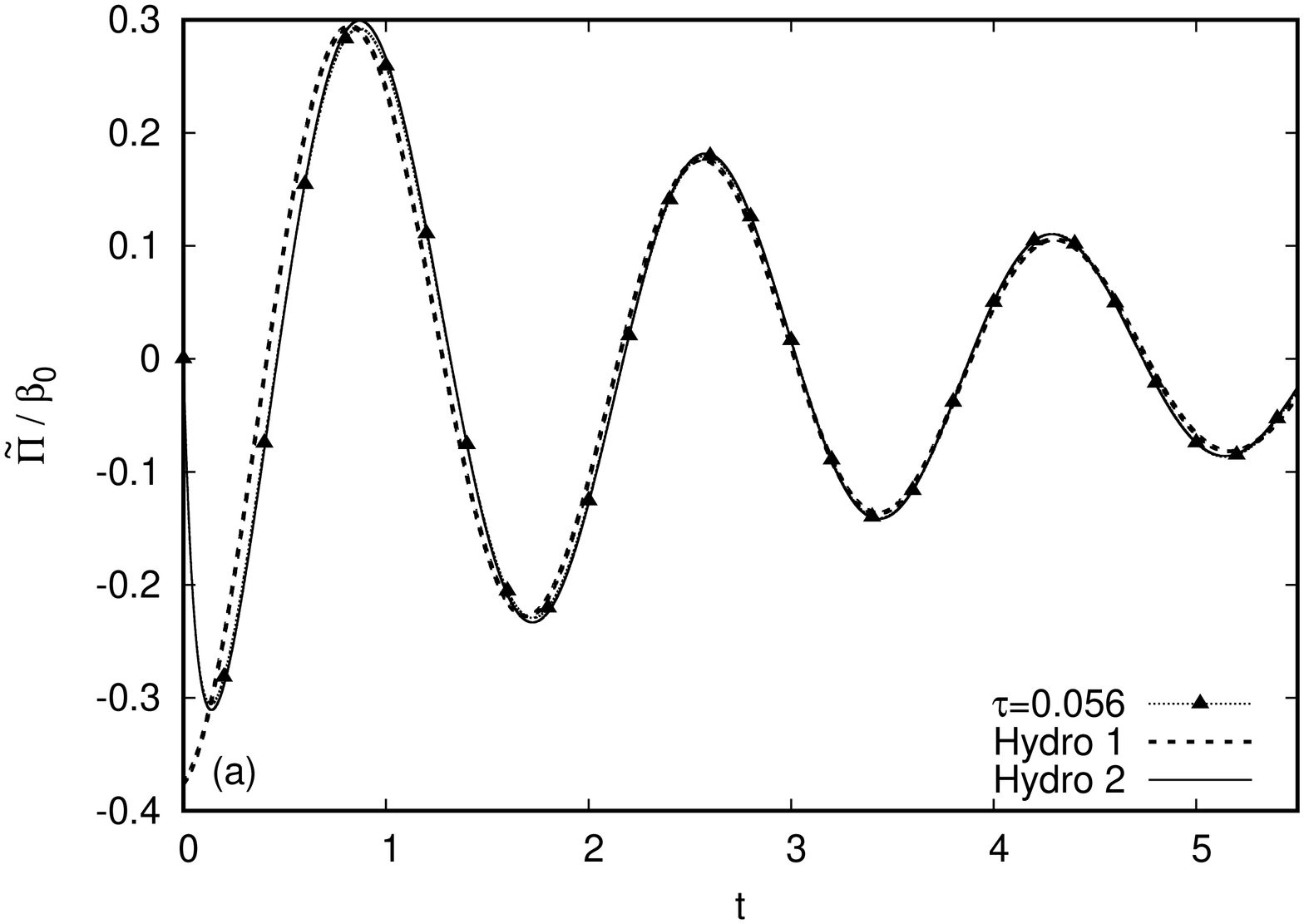} &
\includegraphics[angle=270,width=0.48\linewidth]{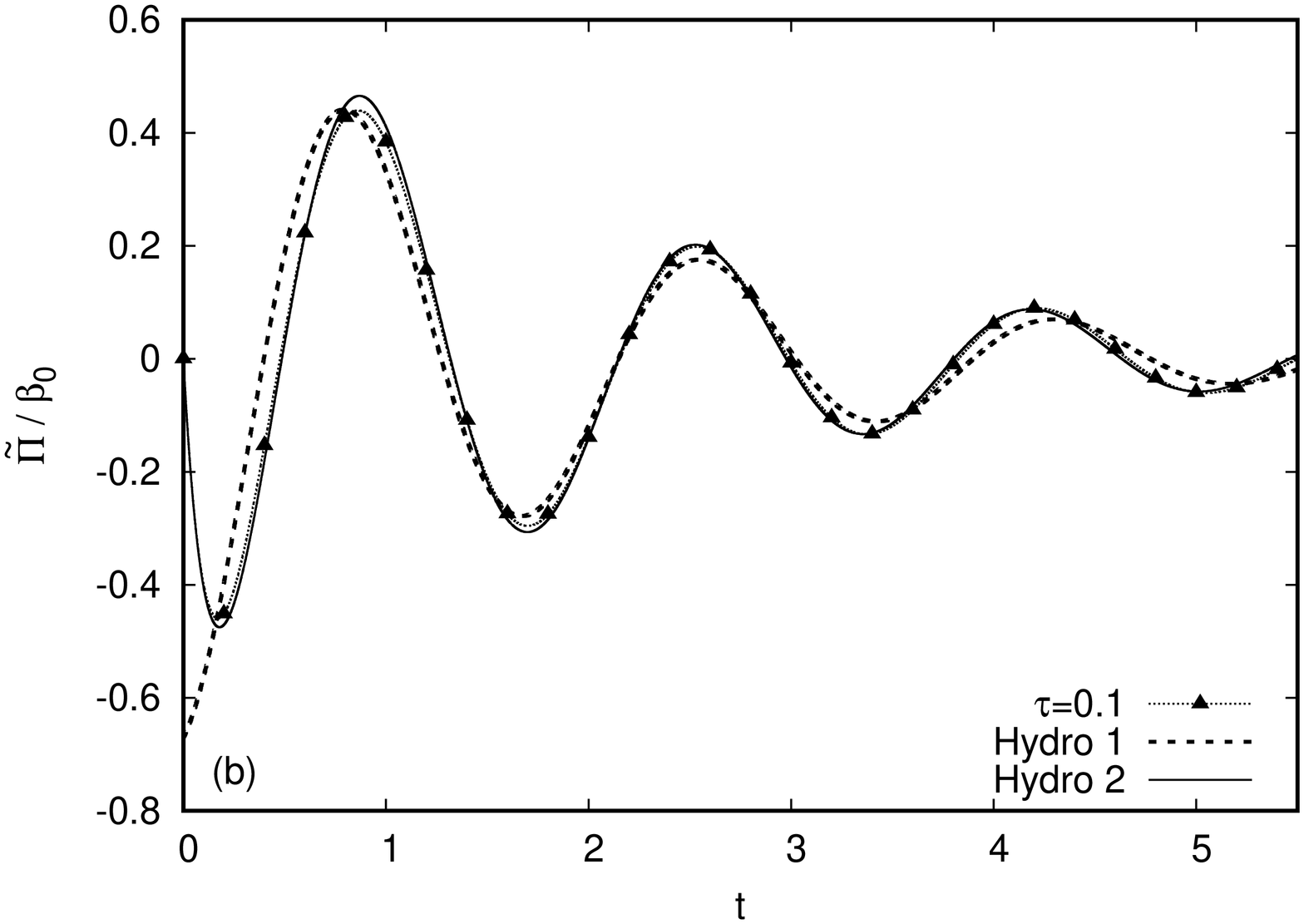}\\ 
\includegraphics[angle=270,width=0.48\linewidth]{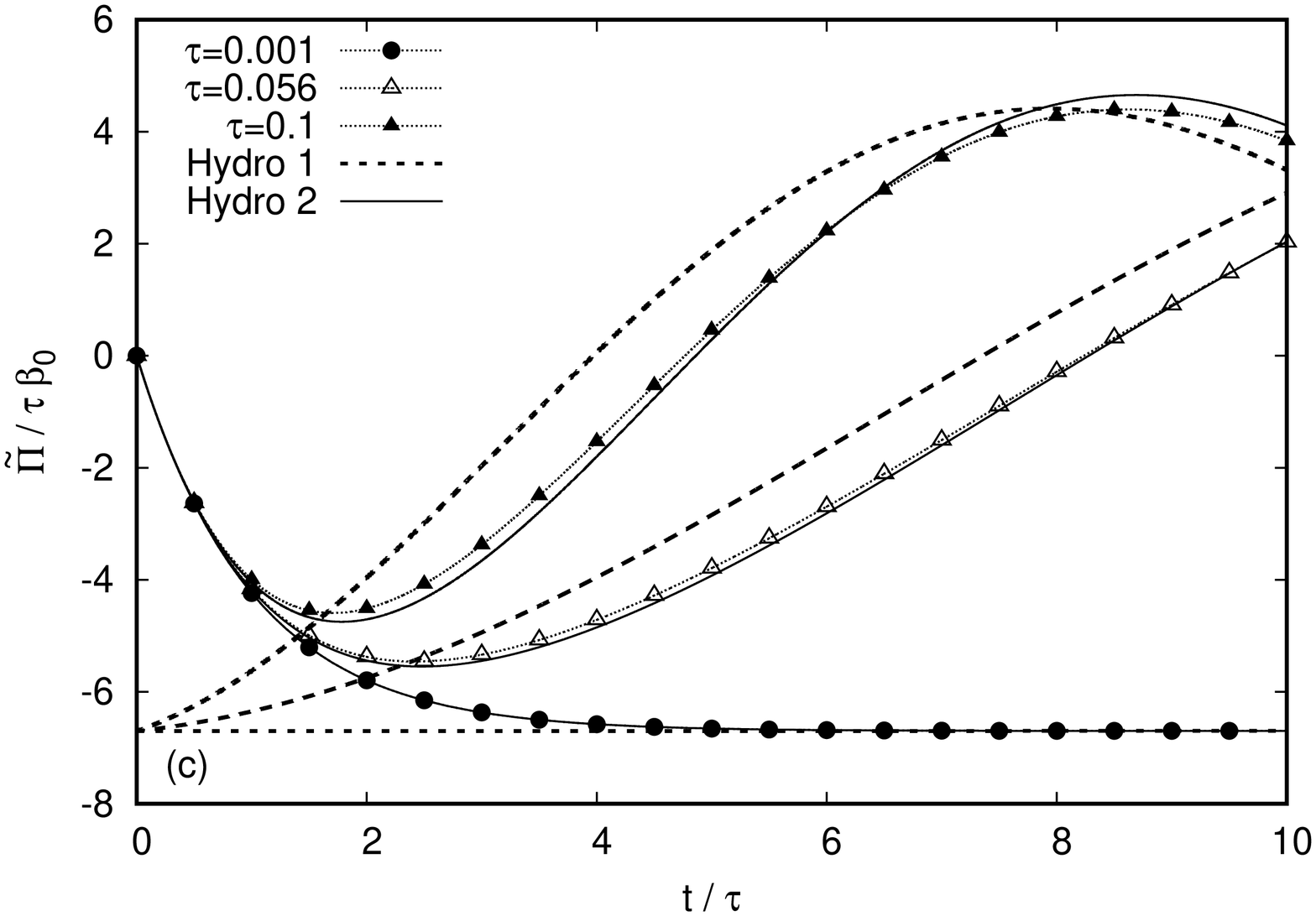} &
\includegraphics[angle=270,width=0.48\linewidth]{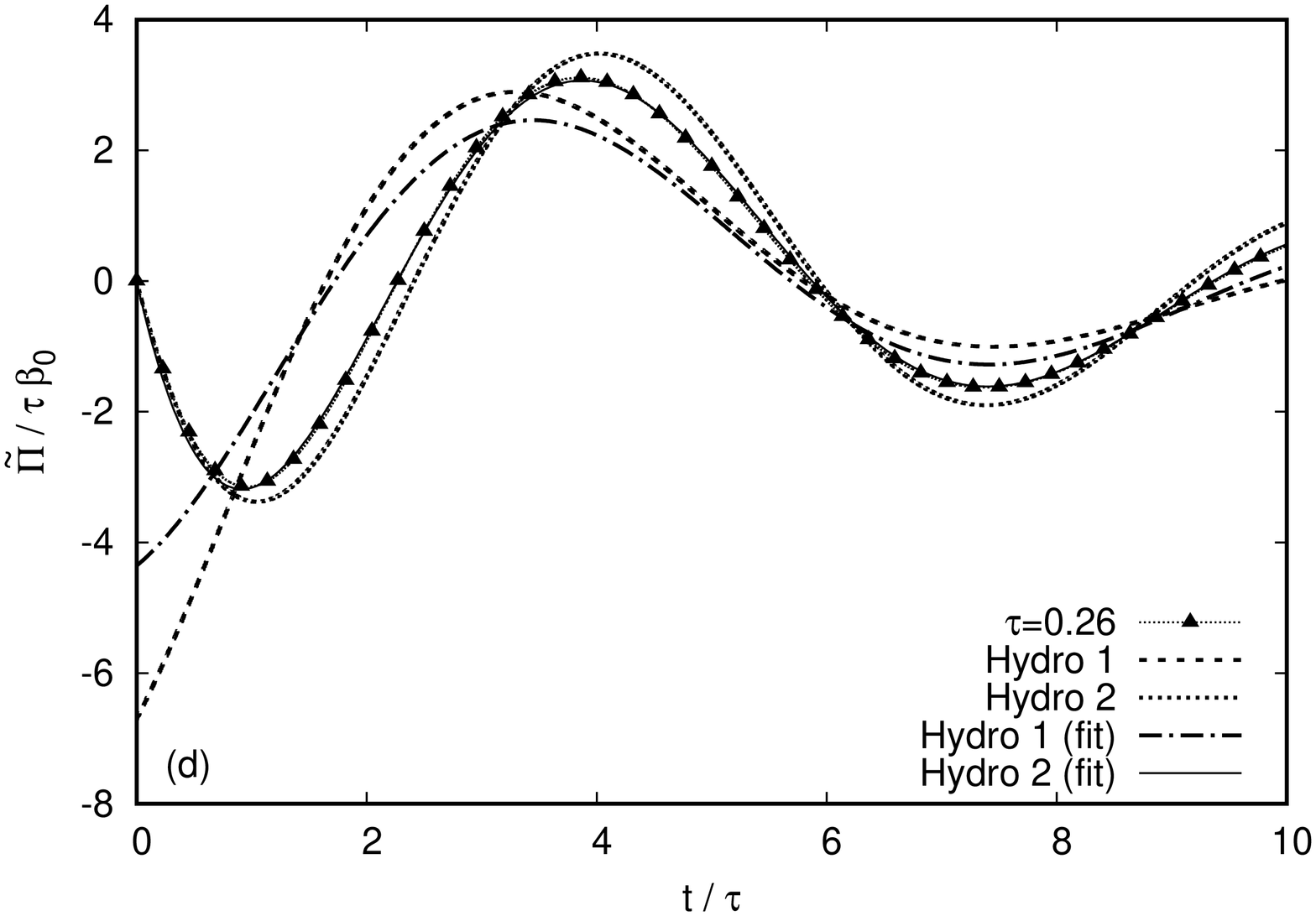} 
\end{tabular}
\end{center}
\caption{
(a),(b) Comparison between our numerical results (dotted lines and points) 
and the analytic expressions \eqref{eq:case1} and \eqref{eq:h2_case1}
obtained in the frame of the first- (dotted lines) and second-order 
(continuous lines) relativistic hydrodynamics for the evolution of 
$\widetilde{\Pi} / \beta_0$ ($\beta_0 = 10^{-3}$) at (a) $\tau = 0.056$  
and (b) $\tau = 0.1$.
(c) Evolution of $\widetilde{\Pi} / \tau \beta_0$ at various values of 
$\tau$ (to ease the comparison, the horizontal axis shows $t / \tau$). 
(d) Evolution of $\widetilde{\Pi} / \tau\beta_0$ at $\tau = 0.26$. The fitted curve 
corresponding to the first-order hydrodynamics 
is obtained by performing a nonlinear fit of Eq.~\eqref{eq:case1} 
using $\alpha_\eta$ and $\alpha_o$ as
free parameters. In the second-order case, the nonlinear fit is performed using Eq.~\eqref{eq:h2_case1}
by considering $\alpha_{\eta, r}$, $\alpha_{\eta, d}$, $\alpha_{\eta, o}$ and the ratio $\eta / \tau_{\Pi}$ 
as free parameters. All non-fitted anayltic curves are obtained using the Chapman-Enskog 
value for $\eta_0$ \eqref{eq:tcoeff_CE} and $\tau_{\Pi,0} = 1$ \eqref{eq:h2_tcoeff_CE}.
}
\label{fig:h2_case1-profiles}
\end{figure*}

\begin{figure}
\begin{center}
\begin{tabular}{c}
\includegraphics[angle=270,width=0.98\linewidth]{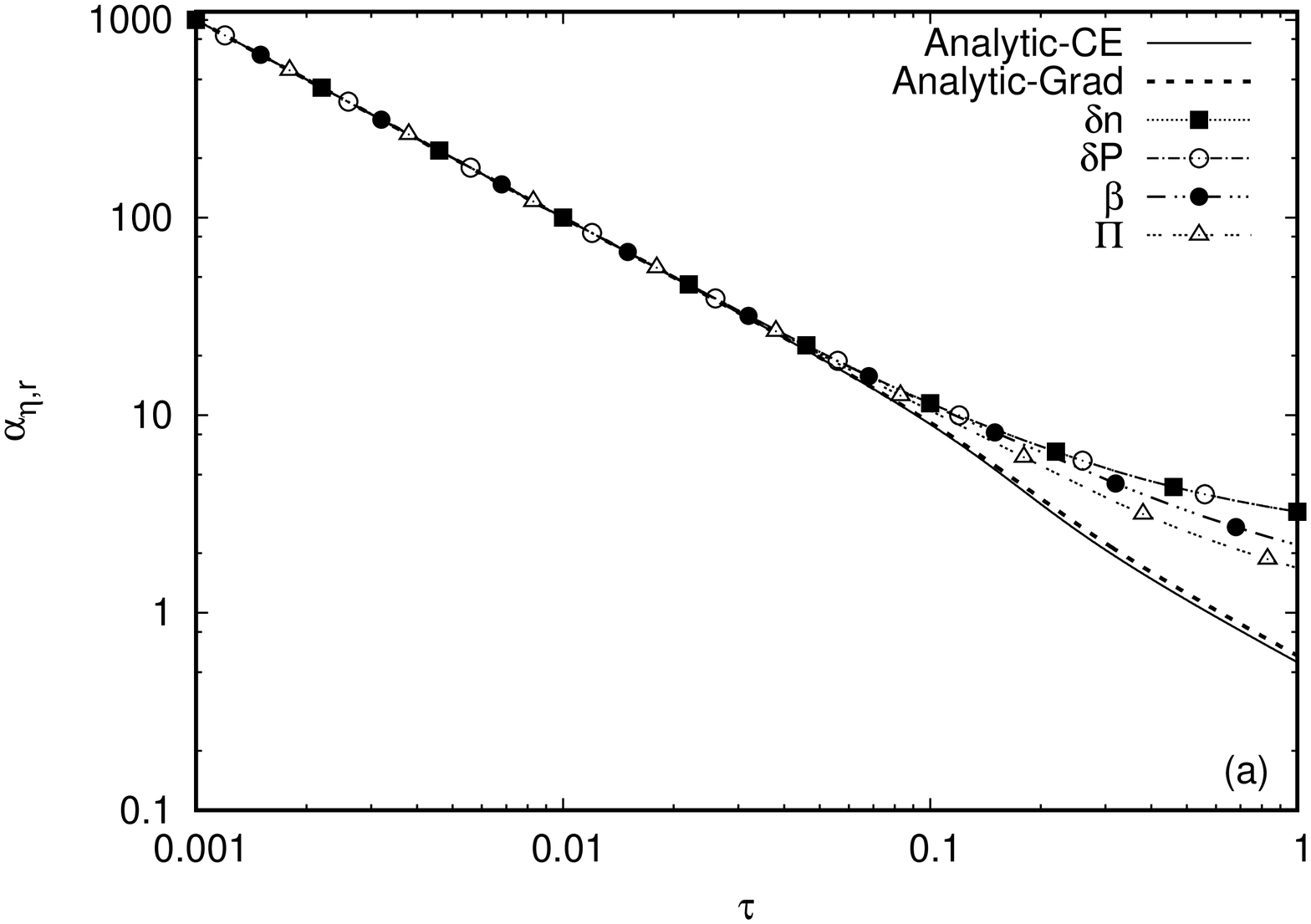} \\
\includegraphics[angle=270,width=0.98\linewidth]{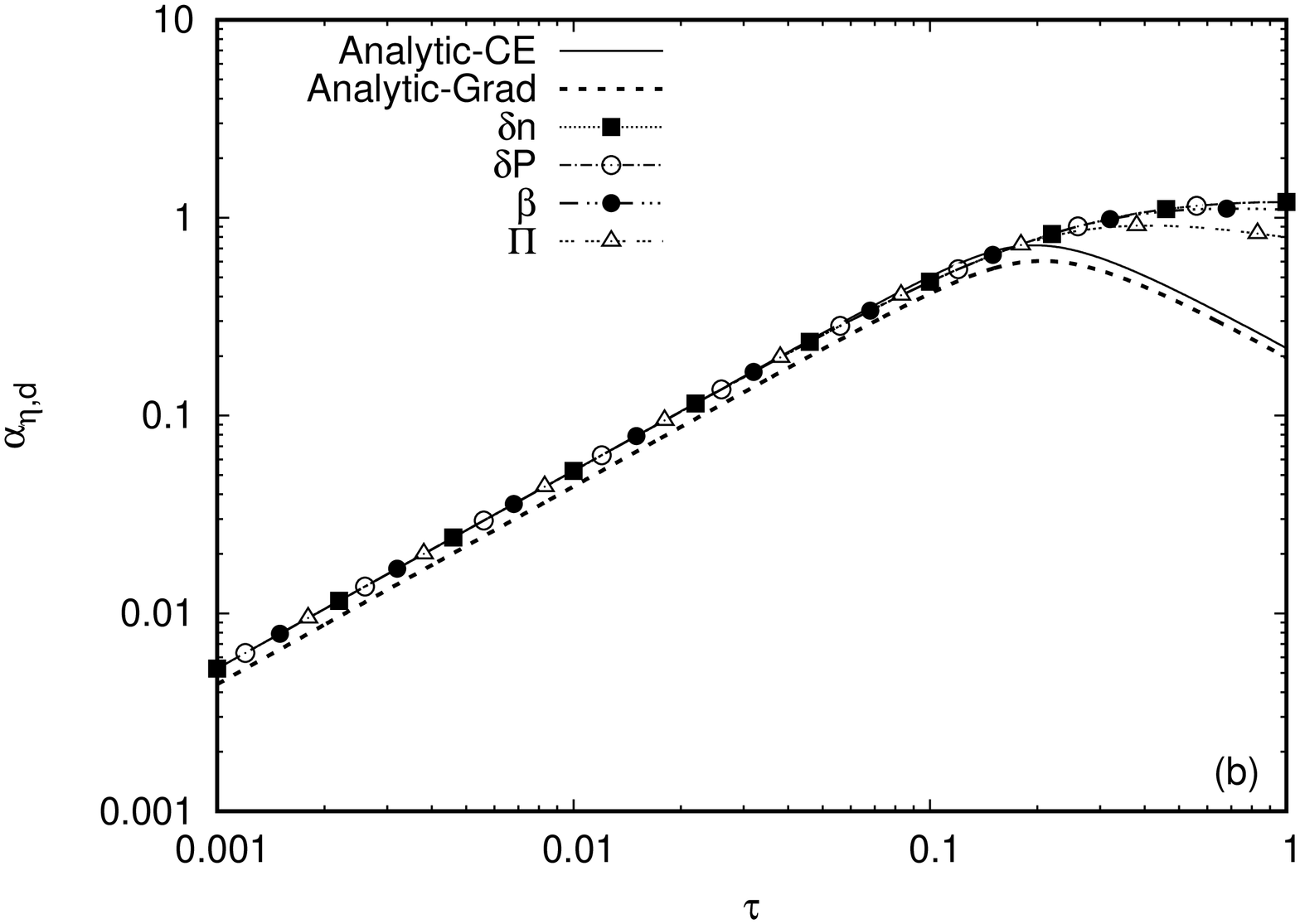} \\
\includegraphics[angle=270,width=0.98\linewidth]{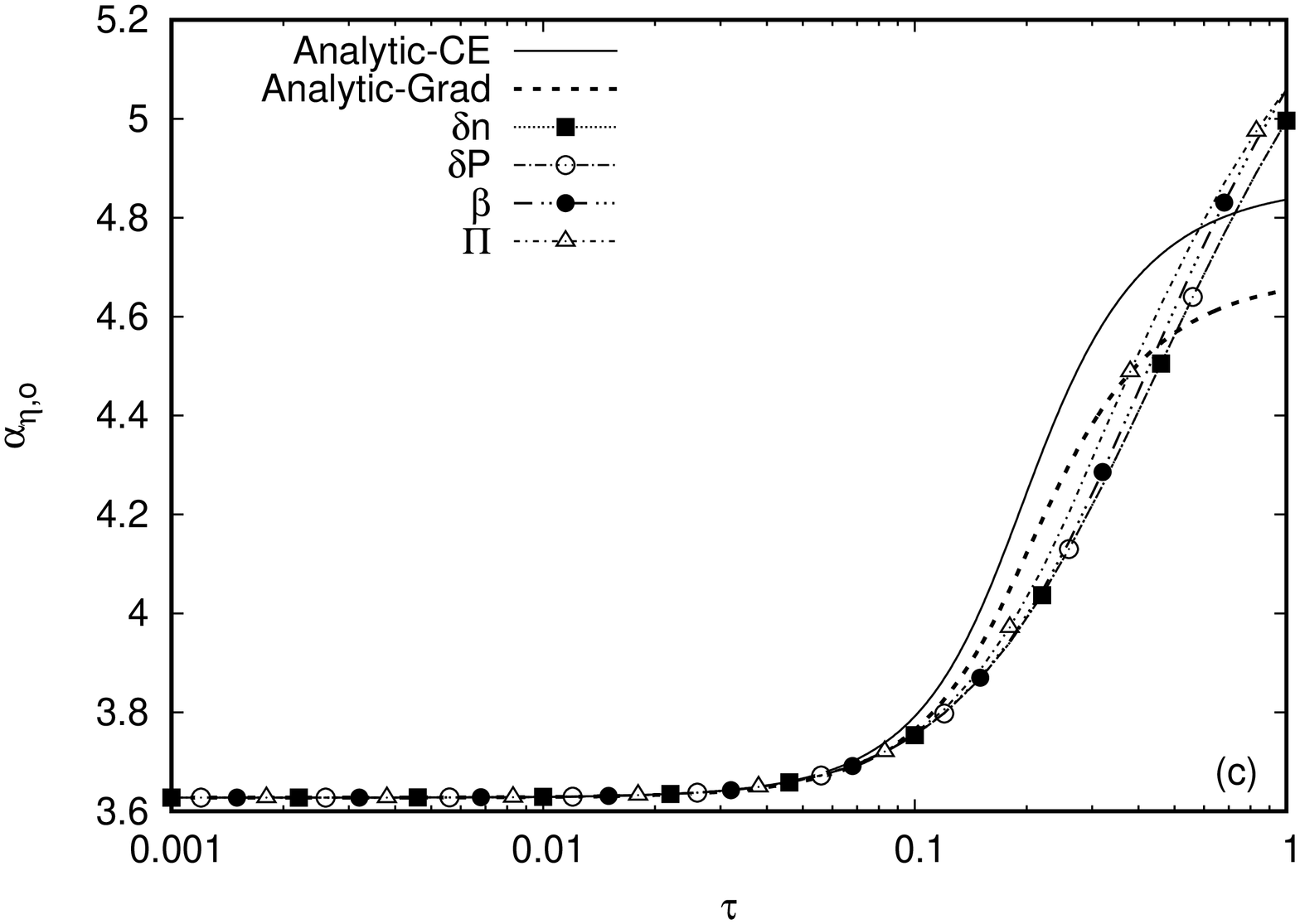} 
\end{tabular}
\end{center}
\caption{
Comparison with respect to $\tau$ between the analytic prediction \eqref{eq:h2_ae} and 
the numerical fit for the coefficients (a) $\alpha_{\eta, r}$, (b) $\alpha_{\eta, d}$ and 
(c) $\alpha_{\eta, o}$.
The analytic curves corresponding to the Chapman-Enskog procedure (continuous lines) and 
Grad moment method (dotted lines) are obtained by using the expressions 
\eqref{eq:tcoeff_CE} and \eqref{eq:tcoeff_G} for $\eta_0$ in Eq.~\eqref{eq:h2_ae}.
The numerical curves (dotted lines and points) are obtained by performing a nonlinear fit on the  
functional forms \eqref{eq:hydro2:sol} of $\widetilde{\beta}$, $\widetilde{\delta n}$, 
$\widetilde{\delta P}$ and $\widetilde{\Pi}$. The simulations were initialised according to 
{\it Case 1}, i.e. $\delta n_0 = \delta P_0 = 0$ and $\beta_0 = 10^{-3}$.
}
\label{fig:h2_case1-alphas}
\end{figure}

\begin{figure}
\begin{center}
\begin{tabular}{c}
\includegraphics[angle=270,width=0.98\linewidth]{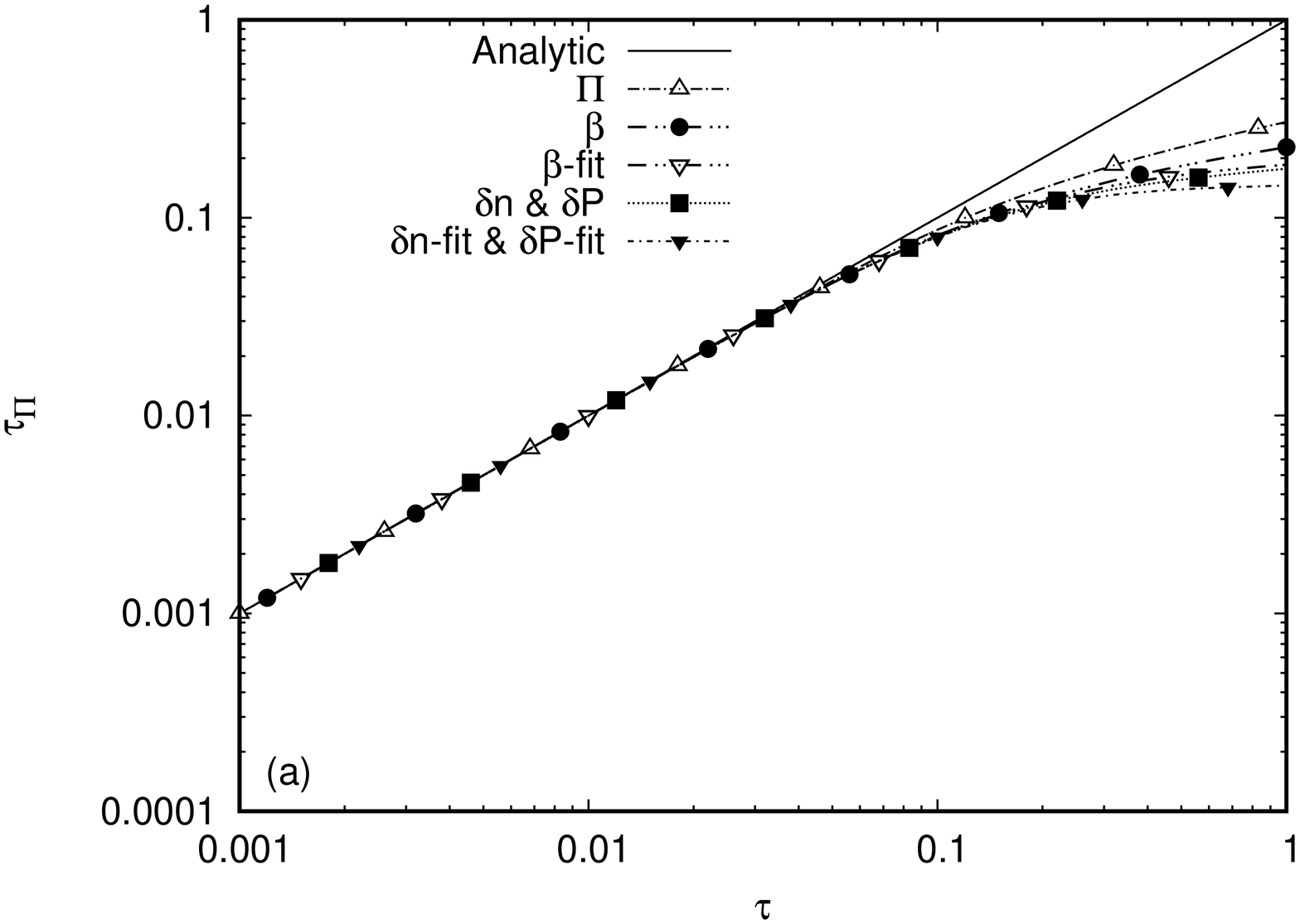} \\
\includegraphics[angle=270,width=0.98\linewidth]{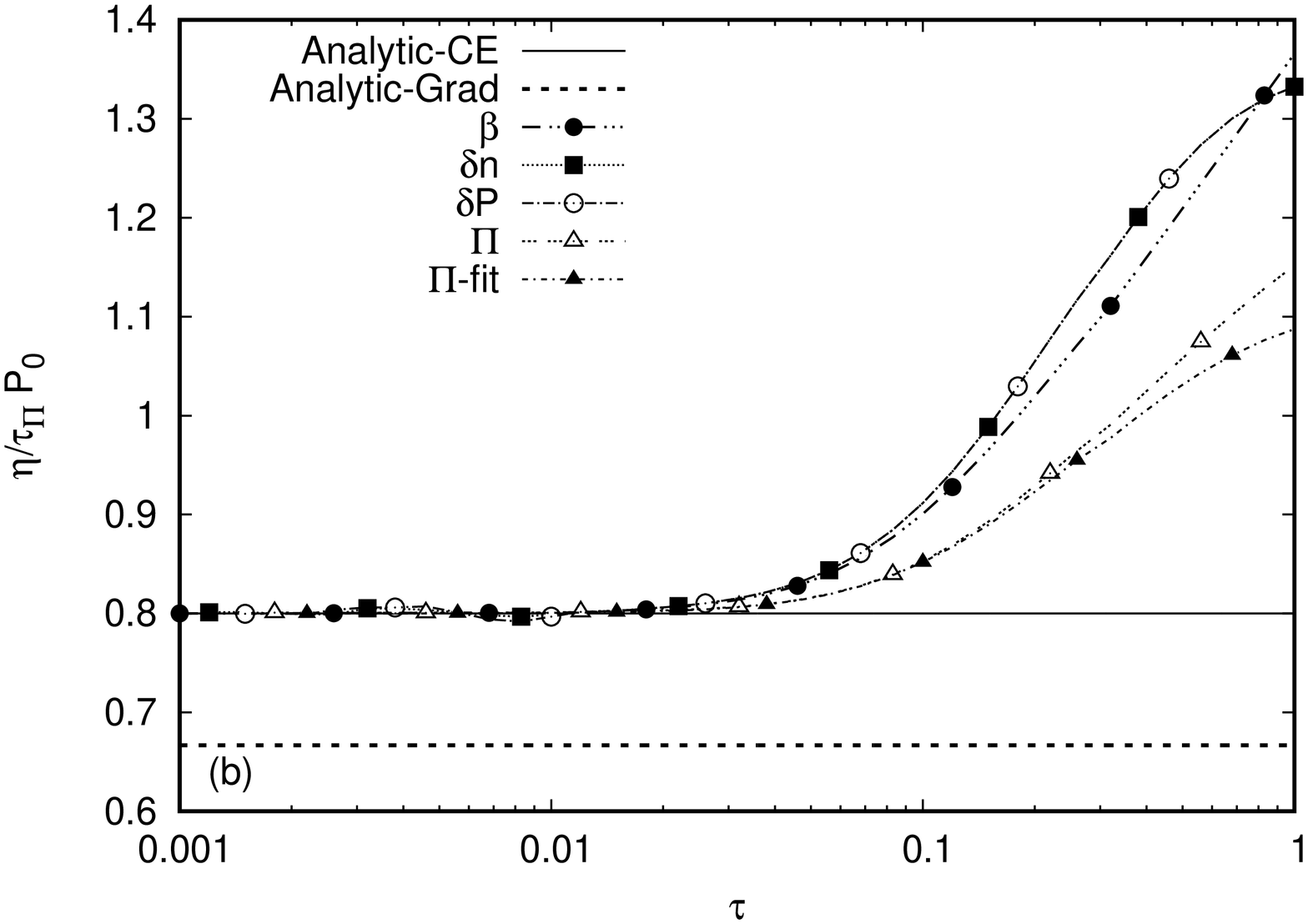} \\
\includegraphics[angle=270,width=0.98\linewidth]{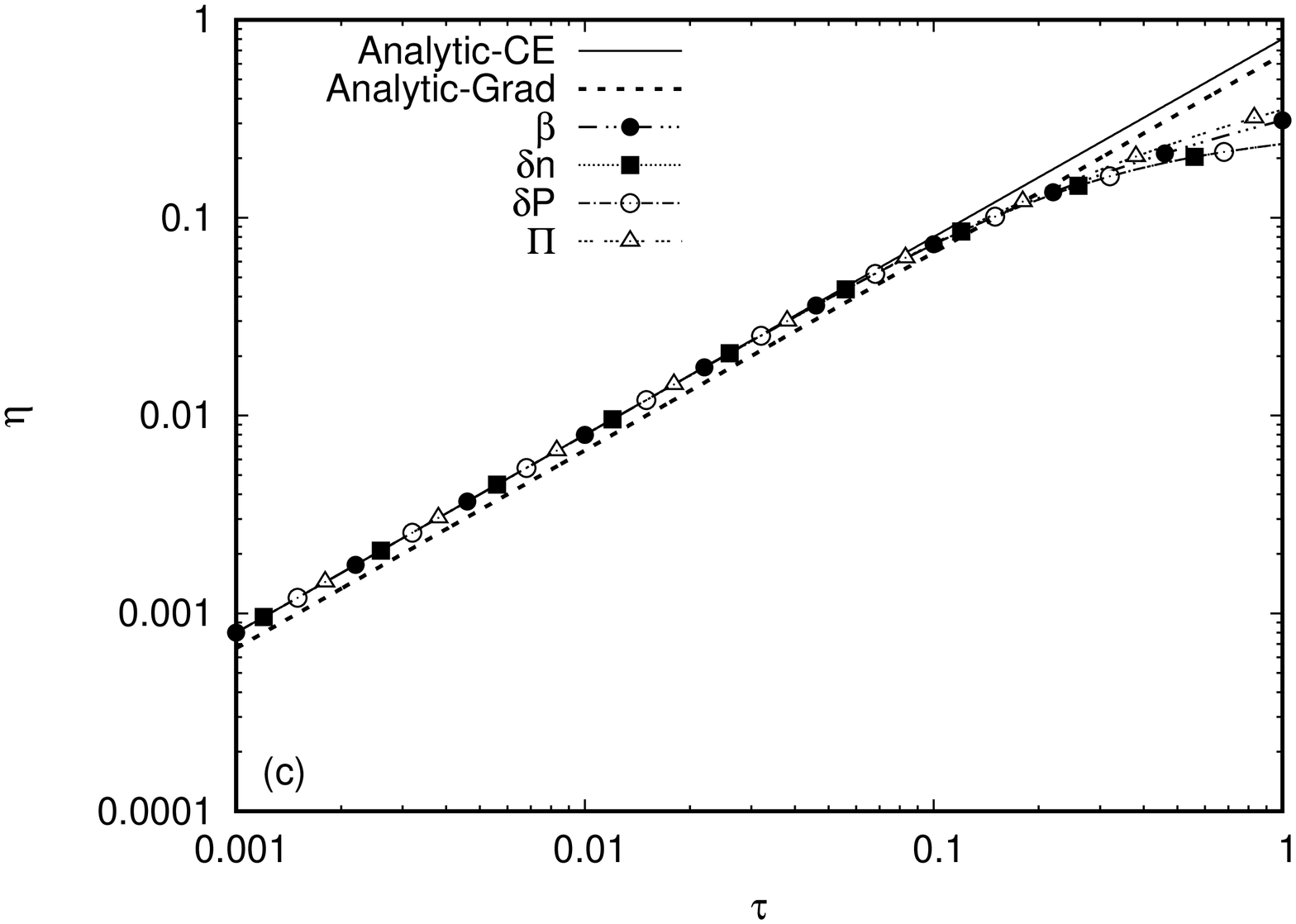} 
\end{tabular}
\end{center}
\caption{
Comparison with respect to $\tau$ between the analytic and numerical predictions for 
(a) the relaxation time $\tau_\Pi$; (b) the ratio $\eta / \tau_\Pi P_0$; (c) 
the coefficient of shear viscosity $\eta$.
The analytic expression for $\tau_{\Pi}$ (continuous line), given in Eq.~\eqref{eq:h2_tcoeff_CE}, is 
$\tau_{\Pi} = \tau$, while the analytic expression for $\eta$ 
is given in Eqs.~\eqref{eq:tcoeff_CE} and \eqref{eq:tcoeff_G} for the Chapman-Enskog (continuous line) 
and Grad (dotted lines) cases. The numerical results (dotted lines and points) are obtained as described in 
Sec.~\ref{sec:hydro2:case1}.
}
\label{fig:h2_case1-tcoeff}
\end{figure}

Setting $\delta n_0 = \delta P_0 = 0$ in Eq.~\eqref{eq:h2_init_soll} yields $\beta_{\lambda, s} = 0$, 
such that $\widetilde{M}_\lambda = 0$ for all $\widetilde{M} \in \{\widetilde{\beta}, \widetilde{\delta n}, 
\widetilde{\delta P}, \widetilde{q}, \widetilde{\Pi}\}$. 
The integration constants $\beta_{\eta, r}$, $\beta_{\eta, c}$ and $\beta_{\eta, s}$ can be found from 
Eq.~\eqref{eq:h2_init_sole} as follows:
\begin{widetext}
\begin{gather}
 \beta_{\eta, r} = - \frac{\alpha_{\eta, r}(1 - \alpha_{\eta, r} \tau_\Pi)}
 {\tau_\Pi[\alpha_{\eta, o}^2 + (\alpha_{\eta, d} - \alpha_{\eta, r})^2]} \beta_0,\qquad
 \beta_{\eta, c} = \beta_0 \left[1 + \frac{\alpha_{\eta, r}(1 - \alpha_{\eta, r} \tau_\Pi)}
 {\tau_\Pi[\alpha_{\eta, o}^2 + (\alpha_{\eta, d} - \alpha_{\eta, r})^2]} \right],\nonumber\\
 \beta_{\eta, s} = \frac{\beta_0}{\alpha_{\eta,o} \tau_\Pi} 
 \frac{(\alpha_{\eta, d}^2 + \alpha_{\eta, o}^2 - \alpha_{\eta, d} \alpha_{\eta, r})(1 - \alpha_{\eta, d} \tau_\Pi) - 
 \alpha_{\eta, r} \alpha_{\eta, o}^2 \tau_\Pi}{\alpha_{\eta, o}^2 + (\alpha_{\eta, d} - \alpha_{\eta, r})^2}.
\end{gather}
Substituting the above expressions in Eq.~\eqref{eq:h2_consteta} yields:
\begin{gather}
 \begin{pmatrix}
  \delta n_{\eta, r} \\
  \delta P_{\eta, r} \\
  \Pi_{\eta, r}
 \end{pmatrix} = -
 \begin{pmatrix}
  \delta n_{\eta, c} \\
  \delta P_{\eta, c} \\
  \Pi_{\eta, c}
 \end{pmatrix} = 
 \begin{pmatrix}
  -k n_0 (1 - \alpha_{\eta, r} \tau_\Pi) \\
  -4 k P_0 (1 - \alpha_{\eta, r} \tau_\Pi) / 3 \\
  4 k \eta \alpha_{\eta, r} / 3
 \end{pmatrix}
 \frac{\beta_0}{\tau_\Pi[\alpha_{\eta, o}^2 + (\alpha_{\eta, d} - \alpha_{\eta, r})^2]},\nonumber\\
 \begin{pmatrix}
  \delta n_{\eta, s}\\
  \delta P_{\eta, s}
 \end{pmatrix} =
 \begin{pmatrix}
  \delta n_{\eta, r}\\
  \delta P_{\eta, r}
 \end{pmatrix}
 \frac{\alpha_{\eta, r} - \alpha_{\eta, d} + (\alpha_{\eta, d}^2 + \alpha_{\eta, o}^2 - 
 \alpha_{\eta, d} \alpha_{\eta, r}) \tau_\Pi}{\alpha_{\eta, o}(1 - \alpha_{\eta, r} \tau_\Pi)}, \qquad
 \Pi_{\eta, s} = -\Pi_{\eta, r} \frac{\alpha_{\eta, d}^2 + \alpha_{\eta, o}^2 - \alpha_{\eta, d} \alpha_{\eta, r}}
 {\alpha_{\eta, o} \alpha_{\eta, r}}.\label{eq:h2_case1}
\end{gather}
\end{widetext}

Figure~\ref{fig:h2_case1-profiles}(a) shows the numerical results for the 
evolution of $\widetilde{\Pi}$ for $\beta_0 = 10^{-3}$ and $\tau = 0.056$, 
compared with the first- and second-order hydrodynamics solutions given in 
Eqs.~\eqref{eq:case1} and \eqref{eq:h2_case1}, respectively. The early 
time disagreement between the first-order hydrodynamics and numerical results 
becomes negligible when $t \gtrsim 2.5$, while the second-order solution is in 
excellent agreement with the numerical results at all values of $t$. 
At $\tau = 0.1$, Fig.~\ref{fig:h2_case1-profiles}(b) shows that the first-order 
hydrodynamics prediction remains in visible disagreement at large times, while 
the initial disagreement between the second-order solution and the numerical 
results becomes negligible for $t \gtrsim 2$.

The early-time validity of the solution \eqref{eq:h2_case1} is inspected in Fig.~\ref{fig:h2_case1-profiles}(c).
At small values of $\tau$,  $\widetilde{\Pi}$ relaxes from its initial vanishing value to the 
value predicted by the first-order expression \eqref{eq:case1} after a time $t \sim 5\tau$. 
For $\tau \gtrsim 0.05$, the first-order approximation becomes non-satisfactory at small values of $t$, 
since it lags behind the numerical solution. At small $\tau$, the second-order approximation is overlapped with the 
numerical solution for all values of $t$. An early-time discrepancy between the 
second-order prediction and the numerical result can be seen when $\tau \gtrsim 0.1$,
which however becomes negligible at large times, as shown in Fig.~\ref{fig:h2_case1-profiles}(b).
At sufficiently small values for $\tau$, 
Fig.~\ref{fig:h2_case1-profiles}(d) shows that this discrepancy arises when the
values of $\eta_0$ and $\tau_{\Pi,0}$ predicted by Eqs.~\eqref{eq:tcoeff_CE} and \eqref{eq:h2_tcoeff_CE} are used 
in the analytic expression. If instead, these values are treated as free parameters,
the functional form \eqref{eq:h2_case1} can be used to accurately represent $\widetilde{\Pi}$,
even at $\tau \simeq 0.26$.

Next, the validity of Eqs.~\eqref{eq:tcoeff_CE} and \eqref{eq:h2_tcoeff_CE} 
pertaining to the Chapman-Enskog expressions for $\eta_0$ and $\tau_{\Pi,0}$
is examined. For this purpose, a nonlinear fit of the functional forms in
Eq.~\eqref{eq:h2_case1} will be performed, where the coefficients 
$\alpha_{\eta, r}$, $\alpha_{\eta, d}$ and $\alpha_{\eta, o}$ are considered 
free parameters. In addition, the solutions for 
$\widetilde{\delta n}$ and $\widetilde{\delta P}$ depend explicitly on $\tau_\Pi$, while 
$\widetilde{\Pi}$ depends explicitly on $\eta / \tau_\Pi$. The inversion of 
Eq.~\eqref{eq:h2_ae} would allow $\eta$ and $\tau_{\Pi}$ to be written in 
terms of $\alpha_{\eta, r}$ and $\alpha_{\eta, d}$, but this operation 
is mathematically intractable. Thus, $\tau_\Pi$ will also be treated as 
a free parameter for $\widetilde{\delta n}$ and $\widetilde{\delta P}$,
while in the case of $\widetilde{\Pi}$, $\eta / \tau_\Pi$ will be considered
as an independent parameter. The results of the numerical fits for 
$\alpha_{\eta, r}$, $\alpha_{\eta, d}$ and $\alpha_{\eta, o}$ are presented in 
Fig.~\ref{fig:h2_case1-alphas}. In the case of the $\alpha_{\eta, r}$ 
coefficient, it can be seen that the analytic expression predicts a sharper 
decrease at large $\tau$ compared to the numerical results.
The shapes of $\alpha_{\eta, d}$ and $\alpha_{\eta, o}$ remain qualitatively 
similar to those obtained for $\alpha_d$ and $\alpha_o$ in the first-order 
theory, which are shown in Fig.~\ref{fig:reg-tau}. 
In the first-order theory, $\alpha_d$ is directly proportional to $\tau$, 
while the numerical results seem to indicate a saturation of $\alpha_d$ for 
$\tau \gtrsim 0.1$. This saturation can be seen also for $\alpha_{\eta, d}$, 
but in this case, the analytic expression predicts that $\alpha_{\eta, d}$ 
decreases with $\tau$. Another notable difference can be seen in the analytic 
prediction for $\alpha_{\eta, o}$, which in the second order case qualitatively 
follows the numerical results ($\alpha_{\eta,o}$ increases at large $\tau$) compared 
to the first-order case, when $\alpha_o$ is predicted to decrease at large $\tau$.

Finally, the analysis of the dependence of the relaxation time $\tau_\Pi$ 
and of the ratio $\eta / \tau_\Pi P_0$ on $\tau$ is presented below. 
The curves shown in Fig.~\ref{fig:h2_case1-tcoeff} 
represent three types of results. The first type corresponds to the various analytic 
predictions, which are represented as follows: 
in Fig.~\ref{fig:h2_case1-tcoeff}(a), $\tau_\Pi$ \eqref{eq:h2_tcoeff_CE} is 
shown using a continuous line; 
in Figs.~\ref{fig:h2_case1-tcoeff}(b) and \ref{fig:h2_case1-tcoeff}(c), 
the Chapman-Enskog and Grad predictions for $\eta / P_0 \tau_\Pi$ and 
$\eta$ are shown using continuous and dotted lines. 
The second type of results are obtained using the nonlinear fit procedure described 
in the previous paragraph for $\tau_\Pi$ (obtained from $\widetilde{\beta}$, 
$\widetilde{\delta n}$ and $\widetilde{\delta P}$) 
and for $\eta / \tau_\Pi P_0$ (obtained from $\widetilde{\Pi}$), being 
labelled using the suffix ``-fit''. 
The third type of results are obtained by numerically finding the values of 
$\tau_\Pi$ and $\eta / \tau_\Pi P_0$ for which the roots of Eq.~\eqref{eq:h2_ae_eq} 
correspond to the values of $\alpha_{\eta, r}$ and $\alpha_{\eta, d}$ obtained 
through the nonlinear fit procedure 
(the value of $\alpha_{\eta, o}$ is not taken into account in this algorithm).
As expected, the relaxation time $\tau_\Pi$ stops increasing linearly with $\tau$ when 
$\tau \gtrsim 0.1$. Figure~\ref{fig:h2_case1-tcoeff}(b) 
shows that the ratio $\eta / \tau_\Pi P_0$ increases with $\tau$, which seems 
to indicate that the increase of the shear viscosity of the gas with respect to 
$\tau$ is of higher order than the linear prediction of the first-order theory
\eqref{eq:tcoeff}. However, Fig.~\ref{fig:h2_case1-tcoeff}(c) 
shows that in fact $\eta$ (obtained by multiplying the results for 
$\eta / \tau_\Pi P_0$ and $\tau_\Pi$ obtained as explained above) 
increases at a sub-linear rate with respect to $\tau$ when $\tau \gtrsim 0.1$. 
This result is consistent with the one obtained in Fig.~\ref{fig:reg-tau} in the 
first-order case. When $\tau \lesssim 0.1$, the numerical results favor the 
Chapman-Enskog prediction for the transport coefficients, as well as 
the relations \eqref{eq:h2_tcoeff_CE}.

\subsection{Non-adiabatic flow}\label{sec:hydro2:case2}

\begin{figure*}
\begin{center}
\begin{tabular}{ccc}
\includegraphics[angle=270,width=0.32\linewidth]{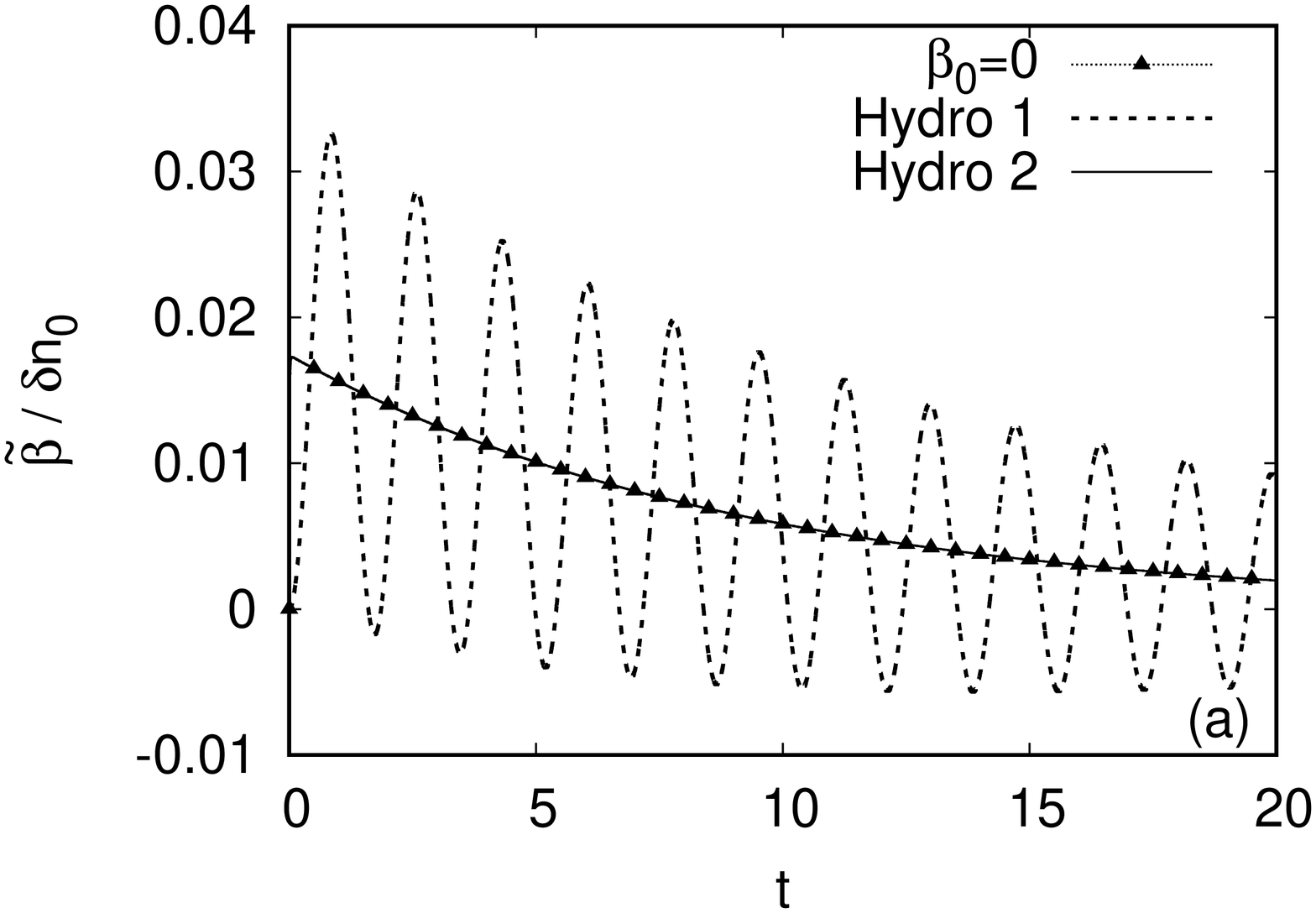} &
\includegraphics[angle=270,width=0.32\linewidth]{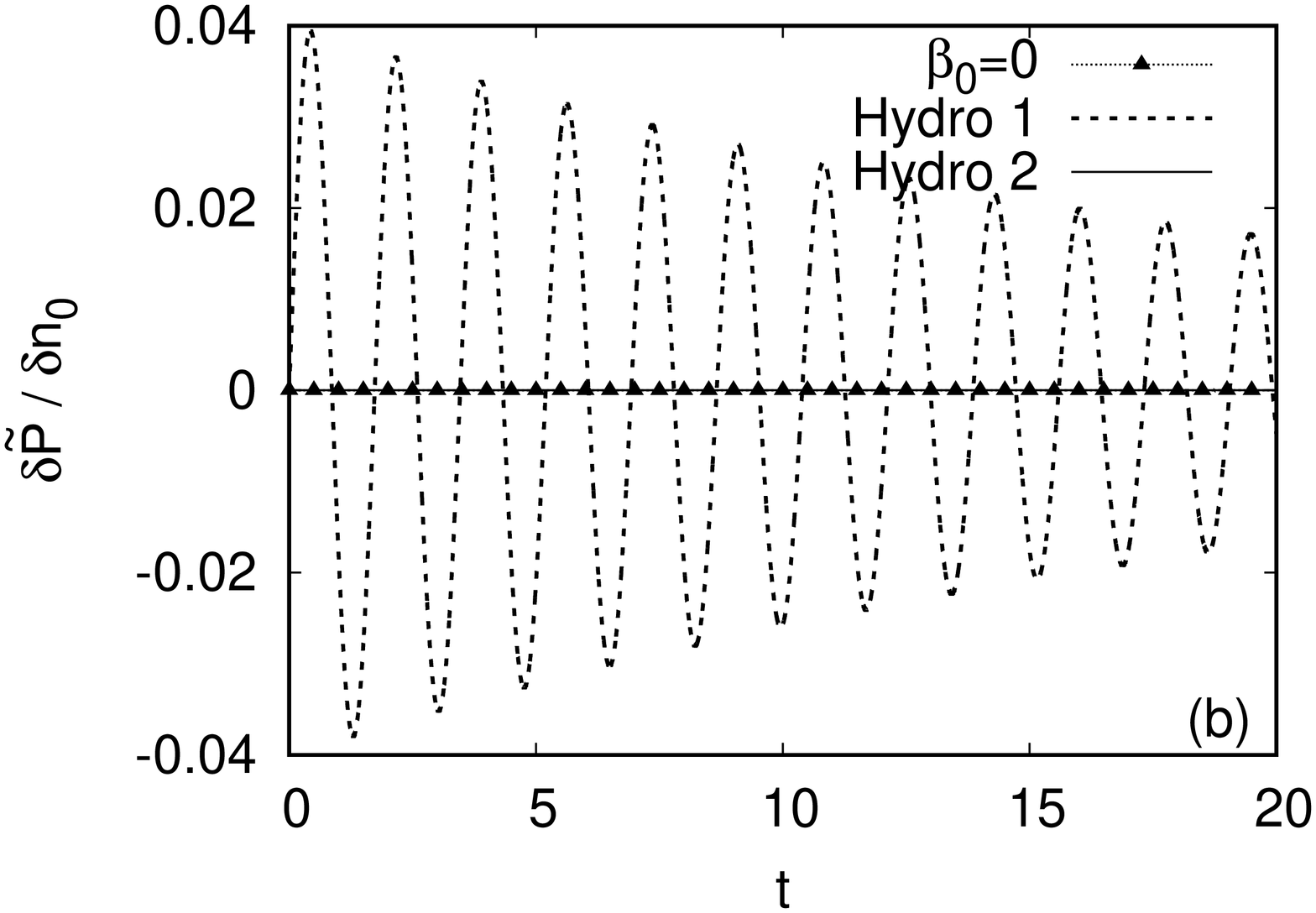} &
\includegraphics[angle=270,width=0.32\linewidth]{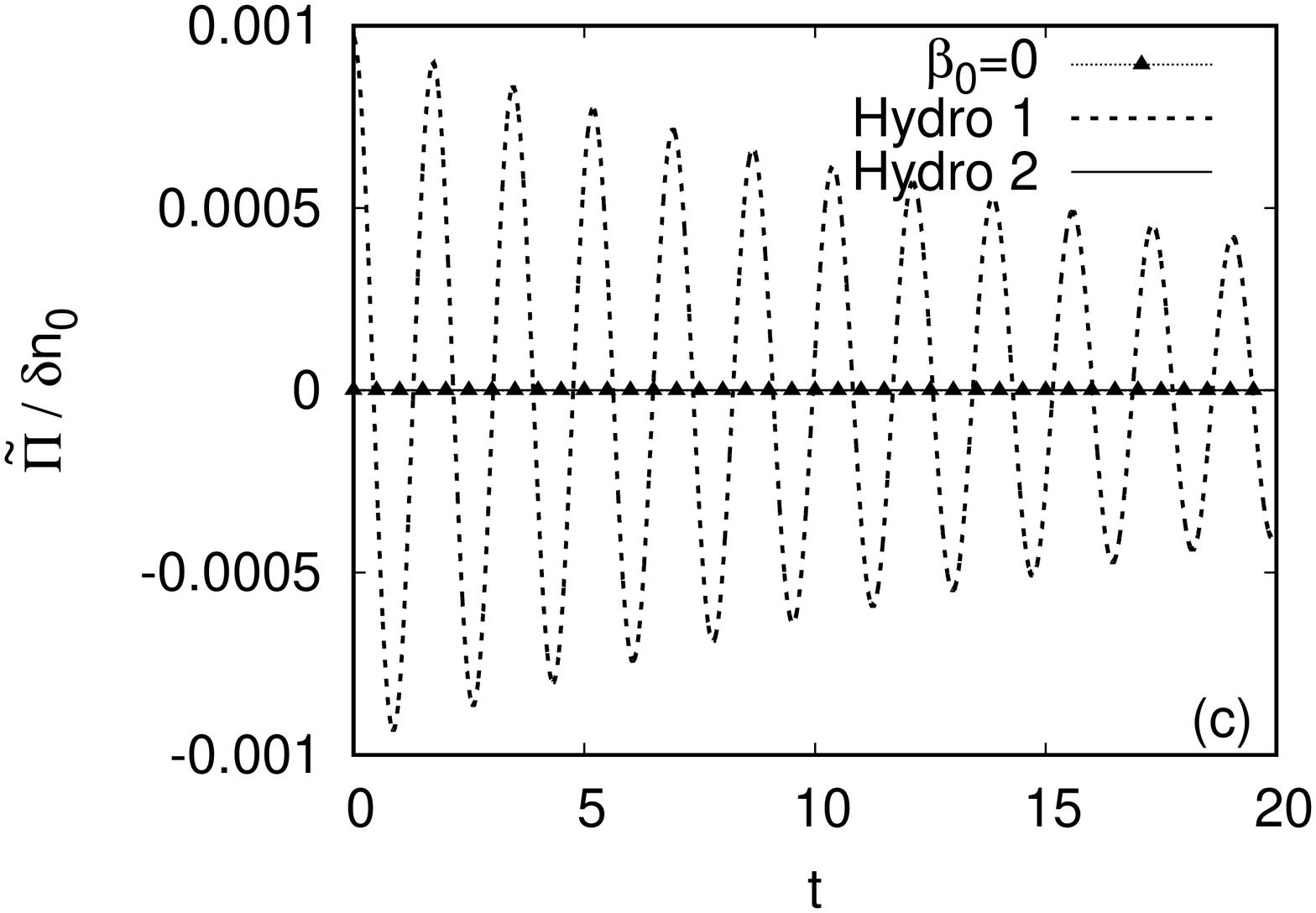} \\
\includegraphics[angle=270,width=0.32\linewidth]{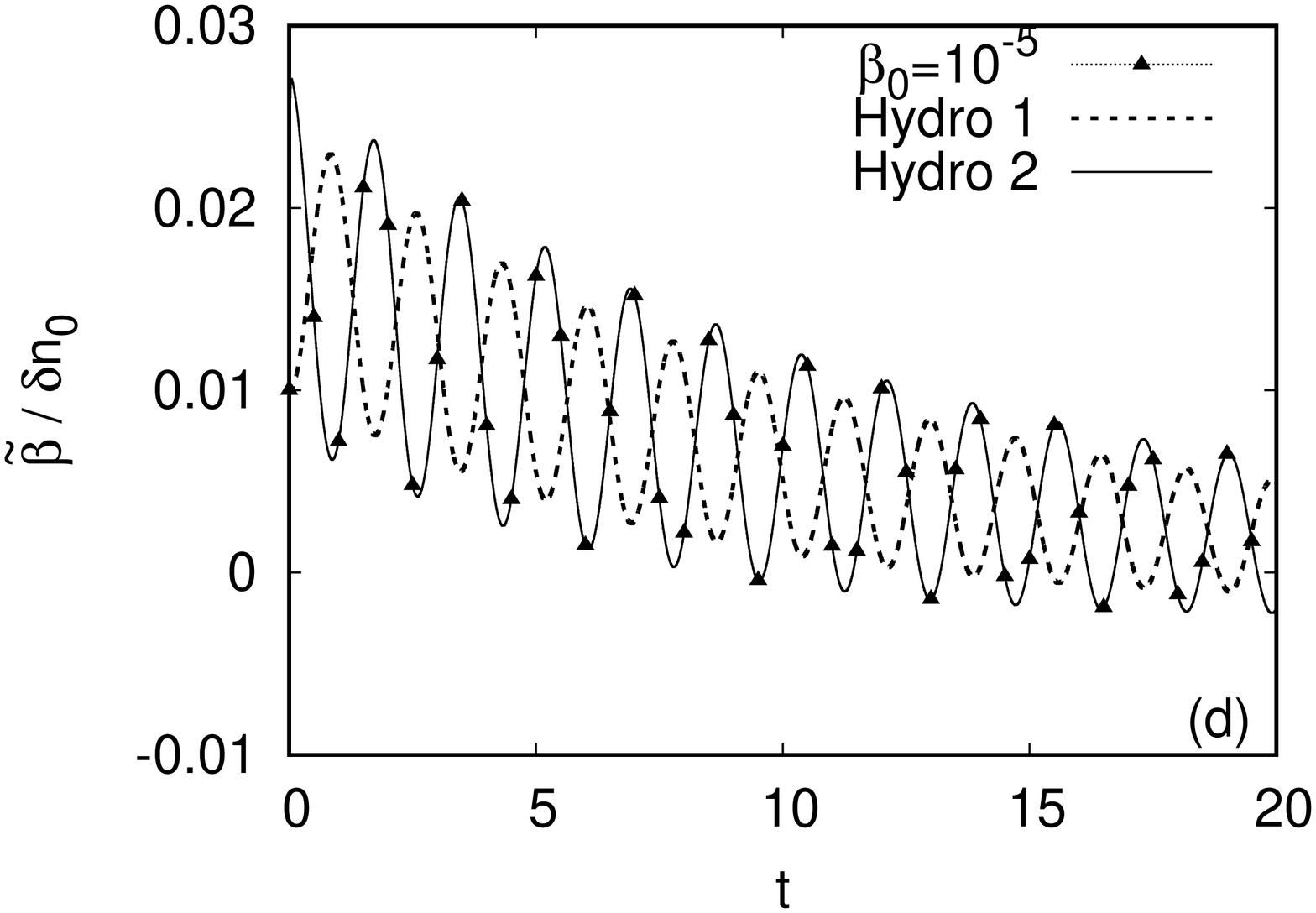} &
\includegraphics[angle=270,width=0.32\linewidth]{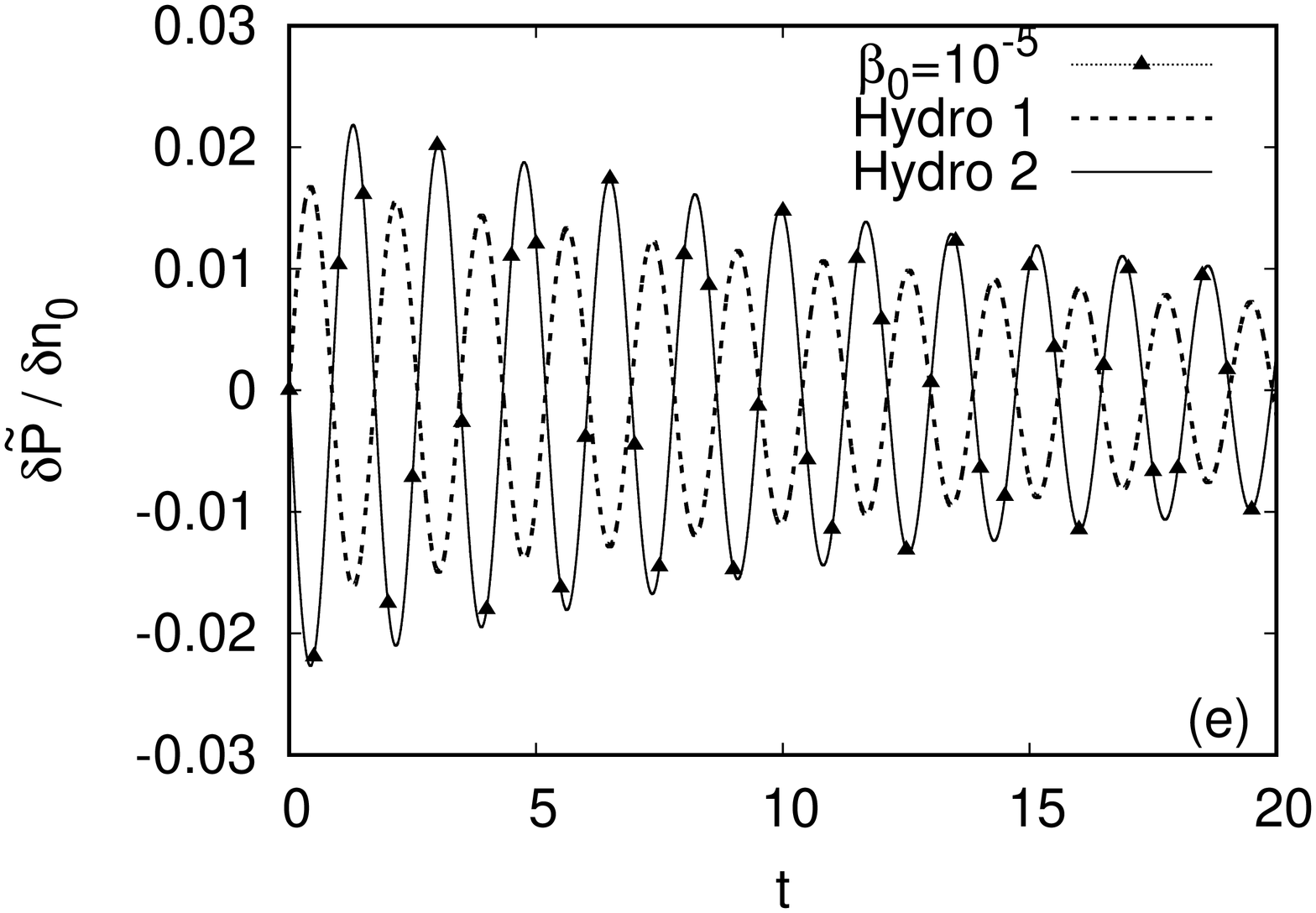} &
\includegraphics[angle=270,width=0.32\linewidth]{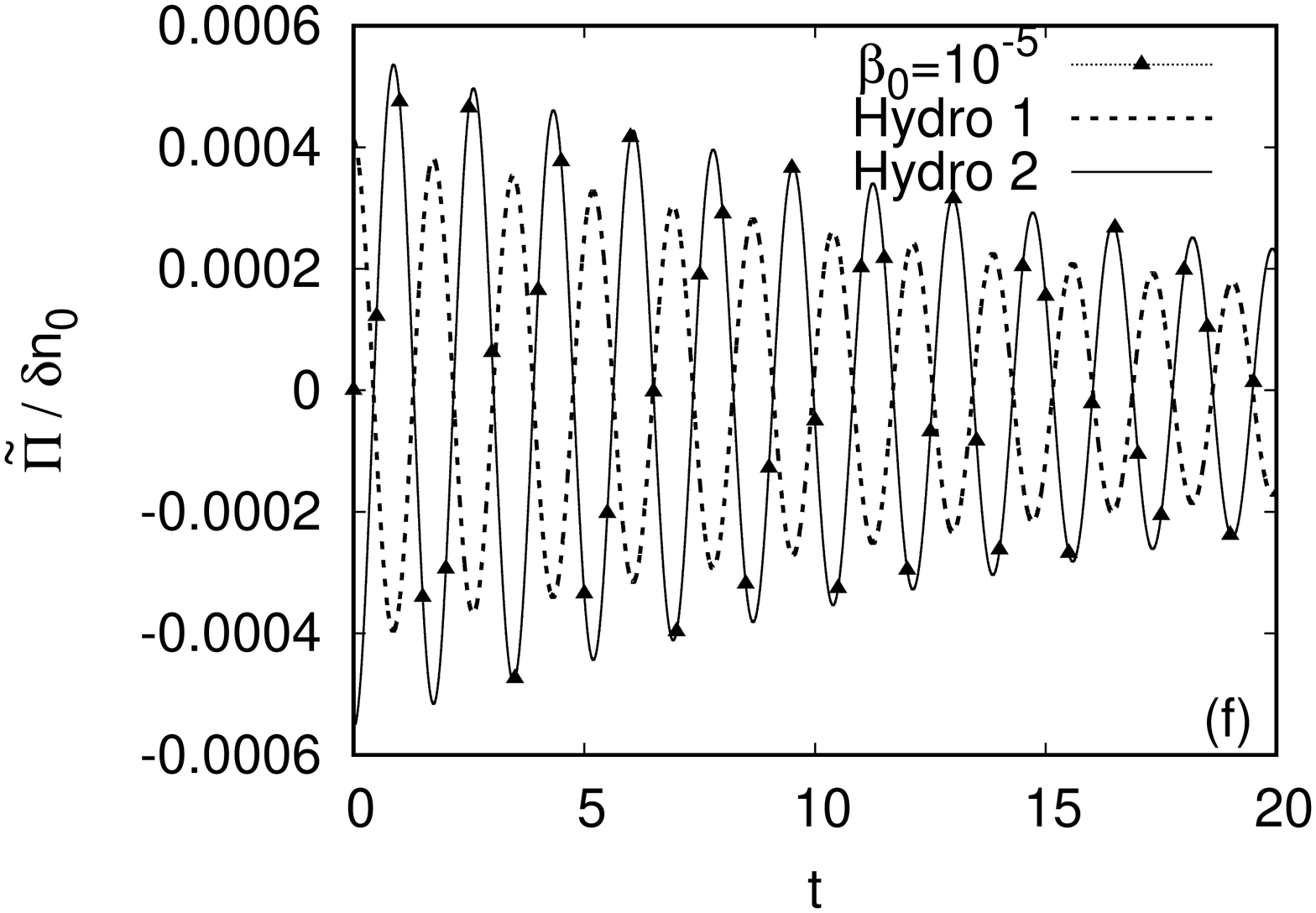} \\
\includegraphics[angle=270,width=0.32\linewidth]{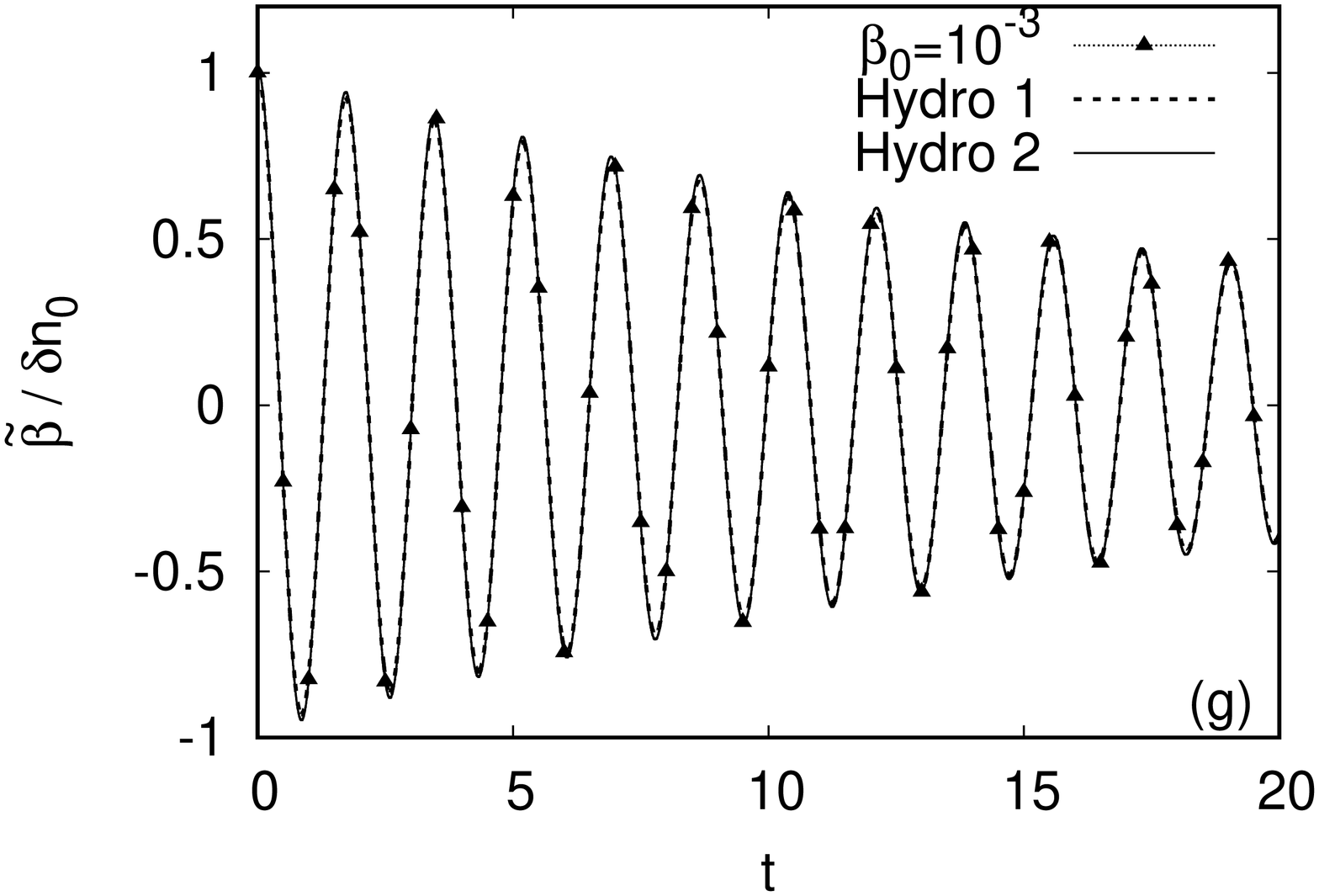} &
\includegraphics[angle=270,width=0.32\linewidth]{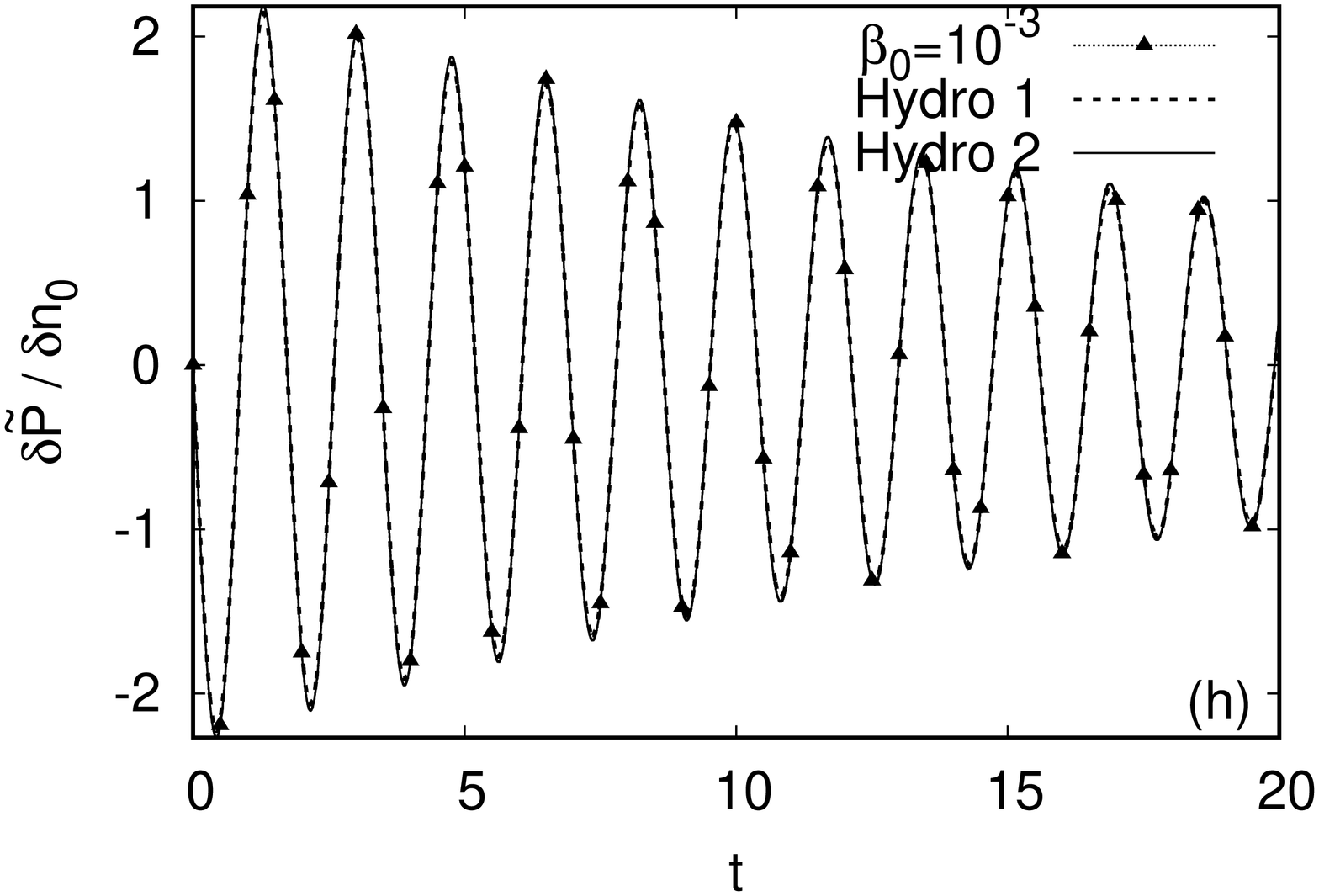} &
\includegraphics[angle=270,width=0.32\linewidth]{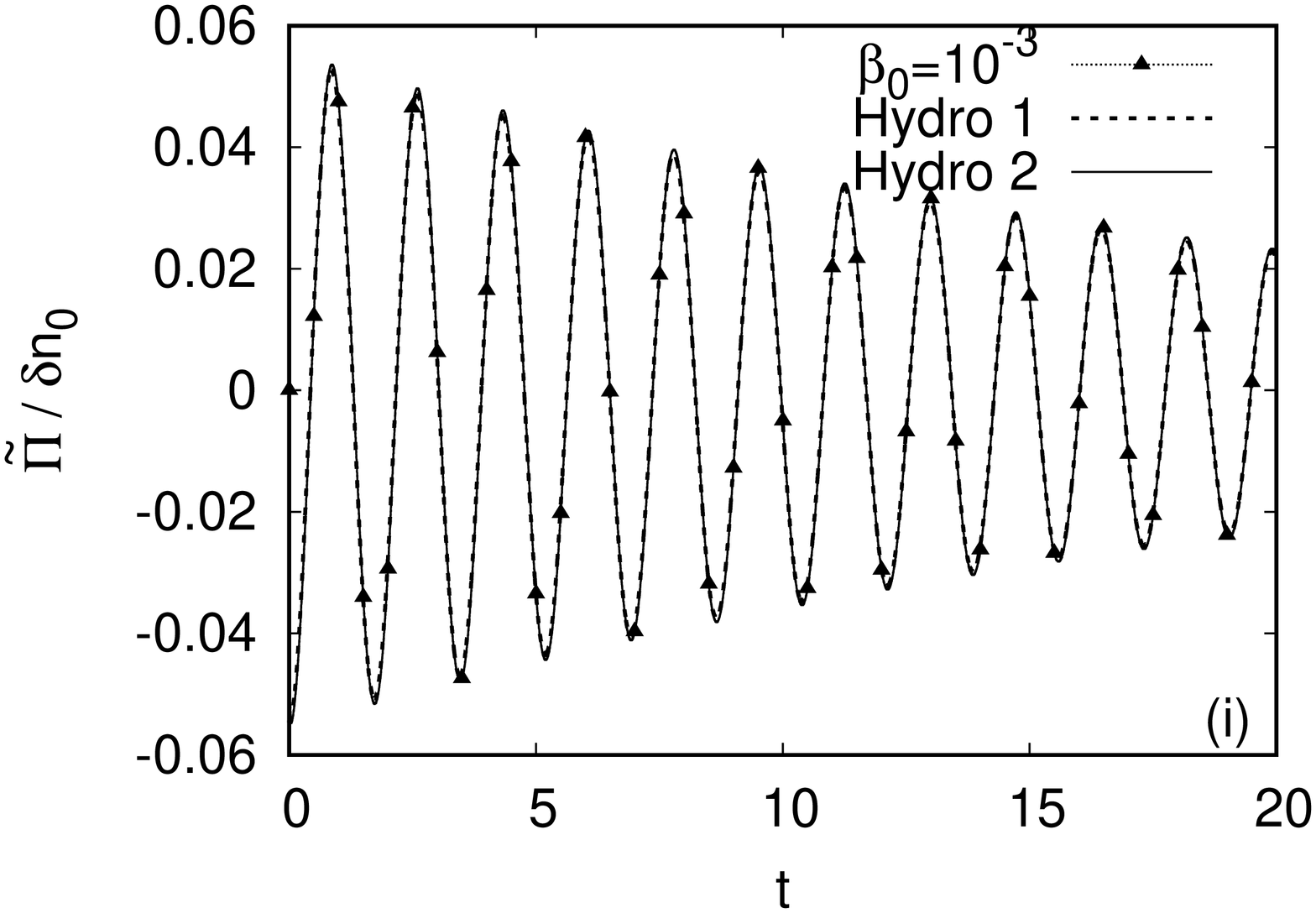} 
\end{tabular}
\end{center}
\caption{
Comparison between the numerical results (dotted lines and points) for 
$\widetilde{\beta}$ (first column), $\widetilde{\delta P}$ (second column) 
and $\widetilde{\Pi}$ (third column) and the corresponding analytic 
predictions of the first-order (dotted lines) and second-order (continuous lines) 
hydrodynamics when $\tau_\Pi = \tau$ and the Chapman-Enskog value 
\eqref{eq:tcoeff_CE} for $\lambda$ is employed.
The initial conditions for the plots on the first line correspond to 
{\it Case 2b}, i.e. $\beta_0 = \delta P_0 = 0$ and $\delta n_0 = 10^{-3}$.
On the second line, $\beta_0$ is increased to $10^{-5}$ and on the third line,
$\beta_0 = 10^{-3}$. All curves are normalised with respect to $\delta n_0 = 10^{-3}$ 
and $\tau = 0.0083$ was used throughout the simulations.
}
\label{fig:h2_case2-profiles1}
\end{figure*}

The non-adiabatic case can be analysed when the system is initialized 
according to {\it Case 2b}, when $\beta_0 = \delta P_0 = 0$. 
This case is particularly simple since, 
according to Eq.~\eqref{eq:h2_init_sole}, $\beta_{\eta, *} = 0$ for all 
$\eta \in \{r, c, s\}$. 
The only non-vanishing integration 
constants are $\beta_{\lambda,s}$ (overdamped case) and 
$\overline{\beta}_{\lambda, s}$ (underdamped case), which can be found from Eq.~\eqref{eq:h2_init_soll}:
\begin{equation}
 \begin{pmatrix}
  \beta_{\lambda, s}\\
  \overline{\beta}_{\lambda, s}
 \end{pmatrix} =
 \begin{pmatrix}
  (\alpha_{\lambda,d}^2 - \alpha_{\lambda,o}^2) / \alpha_{\lambda, o}\\
  (\overline{\alpha}_{\lambda,d}^2 + \overline{\alpha}_{\lambda,o}^2) / \overline{\alpha}_{\lambda, o}
 \end{pmatrix}
 \frac{\delta n_0}{k n_0}.
\end{equation}
Noting that $\widetilde{\delta P} = \widetilde{\Pi} = 0$ by virtue of Eq.~\eqref{eq:hydro2:sol},
the full solution in the overdamped case reads:
\begin{subequations}\label{eq:h2_case2}
\begin{align}
 \widetilde{\beta} =& \frac{\delta n_0}{kn_0}
 \frac{\alpha_{\lambda,d}^2 - \alpha_{\lambda,o}^2}{\alpha_{\lambda, o}} e^{-\alpha_{\lambda, d} t} 
 \sinh \alpha_{\lambda, o} t,\nonumber\\
 \widetilde{\delta n} =& \delta n_0\left(\cosh\alpha_{\lambda, o} t + 
 \frac{\alpha_{\lambda, d}}{\alpha_{\lambda, o}} \sinh \alpha_{\lambda, o} t\right) 
 e^{-\alpha_{\lambda, d} t},\nonumber\\
 \widetilde{q} =& - 4 P_0 \frac{\delta n_0}{kn_0}
 \frac{\alpha_{\lambda,d}^2 - \alpha_{\lambda,o}^2}{\alpha_{\lambda, o}} e^{-\alpha_{\lambda, d} t} 
 \sinh \alpha_{\lambda, o} t,\label{eq:h2_case2OD}
\end{align}
while in the underdamped case, the solution reads:
\begin{align}
 \widetilde{\beta} =& \frac{\delta n_0}{kn_0}
 \frac{\overline{\alpha}_{\lambda,d}^2 + \overline{\alpha}_{\lambda,o}^2}{\overline{\alpha}_{\lambda, o}} 
 e^{-\overline{\alpha}_{\lambda, d} t} \sin \overline{\alpha}_{\lambda, o} t,\nonumber\\
 \widetilde{\delta n} =& \delta n_0\left(\cos\overline{\alpha}_{\lambda, o} t + 
 \frac{\overline{\alpha}_{\lambda, d}}{\overline{\alpha}_{\lambda, o}} 
 \sin \overline{\alpha}_{\lambda, o} t\right) e^{-\overline{\alpha}_{\lambda, d} t},\nonumber\\
 \widetilde{q} =& - 4 P_0 \frac{\delta n_0}{kn_0}
 \frac{\overline{\alpha}_{\lambda,d}^2 + \overline{\alpha}_{\lambda,o}^2}{\overline{\alpha}_{\lambda, o}} 
 e^{-\overline{\alpha}_{\lambda, d} t} \sin \overline{\alpha}_{\lambda, o} t.\label{eq:h2_case2UD}
\end{align}
\end{subequations}

In the first-order theory, Eqs.~\eqref{eq:init_sol} and \eqref{eq:delta_aux3} predict that, 
when $\beta_0 = \delta P_0 = 0$,
$\beta_c$ and $\beta_s$ become proportional to $\delta n_0$, which 
also implies that $\widetilde{\Pi}$ and $\widetilde{\delta P}$ are non-zero. This 
prediction is in contradiction with the second-order theory results. This discrepancy 
holds for any value of $\tau$, hence it cannot be considered a ``higher order'' (rarefaction)
effect. Instead, it represents a fundamental flaw of the first-order theory, which can 
be explained as follows. 

The solution $\widetilde{\Pi} = \widetilde{\delta P} = 0$ is supported in the 
first-order theory only by the $\alpha_\lambda$ mode, as indicated in 
Eq.~\eqref{eq:delta_aux1}. However, according to Eq.~\eqref{eq:delta_aux1}, 
$\delta n_\lambda$ is proportional 
to $\beta_\lambda$, such that requiring that $\beta_\lambda = 0$ automatically 
implies $\delta n_\lambda = 0$. Thus, the initial conditions corresponding to 
{\it Case 2b} cannot be imposed using only the $\alpha_\lambda$ mode, such that the 
$\alpha_\pm$ modes also become excited. The existence of two modes 
$\alpha_{\lambda, \pm}$ in the second-order theory which support 
$\widetilde{\Pi} = \widetilde{\delta P} = 0$ \eqref{eq:h2_solpm} is sufficient to allow 
the initial conditions of {\it Case 2b} to be imposed without exciting the 
$\alpha_{\eta, *}$ modes, such that $\widetilde{\Pi} = \widetilde{P} = 0$ 
throughout the evolution.

The case discussed above is represented in the plots on the first line 
of Fig.~\ref{fig:h2_case2-profiles1}, where 
the time evolution of $\widetilde{\beta}$, $\widetilde{\delta P}$ and $\widetilde{\Pi}$ (normalised with 
respect to $\delta n_0$) is represented 
for the initial conditions $\delta n_0 = 10^{-3}$, $\delta P_0 = 0$ and $\beta_0 = 0$. 
Both the second-order theory and the numerical results indicate that 
$\widetilde{\delta P}$ and $\widetilde{\Pi}$ remain zero throughout the evolution, while the first-order
result predicts oscillations of these quantities. Moreover, in the first-order theory, $\widetilde{\beta}$ 
presents strong oscillations about a decaying exponential, which are not present when the 
second-order theory is employed. It is worth mentioning that the evolution predicted 
by the second-order theory and captured by the numerical method is consistent with the 
evolution equations \eqref{eq:cons_lin}, which reduce for $\delta P = \Pi = 0$ to:
\begin{equation}
 q = -4 P_0 \beta, \qquad 
 \partial_t \delta n + n_0 \partial_z \beta = 0.
 \label{eq:h2_case2b_aux}
\end{equation}
The above relations are also recovered for the $\alpha_\lambda$ mode 
\eqref{eq:delta_aux1} of the first order theory. However, the 
initial conditions $\beta_0 = \delta P_0 / P_0 = 0$ and $\delta n_0 = 10^{-3}$
cannot be imposed using only the $\alpha_\lambda$ mode, since 
according to Eq.~\eqref{eq:delta_aux1}, 
$\delta n_\lambda = k n_0 \beta_\lambda / \alpha_\lambda$ implies that 
$\delta n_0$ is proportional to $\beta_\lambda$. Thus, setting $\beta_0 = 0$ 
and $\delta n_0 \neq 0$ automatically excites the modes $\delta n_\pm$ and 
$\beta_\pm$, which no longer satisfy Eq.~\eqref{eq:h2_case2b_aux}, such that the 
solution becomes contaminated with the addition of oscillatory modes.

The discussion in the preceding paragraph suggests that the first-order theory cannot fully recover the 
hydrodynamic regime (small $\tau$) of the decaying longitudinal wave when the initialisation is 
performed according to {\it Case 2b}, i.e. $\beta_0 = \delta P_0 = 0$ and $\delta n_0 = 10^{-3}$. 
It is instructive to further test if this conclusion holds at non-vanishing values of $\beta_0$,
while keeping $\delta P_0 = 0$. The second and third lines 
of Fig.~\ref{fig:h2_case2-profiles1} show the evolution of $\widetilde{\beta}$, $\widetilde{\delta P}$ and 
$\widetilde{\Pi}$ (again normalised with respect to $\delta n_0$) when $\beta_0 = 10^{-5}$ and 
$\beta_0 = 10^{-3}$, respectively. At $\beta_0 = 10^{-5}$, the first-order theory predicts oscillations which are 
(nearly) in antiphase to the numerical and second-order theory results.  When $\beta_0 = 10^{-3}$, the discrepancy 
between the curves corresponding to the first-order theory and the numerical and second-order theory results is 
no longer visible.

\begin{figure}
\begin{center}
\begin{tabular}{c}
\includegraphics[angle=270,width=0.98\linewidth]{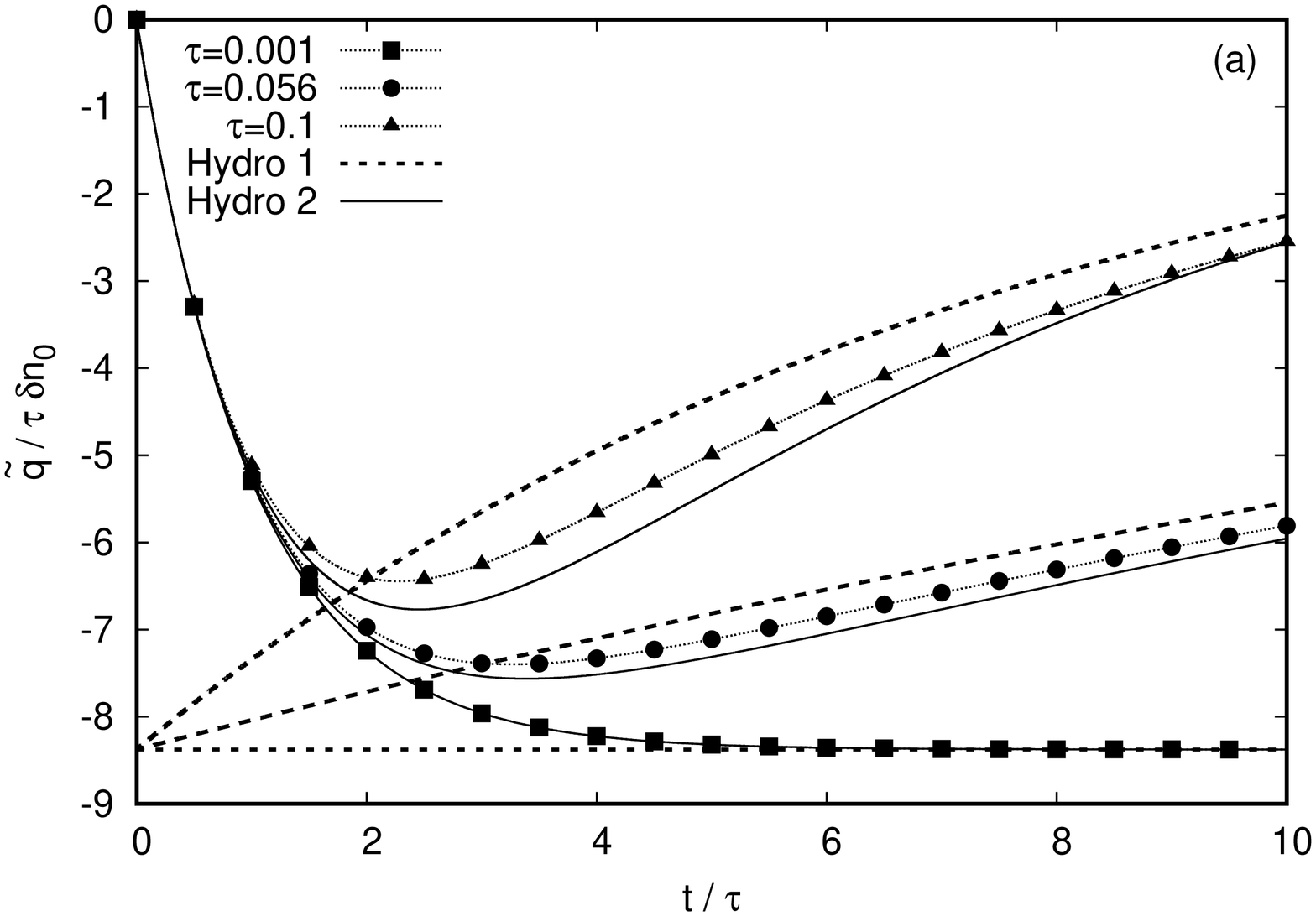} \\
\includegraphics[angle=270,width=0.98\linewidth]{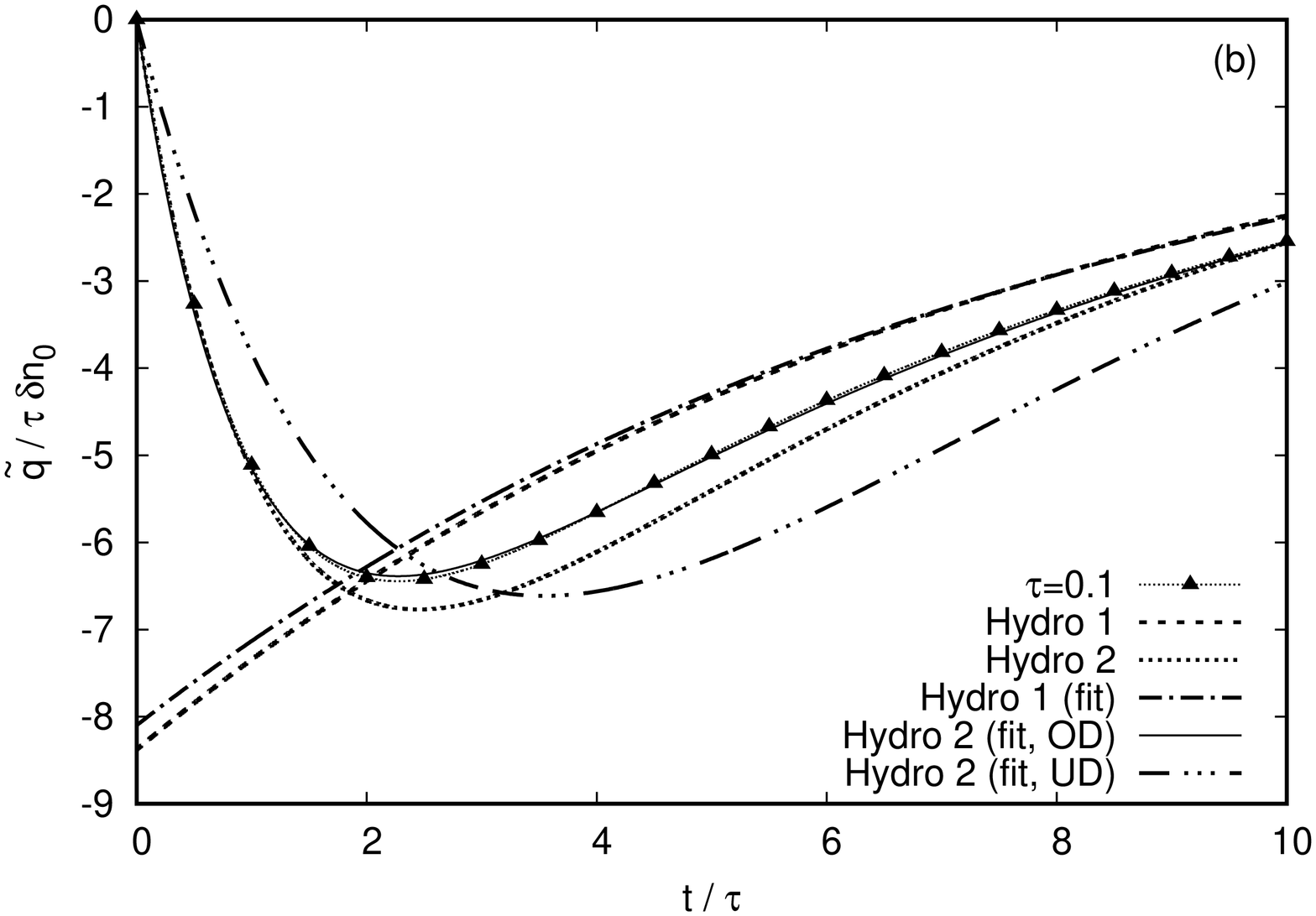} 
\end{tabular}
\end{center}
\caption{
(a) Numerical results (dotted lines and points) for the evolution of $\widetilde{q} / \tau \delta n_0$ 
at various values of $\tau$ for $\delta n_0 = 10^{-3}$ (to ease the comparison, the horizontal axis shows 
$t / \tau$). The analytic curves corresponding to the first 
\eqref{eq:case1} and second \eqref{eq:h2_case1} order hydrodynamics are shown using 
dotted and continuous lines, respectively. (b) Time evolution of $\widetilde{q} / \tau\delta n_0$ at 
$\tau = 0.1$ and $\delta n_0 = 10^{-3}$. The fitted curve corresponding to the first-order hydrodynamics 
is obtained by performing a nonlinear fit of Eq.~\eqref{eq:case2} using $\alpha_\eta$ and $\alpha_o$ as
free parameters. In the second-order case, the nonlinear fit is performed on Eqs.~\eqref{eq:h2_case2OD}
and \eqref{eq:h2_case2UD} for the overdamped and underdamped cases 
by considering $\alpha_{\lambda, d}$ and $\alpha_{\lambda, o}$ 
($\overline{\alpha}_{\lambda, d}$ and $\overline{\alpha}_{\lambda, o}$) 
as free parameters. All non-fitted anayltic curves are obtained 
using the Chapman-Enskog value for $\lambda_0$ \eqref{eq:tcoeff_CE} and $\tau_{q,0} = 1$ \eqref{eq:h2_tcoeff_CE}.
}
\label{fig:h2_case2-profiles2}
\end{figure}

The ability of the second-order hydrodynamics to capture the relaxation of $\widetilde{q}$ from $0$ at initial time 
to the value predicted by the first-order theory is investigated in Fig.~\ref{fig:h2_case2-profiles2}(a).
It can be seen that at small $\tau$, $\widetilde{q}$ relaxes to the value
predicted through the first-order theory with $\lambda$ given by the 
Chapman-Enskog expansion \eqref{eq:tcoeff_CE} after a time $t \sim 5\tau$.
At $\tau \gtrsim 0.1$, the first-order approximation seems to no longer agree 
with the numerical solution (as indicated in Sec.~\ref{sec:hydro1:hydro}), 
while the second-order approximation slowly loses
its validity. In Fig.~\ref{fig:h2_case2-profiles2}(b), the numerical 
result for the evolution of $\widetilde{q}$ at $\tau = 0.1$ is compared to the first \eqref{eq:case2}
and second \eqref{eq:h2_case2OD} order hydrodynamics predictions, specialised to the Chapman-Enskog
case, when $\lambda_0 = 4/3$ and $\tau_{q,0} = 1$. It can be seen that there is significant discrepancy
between the analytic and numerical results. The curve labelled {\it Hydro 1 (fit)} represents the best fit 
of the functional form of the analytic solution \eqref{eq:case2}
to the numerical results, with $\alpha_\lambda$ 
considered as a free parameter. The second-order fits are performed on the two functional forms 
\eqref{eq:h2_case2OD} and \eqref{eq:h2_case2UD}, corresponding to the overdamped and underdamped 
cases, respectively. In the overdamped case, Eq.~\eqref{eq:h2_case2OD} is fitted to the numerical data 
by considering $\alpha_{\lambda, d}$ and $\alpha_{\lambda, o}$ as free parameters. In the underdamped case,
$\overline{\alpha}_{\lambda, d}$ and $\overline{\alpha}_{\lambda, o}$ are found by performing a nonlinear 
fit of Eq.~\eqref{eq:h2_case2UD} with respect to the numerical data. It can be seen in 
Fig.~\ref{fig:h2_case2-profiles2}(b) that, at $\tau = 0.1$, 
the first- and underdamped second-order fits still present significant deviations from the numerical curve.
However, the second-order overdamped fit is in very good agreement with the numerical result, validating the 
functional form \eqref{eq:h2_case2OD} for $\tau \lesssim 0.1$.

\begin{figure}
\begin{center}
\begin{tabular}{c}
\includegraphics[angle=270,width=0.98\linewidth]{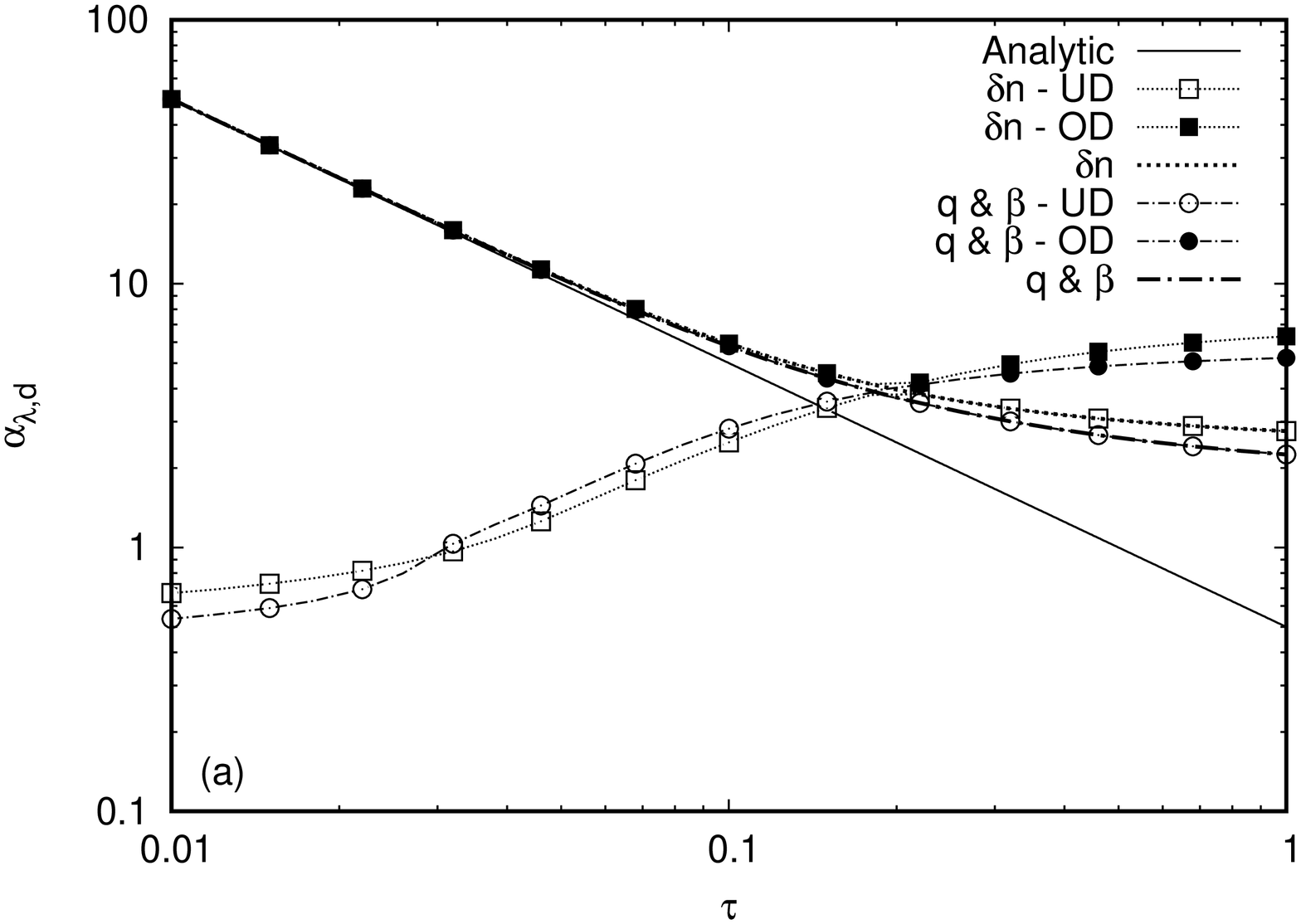} \\
\includegraphics[angle=270,width=0.98\linewidth]{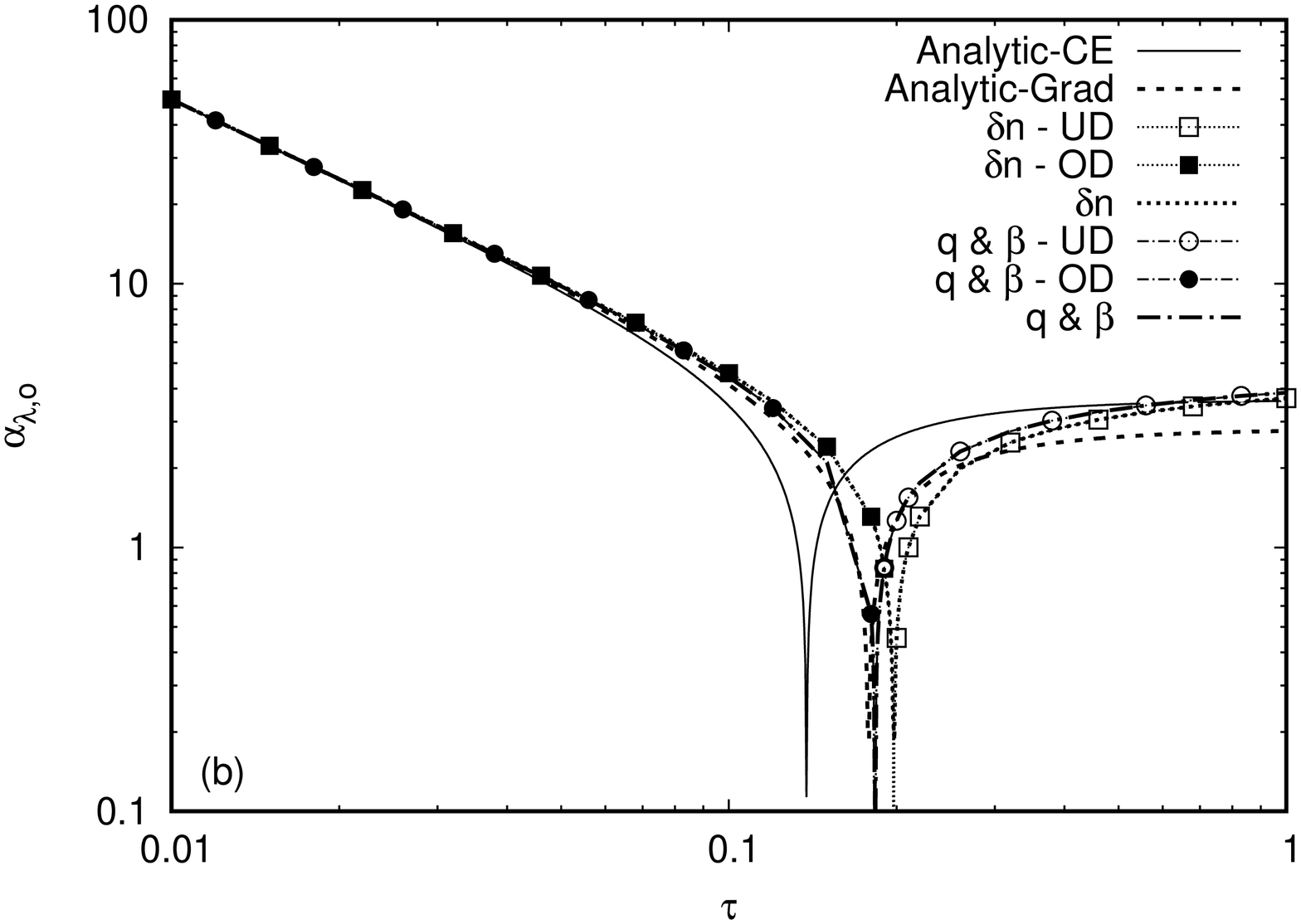}
\end{tabular}
\end{center}
\caption{
Analysis with respect to $\tau$ of (a) $\alpha_{\lambda, d}$ and (b) $\alpha_{\lambda, o}$
for the initial conditions of {\it Case 2b} (i.e. $\beta_0 = \delta P_0 = 0$ and 
$\delta n_0 = 10^{-3}$).
The analytic curve in (a) is $\alpha_{\lambda, d} = 1/2\tau$, while in (b), the 
analytic result \eqref{eq:h2_ald_alo} is represented using the 
Chapman-Enskog (continuous line) and Grad (dashed line) values for $\lambda$, 
while $\tau_q$ was taken equal to $\tau$.
The numerical curves shown with dotted lines and filled symbols are obtained 
by performing a nonlinear fit on the overdamped (OD) solution \eqref{eq:h2_case2OD} 
while considering $\alpha_{\lambda, d}$ and $\alpha_{\lambda, o}$ as free parameters.
The dotted lines with hollow symbols are obtained by fitting the values of 
$\overline{\alpha}_{\lambda, d}$ and $\overline{\alpha}_{\lambda, o}$ to the numerical 
results using the underdamped solution \eqref{eq:h2_case2UD}. 
The numerical curves represented with dashed lines and without points are obtained by 
piecing together the UD and OD results.
The transition from the OD to the UD regime occurs when $\alpha_{\lambda, o} = 0$, as 
indicated by the spikes in (b).
}
\label{fig:h2_case2-alphal}
\end{figure}

Next, Fig.~\ref{fig:h2_case2-alphal} shows an analysis of 
$(\alpha_{\lambda, d}, \alpha_{\lambda, o})$ and 
$(\overline{\alpha}_{\lambda, d}, \overline{\alpha}_{\lambda, o})$ as obtained 
by performing a two-parameter nonlinear fit of 
Eqs.~\eqref{eq:h2_case2OD} and \eqref{eq:h2_case2UD}
for the overdamped (OD) and underdamped (UD) regimes, respectively, to the numerical results obtained 
for $\widetilde{\beta}$, $\widetilde{\delta n}$ and $\widetilde{q}$. Only the initialisation 
corresponding to {\it Case 2b} (i.e. $\beta_0 = \delta P_0 =0$ and $\delta n_0 = 10^{-3}$) 
is considered here. The numerical fit confirms that at small values of $\tau$, 
$\alpha_{\lambda, d}$ and $\alpha_{\lambda, o} \simeq 1/2\tau$.
Furthermore, the results shown in Fig.~\ref{fig:h2_case2-alphal}(a) for $\alpha_\lambda$ indicate that 
the UD regime is not valid when $\tau \lesssim 0.2$, while the OD regime loses applicability 
when $\tau \gtrsim 0.1$. This can also be seen in Fig.~\ref{fig:h2_case2-alphal}(b), where 
the strong spikes indicate the points where $\alpha_{\lambda, o} = 0$, i.e. where 
the transition from the UD to the OD regime occurs. According to Eq.~\eqref{eq:h2_taullim},
this happens when $\tau = \tau_{\lambda, {\rm lim}}^{\rm CE} \simeq 0.138$ 
and $\tau = \tau_{\lambda, {\rm lim}}^{\rm G} \simeq 0.1779$ when the 
Chapman-Enskog \eqref{eq:tcoeff_CE} and Grad \eqref{eq:tcoeff_G} values for $\lambda_0$ 
are used, respectively. The above numerical analysis indicates that 
$\tau_{\lambda, {\rm lim}} \simeq 0.199$ (in the case of $\widetilde{\delta n}$) 
and $\tau_{\lambda, {\rm lim}} \simeq 0.183$ (in the case of $\widetilde{\beta}$
and $\widetilde{q}$), higher than both the Chapman-Enskog and the Grad predictions.

\begin{figure}
\begin{center}
\begin{tabular}{c}
\includegraphics[angle=270,width=0.98\linewidth]{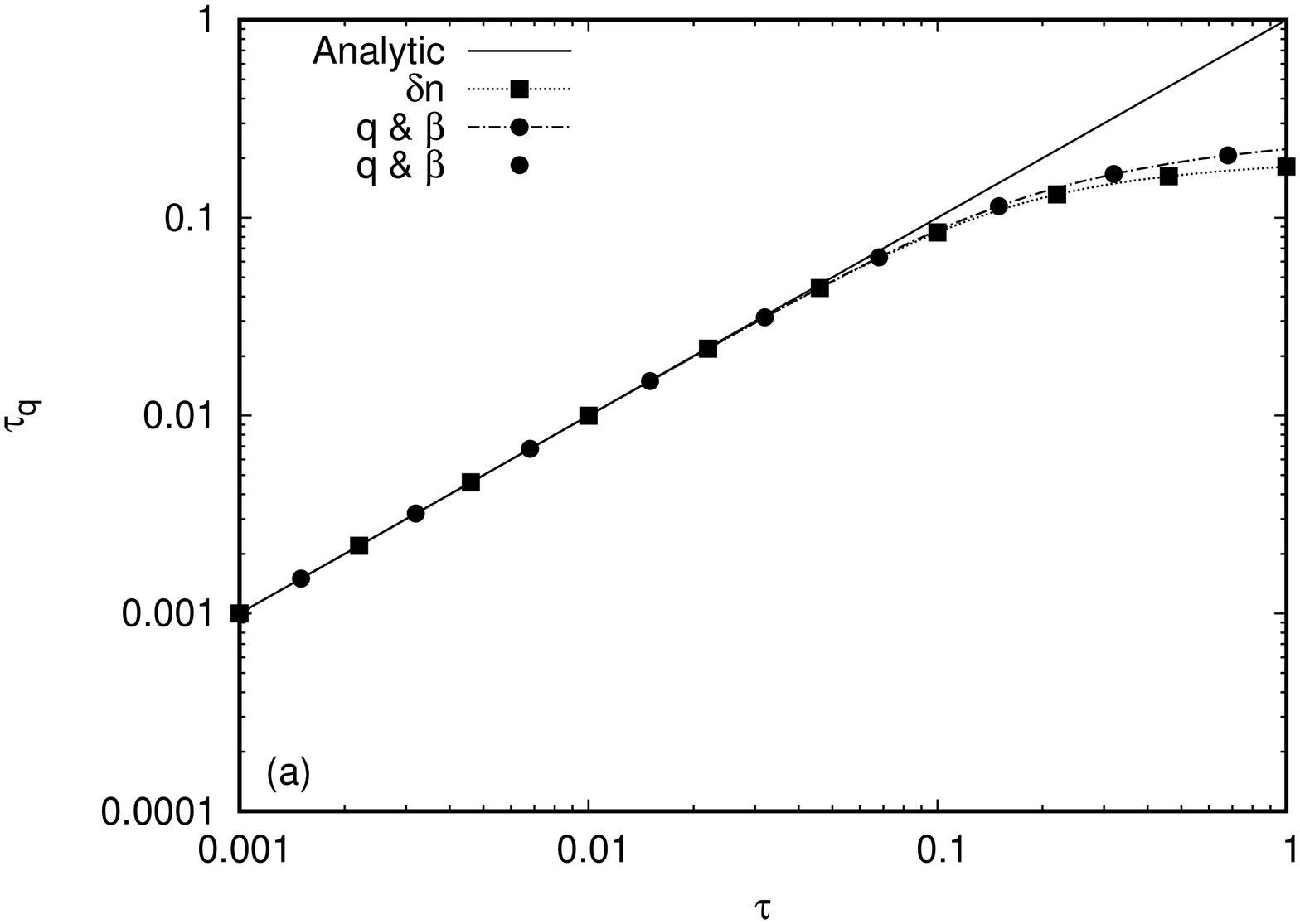} \\ 
\includegraphics[angle=270,width=0.98\linewidth]{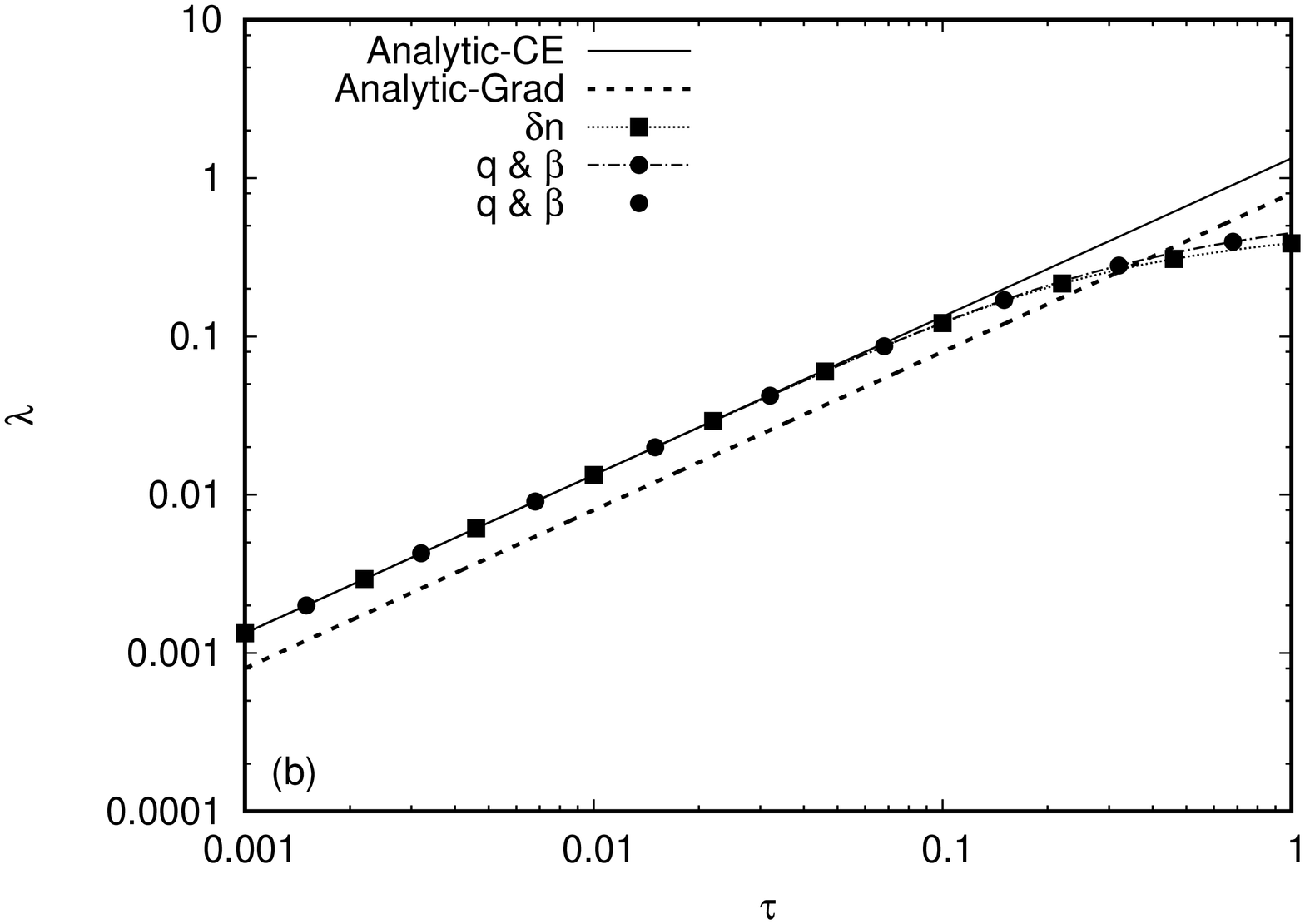} 
\end{tabular}
\end{center}
\caption{
Analysis with respect to $\tau$ of (a) $\tau_q$ and (b) $\lambda$
for the initial conditions of {\it Case 2b} (i.e. $\beta_0 = \delta P_0 = 0$ and 
$\delta n_0 = 10^{-3}$). The analytic curve shown in (a) using a continuous line 
is $\tau_q = \tau$, while 
in (b), the analytic curves correspond to the Chapman-Enskog (continuous line) and 
Grad (dotted line) values for $\lambda$, given in Eqs.~\eqref{eq:tcoeff_CE} and \eqref{eq:tcoeff_G}, 
respectively. The numerical results are obtained as explained in Sec.~\ref{sec:hydro2:case2}.
}
\label{fig:h2_case2-tcoeff}
\end{figure}

The analysis of the relaxation time $\tau_q$ and the heat conductivity $\lambda$
is presented in Fig.~\ref{fig:h2_case2-tcoeff}. Starting from the numerical fits of 
$\alpha_{\lambda, d}$ and $\alpha_{\lambda, o}$, 
where the overdamped (OD) and underdamped (UD) values are taken when 
$\tau < \tau_{\lambda, {\rm lim}}$ and $\tau > \tau_{\lambda, {\rm lim}}$, respectively, 
the values of $\lambda$ and $\tau_q$ are found by requiring that
the roots $\alpha_{\lambda, \pm}$ \eqref{eq:h2_al} satisfy:
\begin{equation}
 \frac{\alpha_{\lambda, +} + \alpha_{\lambda, -}}{2} = \alpha_{\lambda, d}, \qquad 
 \left|\frac{\alpha_{\lambda, +} - \alpha_{\lambda, -}}{2}\right| = \alpha_{\lambda, o},
\end{equation}
where the absolute value is interpreted in the usual sense in the case when 
$\alpha_{\lambda, \pm}$ are complex numbers.
The results for $\tau_q$ and $\lambda$ are shown in Figs.~\ref{fig:h2_case2-tcoeff}(a) 
and \ref{fig:h2_case2-tcoeff}(b), respectively. It can be seen that the numerical 
results for $\tau_q$ 
agree with the theoretical prediction $\tau_q = \tau$ at small values of $\tau$, while for 
larger values of $\tau$, $\tau_q$ seems to reach a plateau. The curve representing the 
numerical results for $\lambda$ is practically overlapped with the Chapman-Enskog prediction 
\eqref{eq:tcoeff_CE}, while at larger values of $\tau$, $\lambda$ seems to reach a plateau, in 
good qualitative agreement with the predictions of the first-order theory presented in 
Fig.~\ref{fig:reg-tau}.

\subsection{Summary}\label{sec:hydro2:summary}

In this section, a particular form of the second-order hydrodynamics equations was 
employed to study the attenuation of a longitudinal wave in the linearised regime.
More precisely, the choice $\alpha_1 = -1/4P_0$ was made for the coupling constant 
between the shear pressure $\Pi$ and heat flux $q$ in the theory presented in 
Refs.~\cite{hiscock83,rezzolla13}, in order to ensure consistency with 
the first-order hydrodynamics analysis presented in Sec.~\ref{sec:hydro1}.

The main aim of this section was to confirm that the second-order hydrodynamics correctly 
describes the relaxation process of $\Pi$ and $q$ from essentially arbitrary initial 
values (both vanish at initial time in the analysis presented in this paper) to some 
non-zero value which 
agrees with the prediction of the first-order theory at small enough values of $\tau$.
This is shown in Figs.~\ref{fig:h2_case1-profiles} and \ref{fig:h2_case2-profiles2},
while the values of the relaxation times $\tau_\Pi$ and $\tau_q$ are analysed with respect to 
the relaxation time $\tau$ of the Anderson-Witting model in Figs.~\ref{fig:h2_case1-tcoeff} and 
\ref{fig:h2_case2-tcoeff}.

During this analysis, a fundamental limitation of the first-order theory was 
pointed out, which can be summarised as follows. 
If at initial time, the velocity and pressure perturbations vanish 
(i.e. $\beta_0 = \delta P_0 = 0$), the pressure $\widetilde{\delta P}$ 
and shear stress $\widetilde{\Pi}$ 
perturbations remain zero at all later times, while the attenuation of $\widetilde{\beta}$ is 
purely evanescent (non-oscillatory). This result is obtained analytically in the second-order 
theory and is confirmed via numerical simulations in Fig.~\ref{fig:h2_case2-profiles1}. 
On the other hand, the first-order theory predicts that $\widetilde{\delta P}$ and $\widetilde{\Pi}$ 
are proportional to $\delta n_0$ and $\widetilde{\beta}$ has an oscillatory component 
the amplitude of which is proportional to $\tau$. This leads to the conclusion 
that the first-order theory cannot correctly 
describe this particular dissipative process, even when the relaxation time is small.


\section{Moment method}\label{sec:mom}

The analysis presented in the previous sections provides indication that 
the correct expresions for the transport coefficients $\lambda$ and $\eta$ 
are obtained using the Chapman-Enskog procedure. Recently \cite{denicol12},
it was shown that one of the main drawbacks of the moment method as originally 
employed by Israel and Stewart \cite{israel79} is that the distribution function 
$f$ is expanded with respect to the nonorthogonal basis formed of powers of the 
particle four-momentum $p^\mu$.
Truncating this series at a finite order discards an infinite number of terms 
of first-order with respect to the Knudsen number ${\rm Kn}$ (in the 
Anderson-Witting model, ${\rm Kn} \sim \tau$).
The solution proposed in Ref.~\cite{denicol12} was to expand $f$ in terms of
irreducible tensors with respect to the particle momentum $p^\mu$. The proposed 
scheme recovers the expressions for the transport coefficients $\lambda$ and $\eta$ 
obtained using the Chapman-Enskog expansion.

In this section, a moment-based method similar to the one introduced in 
Ref.~\cite{denicol12} is considered. Spherical coordinates 
$\{p, \theta, \varphi\}$ are employed in the momentum space and the distribution 
function $f$ is expanded with respect to the generalised Laguerre polynomials for the 
momentum magnitude $p$, the Legendre polynomials for $\xi = \cos \theta$ and 
the trigonometric basis $\{\cos m \varphi, \sin m \varphi\}$ for $\varphi$.
Due to the symmetries of the system, $f$ can be considered independent of $\varphi$,
such that only the $p$ and $\xi$ expansions will be discussed. Truncating the 
system at order $N_L = 1$ and $N_\xi = 2$ with respect to the Laguerre and 
Legendre polynomials, respectively, yields a system of six equations 
for the five hydrodynamic variables $\delta n$, $\beta$, $\delta P$, $q$ and 
$\Pi$, as well as a non-hydrodynamic variable. The importance of this sixth variable 
in establishing the symmetry between the shear stress and heat flux solutions
is illustrated, such that the underdamped (UD) and overdamped (OD) regimes 
discussed in Sec.~\ref{sec:hydro2} are represented unitarily in this new solution.

\subsection{Constitutive relations}\label{sec:mom:const}

In this section, the longitudinal wave problem is again approached, but this time 
by employing a moment method. Instead of performing the standard Grad-like expansion of 
$f$ in terms of polynomials in $p^\mu$, $f$ is expanded following 
Refs.~\cite{blaga17aip,blaga17prc} as follows:
\begin{equation}
 f = \frac{1}{2\pi T_0^3} e^{-p / T_0}\sum_{\ell = 0}^\infty 
 \frac{1}{(\ell + 1)(\ell + 2)} \mathcal{F}_\ell L^{(2)}_\ell(p / T_0),
\end{equation}
where $L^{(2)}_{\ell}(z)$ are the generalised Laguerre polynomial of type $2$ 
and order $\ell$, which satisfy the following orthogonality relation:
\begin{equation}
 \int_{0}^\infty dz\, z^2\, e^{-z} L^{(2)}_\ell(z) L^{(2)}_{\ell'}(z) = 
 (\ell + 1)(\ell + 2) \delta_{\ell, \ell'}.
\end{equation}
Thus, the coefficients $\mathcal{F}_\ell$ can be obtained as follows:
\begin{equation}
 \mathcal{F}_\ell = 2\pi \int_0^\infty dp\, p^2\, f\, L_\ell^{(2)}(p / T_0).\label{eq:mom_Fl}
\end{equation}
Multiplying the Boltzmann equation \eqref{eq:boltz_lin}, valid only in the 
linearised regime of the 
longitudinal wave problem, by $p^2 L_\ell^{(2)}(p / T_0)$ and integrating over 
$p$ yields:
\begin{equation}
 \partial_t \mathcal{F}_\ell + \xi \partial_z \mathcal{F}_\ell =
 -\frac{1}{\tau} (\mathcal{F}_\ell - \mathcal{F}^{({\rm eq})}_\ell),\label{eq:mom_boltz_Fl}
\end{equation}
where the coefficients $\mathcal{F}^{({\rm eq})}_\ell$ corresponding to the 
equilibrium distribution function $\feq$ are defined by analogy to Eq.~\eqref{eq:mom_Fl}:
\begin{equation}
 \mathcal{F}_\ell^{({\rm eq})} = 2\pi \int_0^\infty dp\, p^2\, \feq\, 
 L_\ell^{(2)}(p / T_0).\label{eq:mom_Fleq}
\end{equation}
In the absence of collisions [i.e.~when neglecting the right 
hand side in Eq.~\eqref{eq:mom_boltz_Fl}],
each coefficient $\mathcal{F}_\ell$ evolves independently. Since $\feq$ is constructed only 
in terms of $N^\mu$ and $T^{\mu\nu}$, which can be written entirely in terms 
of $\mathcal{F}_\ell$ with $\ell = 0$ and $\ell = 1$, 
the evolution of $N^\mu$ and $T^{\mu\nu}$ is fully determined by considering 
Eq.~\eqref{eq:mom_boltz_Fl} only for $\ell = 0$ and $\ell = 1$ and neglecting 
all higher $\ell$ terms \cite{blaga17prc}.

The coefficients $\mathcal{F}_\ell$ are further expanded with respect to 
$\xi$ using the complete set of Legendre polynomials $P_s(\xi)$:
\begin{equation}
 \mathcal{F}_\ell = \sum_{s = 0}^\infty \frac{2s + 1}{2} \mathcal{F}_{\ell,s} P_s(\xi),
\end{equation}
where the coefficients $\mathcal{F}_{\ell, s}$ depend only on $z$ and $t$ and are obtained 
using the orthogonality of the Legendre polynomials as follows:
\begin{equation}
 \mathcal{F}_{\ell, s} = \int_{-1}^1 d\xi \, \mathcal{F}_\ell P_s(\xi).
\end{equation}
The coefficients $\mathcal{F}_{\ell,s}^{({\rm eq})}$ corresponding to $\feq$ can be 
defined in a similar manner:
\begin{equation}
 \mathcal{F}_{\ell,s}^{({\rm eq})} = \int_{-1}^1 d\xi \, 
 \mathcal{F}_\ell^{({\rm eq})} P_s(\xi).
\end{equation}

The expansion coefficients $\mathcal{F}_{\ell, s}$ can be linked to 
$N^\mu$ and $T^{\mu\nu}$ as follows:
\begin{gather}
 \mathcal{F}_{0,0} = N^t \simeq n_0 + \delta n, \qquad 
 \mathcal{F}_{0,1} = N^z \simeq n_0 \beta, \nonumber\\
 \mathcal{F}_{1,0} = 3N^t - \frac{1}{T_0} T^{tt} \simeq 3n_0 
 \left(\frac{\delta n}{n_0} - \frac{\delta P}{P_0}\right), \nonumber\\
 \mathcal{F}_{1,1} = 3N^z - \frac{1}{T_0} T^{tz} \simeq -n_0 \beta - \frac{q}{T_0}, \nonumber\\
 3\mathcal{F}_{0,2} - \mathcal{F}_{1,2} \simeq \frac{3 \Pi}{2 T_0}.
 \label{eq:mom_Fl_lin}
\end{gather}
The coefficients $\mathcal{F}_{0,2}$ and $\mathcal{F}_{1,2}$ on their own have no correspondent 
with respect to $N^\mu$ and $T^{\mu\nu}$.
The equilibrium coefficients $\mathcal{F}_{\ell, s}^{({\rm eq})}$ 
($\ell = 0, 1$ and $s = 0, 1, 2$) can be found from Eq.~\eqref{eq:feq_lin}:
\begin{gather}
 \mathcal{F}_{0,0}^{({\rm eq})} \simeq n_0 + \delta n, \qquad 
 \mathcal{F}_{0,1}^{({\rm eq})} \simeq n_0 \beta + \frac{q}{4T_0}, \nonumber\\
 \mathcal{F}_{1,0}^{({\rm eq})} \simeq 3n_0 \left(\frac{\delta n}{n_0} - \frac{\delta P}{P_0}\right), \qquad
 \mathcal{F}_{1,1}^{({\rm eq})} \simeq -n_0 \beta - \frac{q}{4T_0},
 \label{eq:mom_Fleq_lin}
\end{gather}
while $\mathcal{F}_{0,2}^{({\rm eq})} \simeq 0$ and 
$\mathcal{F}_{1,2}^{({\rm eq})} \simeq 0$.

Using the recurrence relation:
\begin{equation}
 \xi P_s(\xi) = \frac{s+1}{2s+1} P_{s+1}(\xi) + \frac{s}{2s+1} P_{s-1}(\xi), 
\end{equation}
Eq.~\eqref{eq:mom_boltz_Fl} can be projected on the space of the Legendre 
polynomials as follows:
\begin{multline}
 \partial_t \mathcal{F}_{\ell, s} + 
 \partial_z \left(\frac{s}{2s+1} \mathcal{F}_{\ell, s-1}
 + \frac{s+1}{2s+1} \mathcal{F}_{\ell, s + 1}\right)\\
 = -\frac{1}{\tau}(\mathcal{F}_{\ell, s} - \mathcal{F}_{\ell, s}^{({\rm eq})}).
\end{multline}
The above procedure produces an infinite system of equations corresponding to various values of $(\ell, s)$,
where knowledge of $\mathcal{F}_{\ell, s+1}$ is required in order to determine the evolution of 
$\mathcal{F}_{\ell, s}$. As also discussed in Ref.~\cite{blaga17prc}, the above system can be closed 
at an order $Q$ by imposing $\mathcal{F}_{\ell, Q} = 0$. This procedure is intimately related to the numerical 
method employed in this paper (described in detail in Ref.~\cite{blaga17prc} and 
also summarised in Appendix~\ref{app:num}). In particular, $Q$ represents 
the quadrature order of the model and the resulting system of equations 
is guaranteed to be hyperbolic. Since only the study of $N^\mu$ and $T^{\mu\nu}$
is of interest in this section, only the case $Q = 3$ will be considered 
henceforth, such that $\mathcal{F}_{0, 3} = \mathcal{F}_{1,3} = 0$. The 
resulting set of equations can be written as:
\begin{subequations}\label{eq:mom_eq_F}
\begin{align}
 \partial_t \mathcal{F}_{0,0} + \partial_z \mathcal{F}_{0,1} =& 
 -\frac{1}{\tau}(\mathcal{F}_{0,0} - \mathcal{F}_{0,0}^{({\rm eq})}), \label{eq:mom_eq_F:00}\\
 \hspace{-10pt}
 \partial_t \mathcal{F}_{0,1} + \frac{1}{3} \partial_z \left(\mathcal{F}_{0,0} +
 2\mathcal{F}_{0,2}\right) =& -\frac{1}{\tau} (\mathcal{F}_{0,1} - \mathcal{F}_{0,1}^{({\rm eq})}),\label{eq:mom_eq_F:01}\\
 \partial_t \mathcal{F}_{0,2} + \frac{2}{5} \partial_z \mathcal{F}_{0,1} =& -\frac{1}{\tau} \mathcal{F}_{0,2},\label{eq:mom_eq_F:02}\\
 \partial_t \mathcal{F}_{1,0} + \partial_z \mathcal{F}_{1,1} =& 
 -\frac{1}{\tau}(\mathcal{F}_{1,0} - \mathcal{F}_{1,0}^{({\rm eq})}), \label{eq:mom_eq_F:10}\\
 \hspace{-10pt}
 \partial_t \mathcal{F}_{1,1} + \frac{1}{3} \partial_z \left(\mathcal{F}_{1,0} +
 2\mathcal{F}_{1,2}\right) =& -\frac{1}{\tau} (\mathcal{F}_{1,1} - \mathcal{F}_{1,1}^{({\rm eq})}),\label{eq:mom_eq_F:11}\\
 \partial_t \mathcal{F}_{1,2} + \frac{2}{5} \partial_z \mathcal{F}_{1,1} =& -\frac{1}{\tau} \mathcal{F}_{1,2}.\label{eq:mom_eq_F:12}
\end{align}
\end{subequations}
The above system is closed. Substituting Eqs.~\eqref{eq:mom_Fl_lin} and \eqref{eq:mom_Fleq_lin}
into Eq.~\eqref{eq:mom_eq_F},
the conservation equations \eqref{eq:cons_lin} can be obtained, together with the following constitutive equations:
\begin{subequations}\label{eq:mom_tcoeff_lin}
\begin{gather}
 \tau \partial_t q + q = -\frac{\tau P_0}{3} \partial_z
 \left(\frac{3\delta P}{P_0} - \frac{4\delta n}{n_0}\right) \nonumber\\
 \phantom{aaaaaaaaaaaaaaaaa}
 + \frac{2P_0 \tau}{3n_0} \partial_z (\mathcal{F}_{0,2} + \mathcal{F}_{1,2}),
 \label{eq:mom_tcoeff_lin:q}\\
 \tau \partial_t \Pi + \Pi = -\frac{16 \tau P_0}{15} 
 \partial_z \left(\beta + \frac{q}{4P_0}\right), 
 \label{eq:mom_tcoeff_lin:Pi}\\
 (\tau\partial_t + 1)(\mathcal{F}_{0,2} + \mathcal{F}_{1,2}) =
 \frac{2n_0 \tau}{5P_0} \partial_z q.\label{eq:mom_tcoeff_lin:sup}
\end{gather}
\end{subequations}
Comparing the above equations to the second-order hydrodynamics constitutive 
equations \eqref{eq:h2_tcoeff_lin}, it can be seen that 
the transport coefficients have the following 
expressions:
\begin{equation}
 \lambda = \frac{4}{3} \tau n_0, \qquad 
 \eta = \frac{4}{5} \tau P_0, \qquad 
 \tau_q = \tau, \qquad
 \tau_\Pi = \tau.
 \label{eq:mom_tcoeff_CE}
\end{equation}
The above relations confirm the Chapman-Enskog prediction \eqref{eq:tcoeff_CE} for 
$\lambda$ and $\eta$ and agree with the second-order hydrodynamics 
values for $\tau_q$ and $\tau_\Pi$ given in Eq.~\eqref{eq:h2_tcoeff_CE}. 
Furthermore, the constitutive equation \eqref{eq:mom_tcoeff_lin:q} contains an 
extra term compared to the second-order hydrodynamics version \eqref{eq:h2_tcoeff_lin:q}.
A simple power counting shows that this term is cubic in the relaxation time $\tau$,
hence it cannot be present in the second-order hydroynamics theory.

\subsection{Longitudinal waves: modes}\label{sec:mom:longalpha}

Considering now the propagation of a wave with wave number $k$,
the ansatz \eqref{eq:ansatz} can be applied to the new 
variables $\mathcal{F}_{0,2}$ and $\mathcal{F}_{1,2}$ as follows:
\begin{equation}
 \mathcal{F}_{0,2} = \widetilde{\mathcal{F}}_{0,2} \cos kz, \qquad
 \mathcal{F}_{1,2} = \widetilde{\mathcal{F}}_{1,2} \cos kz.
\end{equation}
The mode decomposition \eqref{eq:alpha_def} can be applied to $\widetilde{\mathcal{F}}_{0,2}$ 
and $\widetilde{\mathcal{F}}_{1,2}$ as follows:
\begin{equation}
 \begin{pmatrix}
  \widetilde{\mathcal{F}}_{0,2}\\
  \widetilde{\mathcal{F}}_{1,2}
 \end{pmatrix} = \sum_\alpha 
 \begin{pmatrix}
  \mathcal{F}_{0,2; \alpha}\\
  \mathcal{F}_{1,2; \alpha}
 \end{pmatrix} e^{-\alpha t}. 
\end{equation}
Substituting the above expansions into Eqs.~\eqref{eq:mom_eq_F:02} and \eqref{eq:mom_eq_F:12} 
gives:
\begin{align}
 \mathcal{F}_{0,2; \alpha} =& -\frac{2n_0 \tau k}{5(1 - \alpha \tau)} \beta_\alpha,
 \nonumber\\
 \mathcal{F}_{1,2; \alpha} =& \frac{2n_0 \tau k}{5(1 - \alpha \tau)} 
 \left(\beta_\alpha + \frac{q_\alpha}{P_0}\right).
\end{align}

The moment method introduced in this section bears many similarities 
with the second-order hydrodynamics method discussed in Sec.~\ref{sec:hydro2}. 
In particular, since the constitutive equations \eqref{eq:mom_tcoeff_lin:Pi} 
and \eqref{eq:h2_tcoeff_lin:Pi} for $\Pi$ are the same in the two theories, 
Eqs.~\eqref{eq:h2_Pi_alpha}, \eqref{eq:h2_dP_alpha} and \eqref{eq:h2_eig}
remain unchanged. The latter equation again can be solved either by setting 
$q_\alpha = -4P_0 \beta_\alpha$ or by setting the square bracket 
to $0$. In the latter case, the allowed values for $\alpha$,
namely $\alpha_{\eta, r}$ and $\alpha_{\eta, \pm}$ can be written as in 
Sec.~\ref{sec:hydro2:longalpha}, being given in Eq.~\eqref{eq:h2_ae}.
For completeness, these expressions are reproduced below, specialised to the 
values of $\eta$, $\lambda$, $\tau_q$ and $\tau_\Pi$ given in 
Eq.~\eqref{eq:mom_tcoeff_CE}:
\begin{align}
 \alpha_{\eta,r} =& \frac{1}{3\tau}\left[1 + 
 \frac{1}{R_\eta}\left(1 - \frac{9 k^2 \tau^2}{5}\right) +
 R_\eta\right]\nonumber\\
 \simeq& \frac{1}{\tau} - 2\alpha_d + O(\tau^3),\nonumber\\
 \alpha_{\eta,d} =& \frac{1}{3\tau}\left[1 - 
 \frac{1}{2R_\eta}\left(1 - \frac{9k^2 \tau^2}{5} \right) - 
 \frac{R_\eta}{2}\right] \nonumber\\
 \simeq& \alpha_d + O(\tau^3),\nonumber\\
 \alpha_{\eta,o} =& \frac{\sqrt{3}}{6\tau} \left[
 \frac{1}{R_\eta} \left(1 - \frac{9k^2 \tau^2}{5} \right) - R_\eta\right] \nonumber\\
 \simeq& \frac{k}{\sqrt{3}} + O(\tau^2),\label{eq:mom_ae}
\end{align}
where $\alpha_d = 2k^2 \tau / 15$ is the first-order coefficient 
given in Eq.~\eqref{eq:alphad_alphao} 
and $R_\eta$ \eqref{eq:h2_Retadef} becomes:
\begin{equation}
 R_\eta = 
 \begin{cases}
  \left(1 - 3k \tau \sqrt{R_{\eta,{\rm aux}}} + \frac{9}{5} k^2 \tau^2\right)^{1/3}, &
  \tau < \tau_{\eta, {\rm lim}}, \\
  -\left(-1 + 3k \tau \sqrt{R_{\eta,{\rm aux}}} - \frac{9}{5} k^2 \tau^2\right)^{1/3}, &
  \tau > \tau_{\eta, {\rm lim}}.
 \end{cases}
 \label{eq:mom_Redef}
\end{equation}
The function $R_{\eta, {\rm aux}}$ \eqref{eq:h2_Reaux} reduces to:
\begin{equation}
 R_{\eta, {\rm aux}} = 1 - \frac{18}{25} k^2 \tau^2 + 
 \frac{81}{125} k^4 \tau^4.
\end{equation}
Since the roots $(k^2 \tau^2)_\pm = \frac{1}{9}(5 \pm 10i)$ of $R_{\eta, {\rm aux}}$ 
have a non-vanishing imaginary part and 
$R_{\eta, {\rm aux}}(\tau = 0) = 1$, $R_{\eta, {\rm aux}}> 0$ for 
all values of $\tau$. The threshold value $\tau_{\eta,{\rm lim}}$ 
appearing in Eq.~\eqref{eq:mom_Redef} is 
\begin{equation}
 \tau_{\eta, {\rm lim}} = \frac{\sqrt{5}}{3k} \simeq 0.119,\label{eq:mom_tauelim}
\end{equation}
which coincides with Eq.~\eqref{eq:h2_tauelim} when $\eta$ and $\tau_{\Pi}$ 
are replaced according to Eq.~\eqref{eq:mom_tcoeff_CE}.

When $q_\alpha = -4P_0 \beta_\alpha$, Eq.~\eqref{eq:h2_alphal_eq} is replaced by:
\begin{equation}
 \left[\frac{4\tau k^2}{3\alpha} - 4(1 - \alpha \tau) - 
 \frac{16\tau^2 k^2}{15(1 - \alpha\tau)}\right] \beta_\alpha = 0.
\end{equation}
The square bracket cancels when $\alpha \in \{\alpha_{\lambda, r}, \alpha_{\lambda, \pm}\}$, 
where $\alpha_{\lambda, \pm}$ can be written as 
$\alpha_{\lambda, \pm} = \alpha_{\lambda, d} \pm i \alpha_{\lambda, o}$.
The exact expressions for the coefficients $\alpha_{\lambda, r}$, 
$\alpha_{\lambda, d}$ and $\alpha_{\lambda, o}$ read:
\begin{align}
 \alpha_{\lambda,r} =& \frac{1}{3\tau}\left[2 - 
 \frac{1}{R_\lambda}\left(1 - \frac{9 k^2 \tau^2}{5}\right) - R_\lambda\right]\nonumber\\
 \simeq& \alpha_\lambda + O(\tau^3),\nonumber\\
 \alpha_{\lambda,d} =& \frac{1}{3\tau}\left[2 + 
 \frac{1}{2R_\lambda}\left(1 - \frac{9 k^2 \tau^2}{5}\right) + \frac{R_\lambda}{2}\right]
 \nonumber\\
 \simeq& \frac{1}{\tau} - \frac{\alpha_\lambda}{2} + O(\tau^3),\nonumber\\
 \alpha_{\lambda,o} =& \frac{\sqrt{3}}{6\tau}\left[ 
 \frac{1}{R_\lambda}\left(1 - \frac{9 k^2 \tau^2}{5}\right) - R_\lambda\right]\nonumber\\
 \simeq& \frac{2k}{\sqrt{15}} + O(\tau^2),\label{eq:mom_al}
\end{align}
where $\alpha_\lambda = k^2 \tau / 3$ is defined in Eq.~\eqref{eq:alphal} and 
\begin{equation}
 R_\lambda = 
 \begin{cases}
  \left[1 - \frac{6k\tau}{\sqrt{5}} \sqrt{R_{\lambda, {\rm aux}}} 
  + \frac{9k^2 \tau^2}{10}\right]^{1/3}, &
  \tau < \tau^{\rm mom}_{\lambda, {\rm lim}}, \\
  -\left[-1 + \frac{6k\tau}{\sqrt{5}} \sqrt{R_{\lambda, {\rm aux}}} 
  - \frac{9k^2 \tau^2}{10}\right]^{1/3}, &
  \tau > \tau_{\lambda, {\rm lim}}^{\rm mom},
 \end{cases}
 \label{eq:mom_Rldef}
\end{equation}
In the above, $R_{\lambda, {\rm aux}}$ is defined as:
\begin{equation}
 R_{\lambda, {\rm aux}} = 1 - \frac{99}{80} k^2 \tau^2 + \frac{81}{100}k^4 \tau^4
\end{equation}
Since the roots $(k^2\tau^2)_{\pm} = \frac{5}{72}(11 \pm 3i \sqrt{15})$ of 
$R_{\lambda, {\rm aux}}$ are complex, 
$R_{\lambda, {\rm aux}} > 0$ for all values of $\tau$.
The parameter $\tau^{\rm mom}_{\lambda, {\rm lim}}$ is defined as the value of $\tau$ 
at which the expression under the 
cubic root in Eq.~\eqref{eq:mom_Rldef} vanishes. It is given by
\begin{equation}
 \tau^{\rm mom}_{\lambda, {\rm lim}} = \frac{\sqrt{5}}{3k},
 \label{eq:mom_taullim}
\end{equation}
being identical to $\tau_{\eta, {\rm lim}}$ \eqref{eq:mom_tauelim}.
The definition \eqref{eq:mom_Rldef} of $R_\lambda$ ensures that the coefficients $\alpha_{\lambda, *}$ ($* \in \{r, d, o\}$), 
defined in Eq.~\eqref{eq:mom_al}, are real for all positive values of $\tau$.

\subsection{Longitudinal waves: solution}\label{sec:mom:longsol}

The solution can be split as in Eq.~\eqref{eq:h2_solM}, i.e. 
$\widetilde{M} = \widetilde{M}_\lambda + \widetilde{M}_\eta$.
In this case, the $\lambda$ and $\eta$ sectors of the solution 
have symmetric expressions, i.e.:
\begin{align}
 \widetilde{M}_\lambda =& M_{\lambda, r} e^{-\alpha_{\lambda, r} t} + 
 \left(M_{\lambda, c} \cos \alpha_{\lambda, o} + 
 M_{\lambda, s} \sin \alpha_{\lambda, o}\right) e^{-\alpha_{\lambda, d} t},\nonumber\\
 \widetilde{M}_\eta =& M_{\eta, r} e^{-\alpha_{\eta, r} t} + 
 \left(M_{\eta, c} \cos \alpha_{\eta, o} + 
 M_{\eta, s} \sin \alpha_{\eta, o}\right) e^{-\alpha_{\eta, d} t}.
 \label{eq:mom_solM}
\end{align}
The coefficients $M_{\times, *}$ 
(where $\times \in \{\lambda, \eta\}$ and $* \in \{r, c, s\}$)
for $M \in \{\delta n, \mathcal{F}_{0, 2}\}$ are
\begin{align}
 \delta n_{\times, c} =& 
 kn_0 \frac{\alpha_{\times, d} \beta_{\times, c} + 
 \alpha_{\times, o} \beta_{\times, s}}
 {\alpha_{\times,d}^2 + \alpha_{\times,o}^2}, \nonumber\\
 \delta n_{\times, s} =& 
 kn_0 \frac{\alpha_{\times, d} \beta_{\times, s} - 
 \alpha_{\times, o} \beta_{\times, c}}
 {\alpha_{\times,d}^2 + \alpha_{\times,o}^2},\nonumber\\
 \delta n_{\times, r} =& k n_0 \frac{\beta_{\times, r}}{\alpha_{\times, r}}, \nonumber\\ 
 \mathcal{F}_{0,2; \times, c} =& -\frac{2k n_0 \tau}{5} 
 \frac{(1 - \alpha_{\times, d} \tau) \beta_{\times, c} - 
 \alpha_{\times, o} \tau \beta_{\times, s}}
 {(1 - \alpha_{\times, d} \tau)^2 + (\alpha_{\times, o} \tau)^2},\nonumber\\ 
  \mathcal{F}_{0,2; \times, s} =& -\frac{2k n_0 \tau}{5} 
 \frac{(1 - \alpha_{\times, d} \tau) \beta_{\times, s} + 
 \alpha_{\times, o} \tau \beta_{\times, c}}
 {(1 - \alpha_{\times, d} \tau)^2 + (\alpha_{\times, o} \tau)^2},\nonumber\\
 \mathcal{F}_{0,2; \times, r} =& -\frac{2kn_0 \tau}{5} 
 \frac{\beta_{\times, r}}{1 - \alpha_{\times, r} \tau}.
\end{align}
On the $\lambda$ sector, $\delta P_{\lambda, *} = \Pi_{\lambda, *} = 0$, while 
\begin{equation}
 q_{\lambda, *} = -4P_0 \beta_{\lambda, *}, \qquad
 \mathcal{F}_{1,2; \lambda, *}  = 3\mathcal{F}_{0,2; \lambda, *}.
\end{equation}
On the $\eta$ sector, $q_{\eta, *} = 0$ and 
\begin{gather}
 \delta P_{\eta, *} = \frac{4 P_0}{3 n_0} \delta n_{\eta, *}, \qquad
 \mathcal{F}_{1,2; \eta, *} = -\mathcal{F}_{0,2; \eta, *}, \nonumber\\
 \Pi_{\eta, *} = \frac{8P_0}{3n_0} \mathcal{F}_{0,2; \eta, *}.
\end{gather}

The initial conditions \eqref{eq:init} and \eqref{eq:init2} refer only 
to $\widetilde{\beta}$, $\widetilde{\delta n}$, $\widetilde{\delta P}$, 
$\widetilde{q}$ and $\widetilde{\Pi}$. In the moment approach 
considered in this section, the coefficients $\mathcal{F}_{0, 2}$ and 
$\mathcal{F}_{1, 2}$ are also free to evolve [in fact, they contribute only one 
degree of freedom, since $\Pi = 2T_0 (\mathcal{F}_{0,2} - 
\frac{1}{3} \mathcal{F}_{1,2})$ is taken as an indepedent variable]. 
Since at $t =0$, the system is initialised with the equilibrium distribution 
$\feq$, the initial conditions for $\mathcal{F}_{0,2}$ and $\mathcal{F}_{1,2}$ 
can be read from Eq.~\eqref{eq:mom_Fleq_lin}:
\begin{equation}
 \widetilde{\mathcal{F}}_{0,2}(t = 0) = 0, \qquad 
 \widetilde{\mathcal{F}}_{1,2}(t = 0) = 0. 
 \label{eq:init3}
\end{equation}

Imposing the initial conditions \eqref{eq:init}, \eqref{eq:init2} and \eqref{eq:init3}
on the solution \eqref{eq:mom_solM} yields the following solution for the 
integration constants $\beta_{\lambda, *}$:
\begin{align}
 \beta_{\lambda, r} =& \frac{\alpha_{\lambda, r}(1 - \alpha_{\lambda, r} \tau) 
 (\alpha_{\lambda, d}^2 + \alpha_{\lambda, o}^2)}{4k[\alpha_{\lambda,o}^2 + 
 (\alpha_{\lambda, d} - \alpha_{\lambda, r})^2]}
 \left(\frac{4\delta n_0}{n_0} - \frac{3\delta P_0}{P_0}\right), \nonumber\\
 \beta_{\lambda, s} =& \frac{\alpha_{\lambda, d}^2 + \alpha_{\lambda, o}^2}
 {4k \alpha_{\lambda, o}} \left(\frac{4\delta n_0}{n_0} - \frac{3\delta P_0}{P_0}\right)
 \nonumber\\
 &- \frac{\alpha_{\lambda, d}^2 + \alpha_{\lambda, o}^2 - 
 \alpha_{\lambda, d} \alpha_{\lambda, r}}{\alpha_{\lambda, r} \alpha_{\lambda, o}} 
 \beta_{\lambda, r},\label{eq:mom_init_sol}
\end{align}
while $\beta_{\lambda, c} = -\beta_{\lambda, r}$. The coefficients
$\beta_{\eta, *}$ are the same as in Eq.~\eqref{eq:h2_init_sole},
which were obtained in the frame of the second-order theory.

Since the integration constants $\beta_{\eta, *}$ obtained in the moment 
approach coincide with those obtained in the second-order theory, the analytic 
solutions for $\widetilde{\delta P}$ and $\widetilde{\Pi}$ are the same in these 
two approaches. Moreover, for the initial conditions corresponding to 
{\it Case 1} (i.e.~$\delta n_0 = \delta P_0 = 0$), when $\beta_{\lambda, *} = 0$,
the full analytic solution is identical in the moment approach and in the second 
order theory, being given in Eq.~\eqref{eq:h2_case1}. {\it Case 1} will therefore 
not be analysed in this section. Instead, {\it Case 2b} will be analysed 
in the following subsection.

\subsection{Numerical results ({\it Case 2b})}\label{sec:mom:num}

\begin{figure*}
\begin{center}
\begin{tabular}{cc}
\includegraphics[angle=270,width=0.49\linewidth]{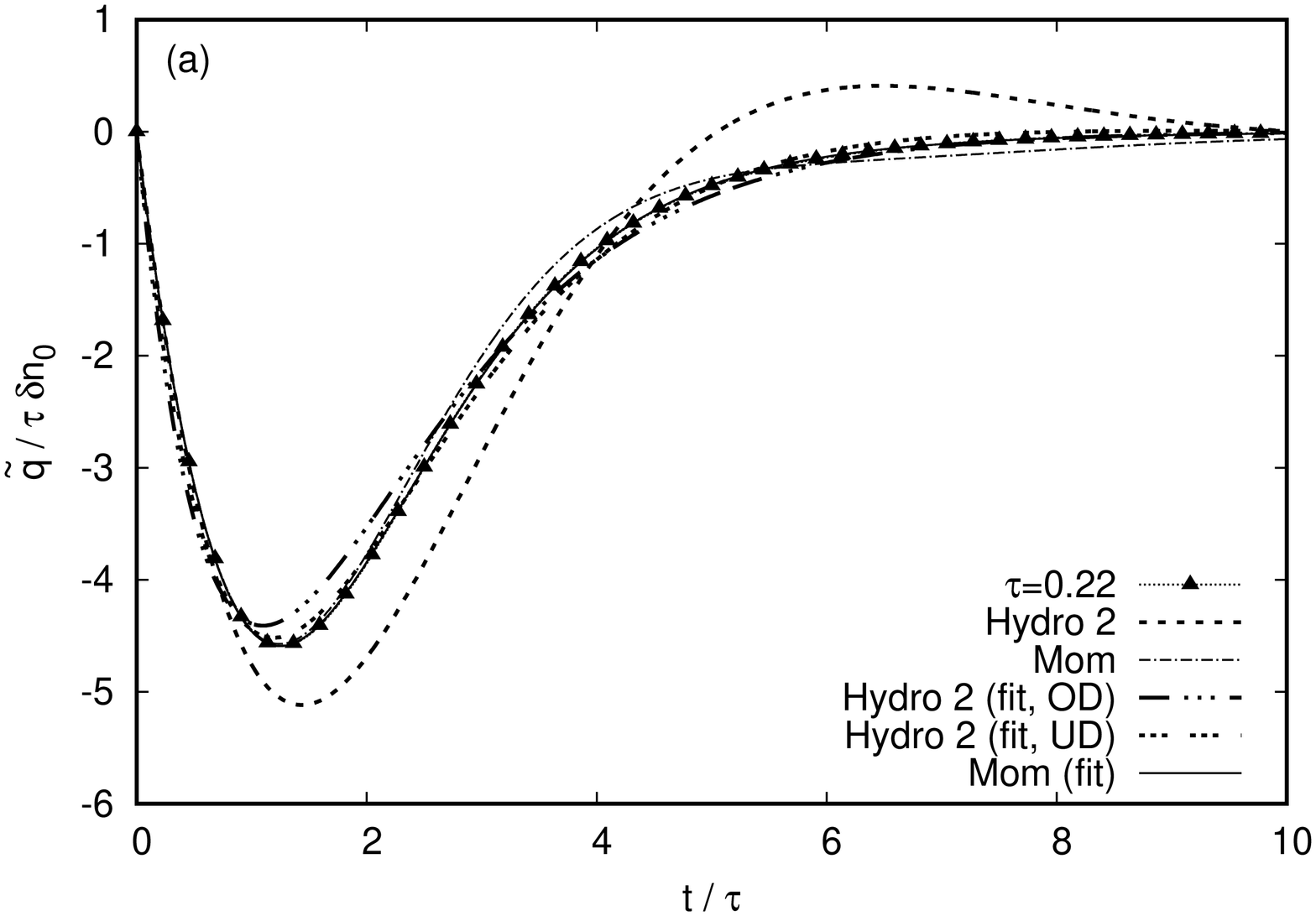} &
\includegraphics[angle=270,width=0.49\linewidth]{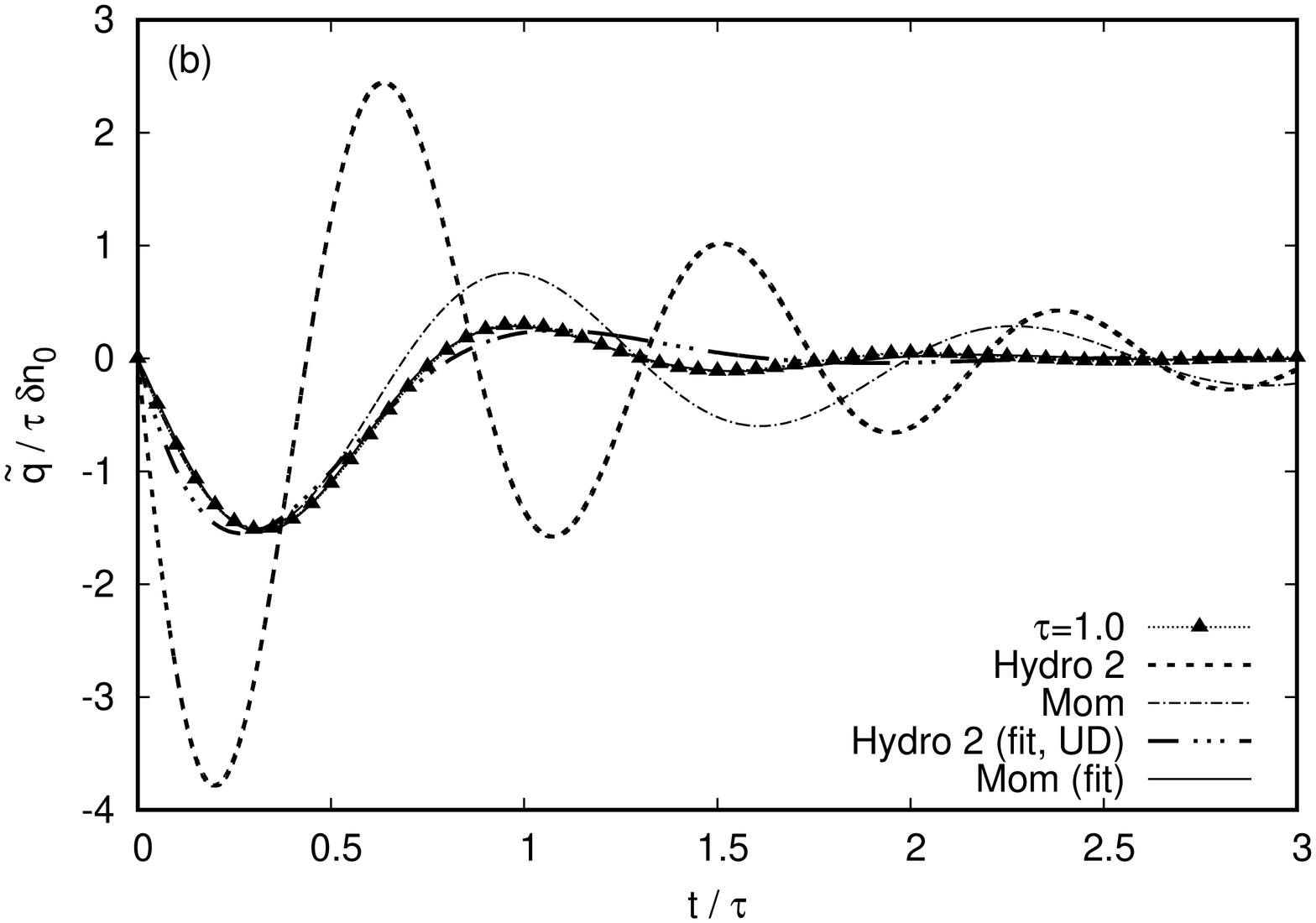} 
\end{tabular}
\end{center}
\caption{
Time evolution of $\widetilde{q} / \tau\delta n_0$ at 
(a) $\tau = 0.22$ and (b) $\tau = 1.0$ for $\delta n_0 = 10^{-3}$. 
Since for the Chapman-Enskog value of $\lambda_0$ \eqref{eq:tcoeff_CE} and 
$\tau_{q,0} = 1$ \eqref{eq:h2_tcoeff_CE}, in both cases $\tau > 
\tau_{\lambda, {\rm lim}} \simeq 0.14$ 
\eqref{eq:h2_taullim}, such that the curve corresponding to the second 
order hydrodynamics theory is given by the underdamped (UD) solution \eqref{eq:h2_case2UD}. 
The analytic solution corresponding to the moment method is given in Eq.~\eqref{eq:mom_case2}.
The fitted curves are obtained as explained in Subsec.~\ref{sec:mom:num}
}
\label{fig:mom_case2-profiles}
\end{figure*}

\begin{figure*}
\begin{center}
\begin{tabular}{cc}
\includegraphics[angle=270,width=0.49\linewidth]{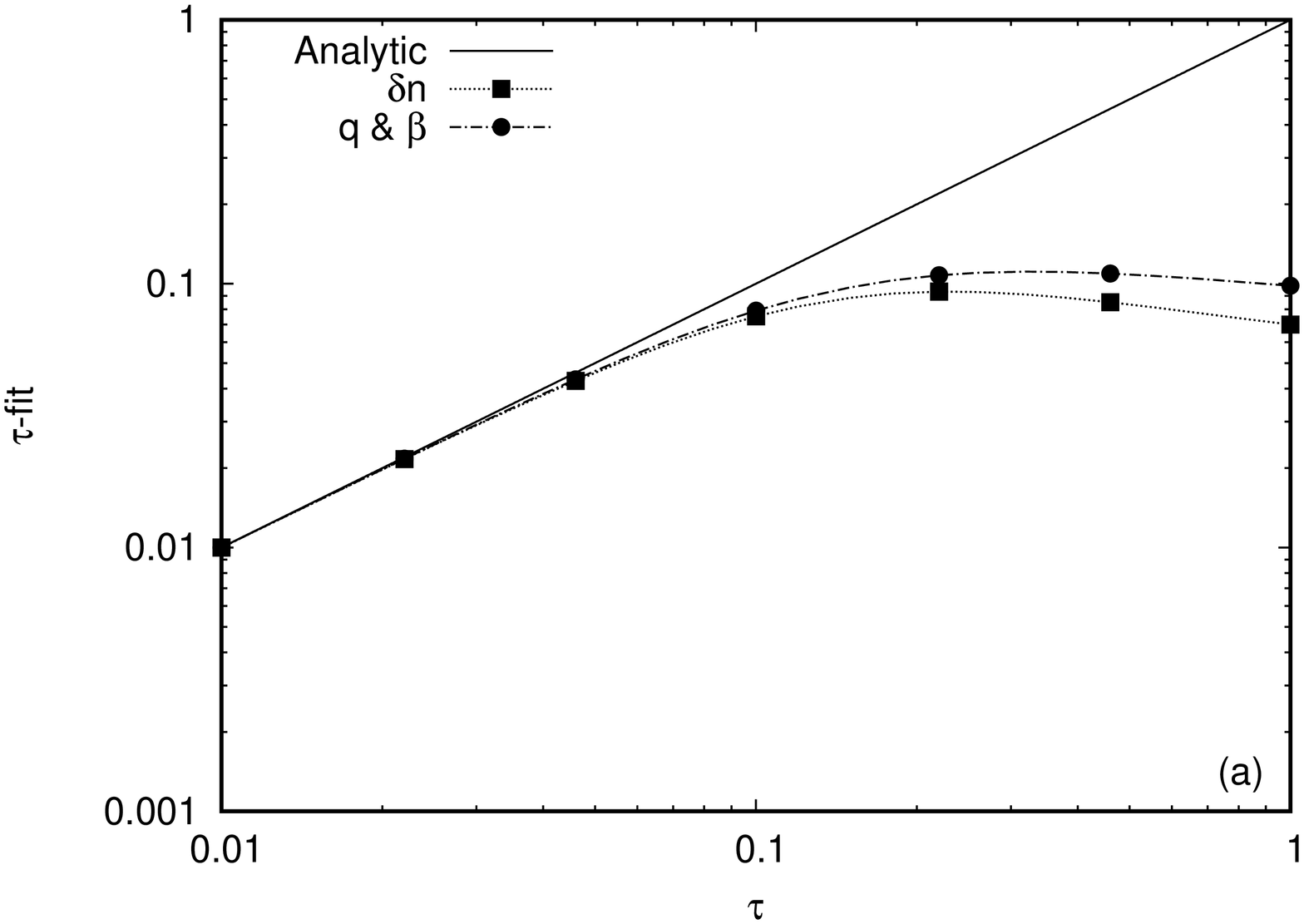} &
\includegraphics[angle=270,width=0.49\linewidth]{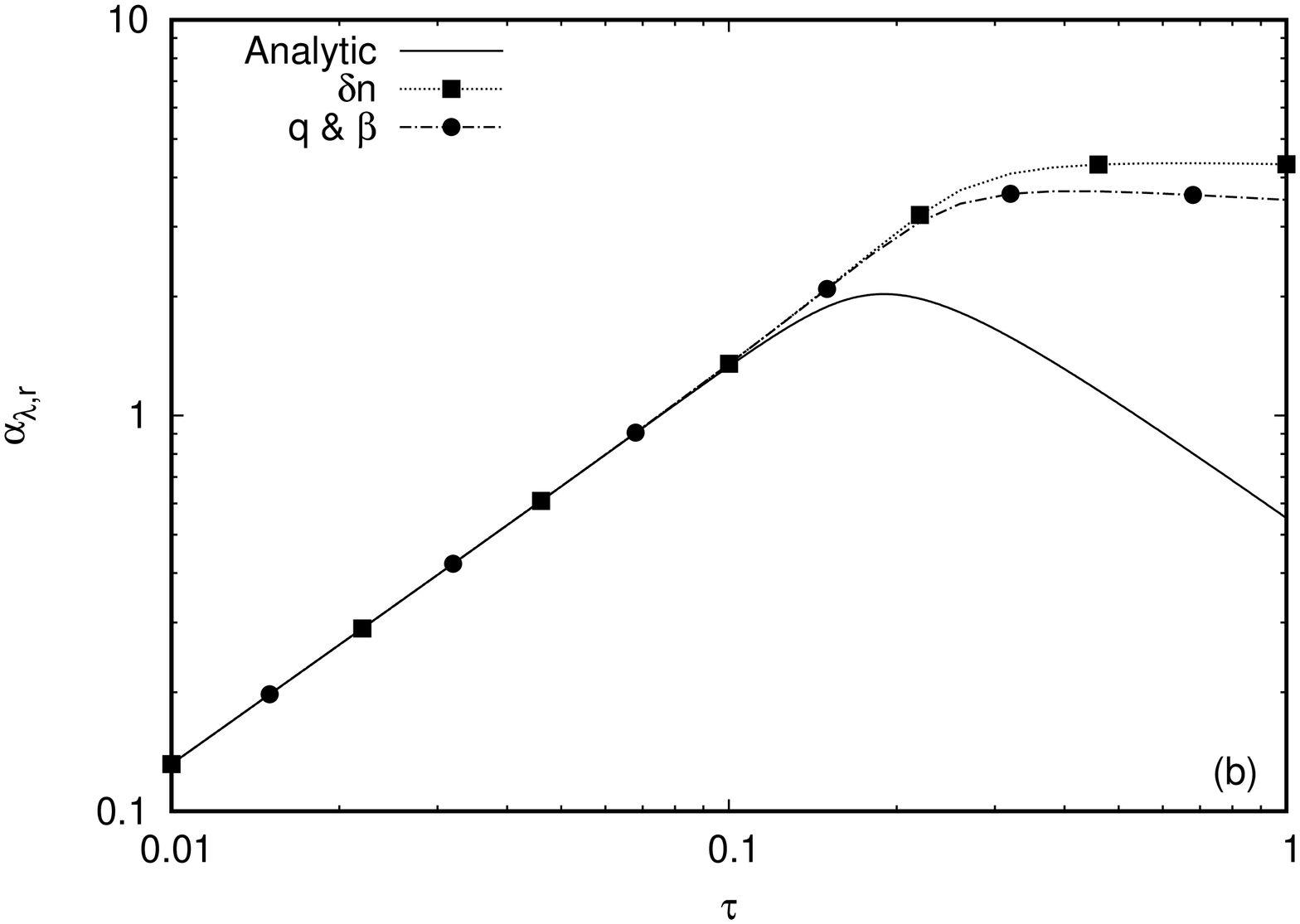} \\
\includegraphics[angle=270,width=0.49\linewidth]{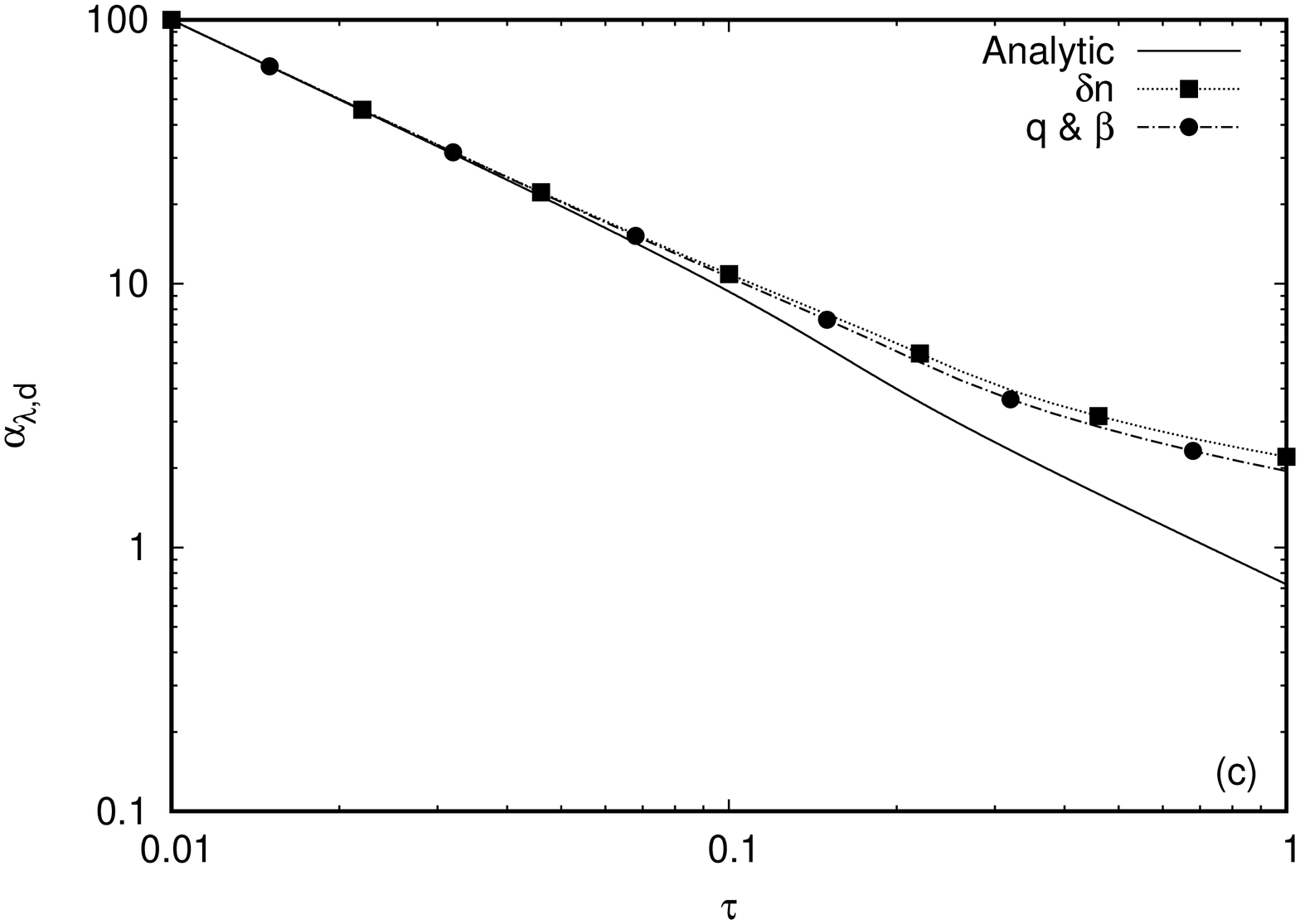} &
\includegraphics[angle=270,width=0.49\linewidth]{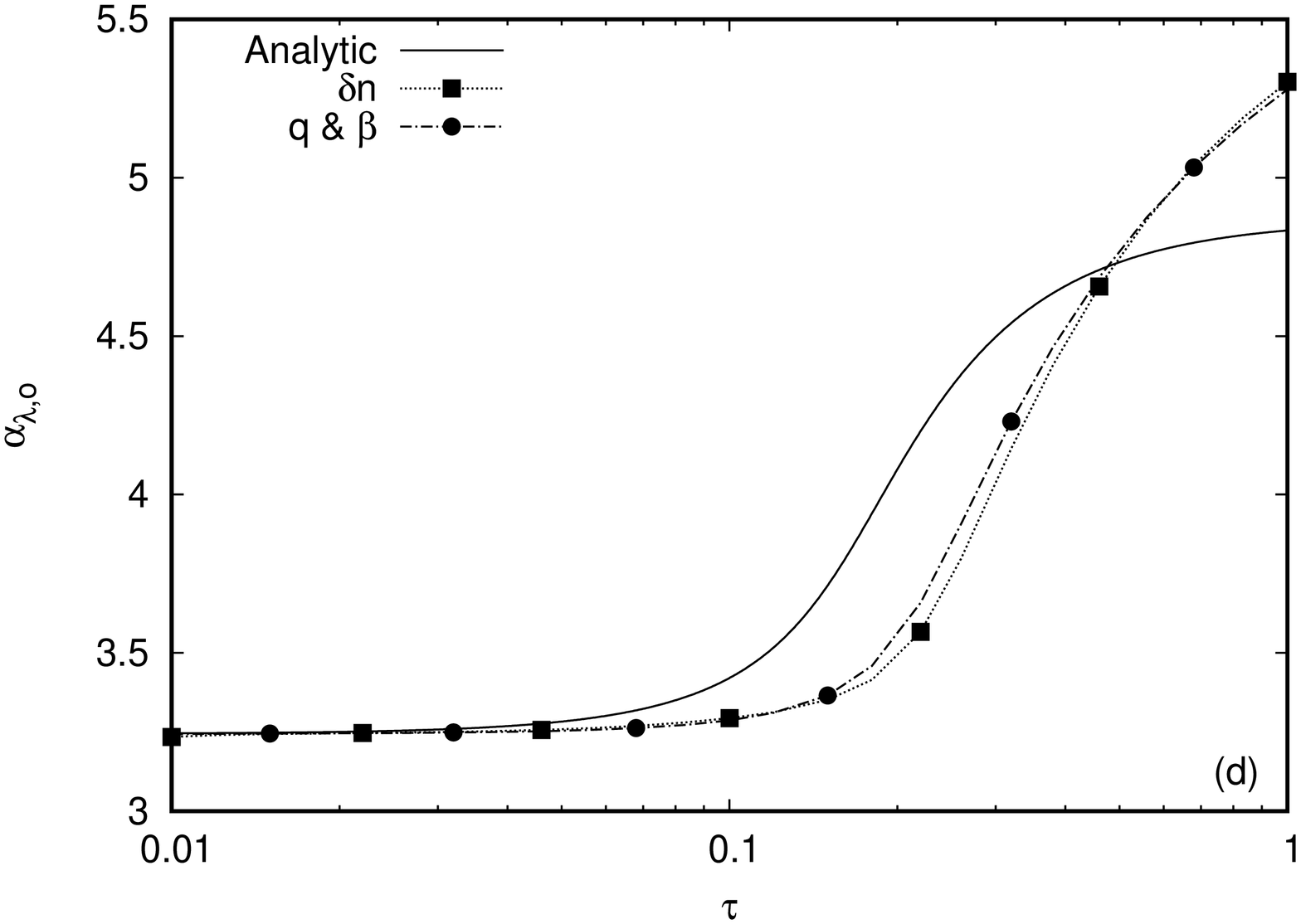} 
\end{tabular}
\end{center}
\caption{
Analysis with respect to $\tau$ of (a) the parameter $\tau$, 
(b) $\alpha_{\lambda, r}$, (c) $\alpha_{\lambda, d}$ and (d) $\alpha_{\lambda, o}$
for the initial conditions of {\it Case 2b} (i.e. $\beta_0 = \delta P_0 = 0$ and 
$\delta n_0 = 10^{-3}$).
The analytic curves are given by Eq.~\eqref{eq:mom_al} in (b)--(d), while 
in (a), the analytic curve represents $\tau$.
The numerical curves shown with dotted lines and symbols are obtained 
by performing a nonlinear fit, as described in Subsec.~\ref{sec:mom:num}.
}
\label{fig:mom_case2-alphal}
\end{figure*}

Setting $\beta_0 = \delta P_0 = 0$ in Eq.~\eqref{eq:mom_init_sol} yields:
\begin{widetext}
\begin{align}
 \beta_{\lambda, r} =& \frac{\alpha_{\lambda, r}(1 - \alpha_{\lambda, r} \tau) 
 (\alpha_{\lambda, d}^2 + \alpha_{\lambda, o}^2)}{k[\alpha_{\lambda,o}^2 + 
 (\alpha_{\lambda, d} - \alpha_{\lambda, r})^2]}
 \frac{\delta n_0}{n_0}, \nonumber\\
 \beta_{\lambda, s} =& \frac{\beta_{\lambda, r}}
 {\alpha_{\lambda, o}(1 - \alpha_{\lambda, r} \tau)} 
 [\alpha_{\lambda, r} - \alpha_{\lambda,d} + 
 (\alpha_{\lambda, d}^2 + \alpha_{\lambda, o}^2 - 
 \alpha_{\lambda, d} \alpha_{\lambda, r}) \tau],\label{eq:mom_case2b} 
\end{align}
while $\beta_{\lambda, c} = -\beta_{\lambda, r}$. Noting that $\widetilde{\beta}_{\eta} = 0$, 
the heat flux is given simply by $\widetilde{q} = -4P_0 \widetilde{\beta}$, while
\begin{align}
 \delta n_{\lambda, r} =& \delta n_0 \frac{(1-\alpha_{\lambda, r} \tau)(\alpha_{\lambda, d}^2 + \alpha_{\lambda, o}^2)}
 {\alpha_{\lambda, o}^2 + (\alpha_{\lambda, d} - \alpha_{\lambda, r})^2}, \nonumber\\
 \delta n_{\lambda, c} =& \delta n_0 \alpha_{\lambda, r} \frac{(\alpha_{\lambda, d}^2 + \alpha_{\lambda, o}^2)\tau +
 \alpha_{\lambda, r} - 2\alpha_{\lambda, d}}{\alpha_{\lambda, o}^2 + (\alpha_{\lambda, d} - \alpha_{\lambda, r})^2} 
 \delta n_0,\nonumber\\
 \delta n_{\lambda,s} =& \delta n_0 \frac{\alpha_{\lambda, r}}{\alpha_{\lambda, o}} \frac{\alpha_{\lambda, o}^2 + 
 (\alpha_{\lambda, d} - \alpha_{\lambda, r})[(\alpha_{\lambda, d}^2 + \alpha_{\lambda, o}^2) \tau - \alpha_{\lambda, d}]}
 {\alpha_{\lambda, o}^2 + (\alpha_{\lambda, d} - \alpha_{\lambda, r})^2}.
\end{align}
The full solution can be written as:
\begin{align}
 \widetilde{\beta} =& \frac{\alpha_{\lambda, r}(\alpha_{\lambda, d}^2 + \alpha_{\lambda, o}^2)}
 {\alpha_{\lambda,o}^2 + 
 (\alpha_{\lambda, d} - \alpha_{\lambda, r})^2}
 \frac{\delta n_0}{k n_0} \left\{(1 - \alpha_{\lambda, r} \tau) e^{-\alpha_{\lambda, r} t} -
 \left[(1 - \alpha_{\lambda, r} \tau) \cos \alpha_{\lambda, o} t 
 \right.\right.\nonumber\\
 &\left.\left.-
 \frac{1}{\alpha_{\lambda, o}}
 [(\alpha_{\lambda, d}^2 + \alpha_{\lambda, o}^2 - \alpha_{\lambda, d} \alpha_{\lambda, r}) \tau + 
 \alpha_{\lambda, r} - \alpha_{\lambda, d}]
 \sin \alpha_{\lambda, o} t\right] e^{-\alpha_{\lambda, d} t}\right\},\nonumber\\
 \widetilde{\delta n} =& \frac{\delta n_0}
 {\alpha_{\lambda,o}^2 + (\alpha_{\lambda, d} - \alpha_{\lambda, r})^2}
 \left\{(1 - \alpha_{\lambda, r} \tau)(\alpha_{\lambda, d}^2 + \alpha_{\lambda, o}^2)
 e^{-\alpha_{\lambda, r} t} 
 + \alpha_{\lambda, r} \Bigg(\left[(\alpha_{\lambda, d}^2 + \alpha_{\lambda, o}^2)\tau + \alpha_{\lambda, r} 
 - 2\alpha_{\lambda, d}\right] \cos \alpha_{\lambda, o} t \right.\nonumber\\
 &\left. +
 \frac{1}{\alpha_{\lambda, o}} \left\{\alpha_{\lambda, o}^2 + 
 (\alpha_{\lambda, d} - \alpha_{\lambda, r})\left[(\alpha_{\lambda, d}^2 + \alpha_{\lambda o}^2)\tau 
 -\alpha_{\lambda, d}\right]\right\} \sin \alpha_{\lambda, o} t\Bigg) e^{-\alpha_{\lambda, d} t}\right\},\label{eq:mom_case2}
\end{align}
\end{widetext}
while $\widetilde{q} = -4P_0 \widetilde{\beta}$ and $\widetilde{\delta P} = 
\widetilde{\Pi} = 0$.

The functional form of $\widetilde{q}$ obtained using the moment method is more convenient to use compared to the one 
given in the second-order hydrodynamics case \eqref{eq:h2_case2}, since in the former case, there is no 
distinction between the overdamped and the underdamped regimes. In Fig.~\ref{fig:mom_case2-profiles},
the validity of the solution \eqref{eq:mom_case2} for $\widetilde{q}$ is 
tested at $\tau = 0.22$ and $\tau = 1.0$.
Since, according to Eq.~\eqref{eq:h2_taullim}, $\tau > \tau_{\lambda, {\rm lim}} \simeq 0.14$ 
(the Chapman-Enskog value $\lambda_0 = 4 / 3$ and $\tau_{q, 0} = 1$ were used), 
Eq.~\eqref{eq:h2_case2UD} is used to represent the analytic solution obtained in the frame of the second-order 
hydrodynamics theory. At $\tau = 0.22$, the solution corresponding to the moments method is much closer 
to the numerical result than the second-order hydrodynamics one. When $\tau = 1.0$, both theories give
solutions which deviate considerably from the numerical results. In this regime, 
the validity of the functional form of the analytic solutions discussed above
can be further tested. In the second-order hydrodynamics case, 
a nonlinear fit of the solutions \eqref{eq:h2_case2OD} and \eqref{eq:h2_case2UD} 
is performed by considering the coefficients $\alpha_{\lambda, d}$ and 
$\alpha_{\lambda, o}$ ($\overline{\alpha}_{\lambda, d}$ and 
$\overline{\alpha}_{\lambda, o}$) as free parameters. 
Fig.~\ref{fig:mom_case2-profiles} shows that, 
at $\tau = 0.22$, the fit corresponding to the overdamped form 
\eqref{eq:h2_case2OD} is less accurate 
than the fit corresponding to the underdamped form \eqref{eq:h2_case2UD}. 
At $\tau = 1.0$, the fit corresponding to 
the UD form also starts to present visible deviations from the numerical result. 
In the moment method solution \eqref{eq:mom_case2}, the nonlinear 
fit is performed by considering $\tau$, $\alpha_{\lambda, r}$, 
$\alpha_{\lambda, d}$ and $\alpha_{\lambda, o}$ as free parameters. 
The resulting fit is in much better agreement with the numerical results.

Next, the $\tau$ dependence of the coefficients $\tau$, $\alpha_{\lambda, r}$, 
$\alpha_{\lambda, d}$ and $\alpha_{\lambda, o}$ as obtained 
by performing a nonlinear fit of Eq.~\eqref{eq:mom_case2} to the numerical data
is considered.
Figure~\ref{fig:mom_case2-alphal}(a) shows that $\tau$-fit [i.e., the best fit value for 
the parameter $\tau$ appearing in Eq.~\eqref{eq:mom_case2}] depends non-monotonically 
on $\tau$, i.e. it reaches a maximum value around $\tau \simeq 0.22$ and $\tau \simeq 0.32$
when considering the evolution of $\widetilde{\delta n}$ and $\widetilde{q}$, respectively, 
after which it decreases with $\tau$. 

The analytic expression for $\alpha_{\lambda, r}$ \eqref{eq:mom_al}, 
represented using a continuous line in Fig.~\ref{fig:mom_case2-alphal}(b), 
reduces at small values of $\tau$ to $\alpha_\lambda$ \eqref{eq:alphal}
defined within the first-order theory. While the first-order theory predicts 
a linear increase of $\alpha_\lambda$ with $\tau$, the moment method predicts 
a maximum of $\alpha_{\lambda, r}$ at $\tau \simeq 1.19/k \simeq 0.19$, 
after which it decreases according to the asymptotic behaviour 
$\lim_{\tau \rightarrow \infty} \alpha_{\lambda, r} = 5/9\tau$.
The fitted values also are non-monotonic, exhibiting a slight decreasing trend when 
$\tau \gtrsim 0.38$ in the case of $\widetilde{q}$ and $\widetilde{\beta}$ and 
$\tau \gtrsim 0.68$ when the nonlinear fit is performed on $\widetilde{\delta n}$.

The coefficient $\alpha_{\lambda, d}$ is analysed in Fig.~\ref{fig:mom_case2-alphal}(c).
The analytic expression \eqref{eq:mom_al} predicts a monotonic decrease of 
$\alpha_{\lambda, d}$. The curve corresponding to the nonlinear fit also 
decreases monotonically over the range of $\tau$ considered 
in Fig.~\ref{fig:mom_case2-alphal}(c), but at a lesser rate compared to the analytic prediction.

Finally, the oscillation frequency $\alpha_{\lambda, o}$ 
(which has no analogue in the first-order theory) 
is represented in Fig.~\ref{fig:mom_case2-alphal}(d). It can be seen that 
both the analytic and the 
numerical curves indicate that $\alpha_{\lambda, o}$ starts to 
increase when $\tau \gtrsim 0.1$.

\subsection{Summary}\label{sec:mom:summary}

In this section, a moment-based method was employed to study the attenuation of 
a longitudinal wave. The moment equations were obtained by projecting the 
distribution function $f$ on the space of the generalised Laguerre and Legendre polynomials.
The same expressions for the transport coefficients as 
those obtained through the Chapman-Enskog expansion are found. The difference between 
the approach taken in this section and the traditional Grad moment method introduced by
Israel and Stewart \cite{israel79} is that the truncation of $f$ is performed 
with respect to orthogonal polynomials, while in the latter approach, a 
nonorthogonal polynomial basis is employed \cite{denicol12}. In this sense, the 
present approach is similar to that employed in Ref.~\cite{denicol12}.

The minimal set of moment equations which gives access to the evolution of the 
macroscopic four-flow $N^\mu$ and stress-energy tensor $T^{\mu\nu}$ is obtained 
by retainig in the expansion of $f$ the zeroth and first order terms with respect to 
$p$ (expanded using generalised Laguerre polynomials) and zeroth, first and second 
order terms with respect to $\xi = p^z / p$ (expanded using Legendre 
polynomials).
We note that an expansion with respect to $p$ and $\xi$ is also performed in 
\cite{denicol16}. The system 
contains 6 equations for the five hydrodynamic variables $\delta n$, $\beta$, 
$\delta P$, $q$ and $\Pi$, as well as for a non-hydrodynamic variable not present 
in the second-order hydrodynamics theory. 

The solution of the set of moment equations is identical 
on the shear stress sector with the one obtained from the second 
order hydrodynamics equations discussed in Sec.~\ref{sec:hydro2}.
On the heat flux sector, the second-order hydrodynamics solution 
is improved in the moment method approach, where the functional 
form of $q$ allows for a smooth transition from the overdamped 
to the underdamped regimes highlighted in Sec.~\ref{sec:hydro2:case2}.
Furthermore, the range of validity of the analytical solution of 
the moment equations is larger than the one corresponding to the 
second-order hydrodynamics equations. Moreover, the functional form 
of the former can be fitted to the numerical 
data with remarkable accuracy even at $\tau = 1$.

\section{The ballistic limit}\label{sec:bal}

This section ends the analysis of the longitudinal wave problem by 
considering the free-streaming limit. In this case, the relativistic 
Boltzmann equation \eqref{eq:boltz} reduces to:
\begin{equation}
 \partial_t f + \xi \partial_z f = 0,\label{eq:boltz_bal}
\end{equation}
where $\xi = p^z / p$. The solution of Eq.~\eqref{eq:boltz_bal} is 
$f(z, \xi, t) = f(z - \xi t)$, 
subject to the following initial condition:
\begin{equation}
 f(z, \xi, t = 0) = \frac{n(z)}{8 \pi T^3(z)} \exp\left\{-\frac{p \gamma(z)}{T(z)}[1 - \xi\,\beta(z)]\right\}.
 \label{eq:fin}
\end{equation}

In the case of the longitudinal wave, the initial conditions for the 
macroscopic fields are:
\begin{gather}
 n(z, t = 0) = n_0 + \delta n_0 \cos kz,\nonumber\\
 P(z, t = 0) = P_0 + \delta P_0 \cos kz, \nonumber\\
 \beta(z, t = 0) = \beta_0 \sin kz.
\end{gather}
In the above, $k = 2\pi / L$ represents the wave number 
for a longitudinal wave having the wavelength equal to $L$.
Assuming that $\delta n_0$, $\delta P_0$ and $\beta_0$ are small, 
Eq.~\eqref{eq:fin} can be linearised as follows:
\begin{multline}
 f(z, \xi, t = 0) \simeq \frac{n_0}{8\pi T_0^3} e^{-p/T_0} 
 \left\{1 + \frac{p \xi}{T_0} \beta_0 \sin kz\right.\\
 \left.+\left[\frac{4 \delta n_0}{n_0} - \frac{3\delta P_0}{P_0} + \frac{p}{T_0} 
 \left(\frac{\delta P_0}{P_0} - \frac{\delta n_0}{n_0}\right)\right] \cos kz \right\}.
 \label{eq:bal_f_t0}
\end{multline}
The solution at $t > 0$ is given by replacing the product $kz$ in Eq.~\eqref{eq:bal_f_t0} 
by $k(z - \xi t)$.

\begin{figure*}[!ht]
\begin{center}
\begin{tabular}{cc}
\includegraphics[angle=270,width=0.49\linewidth]{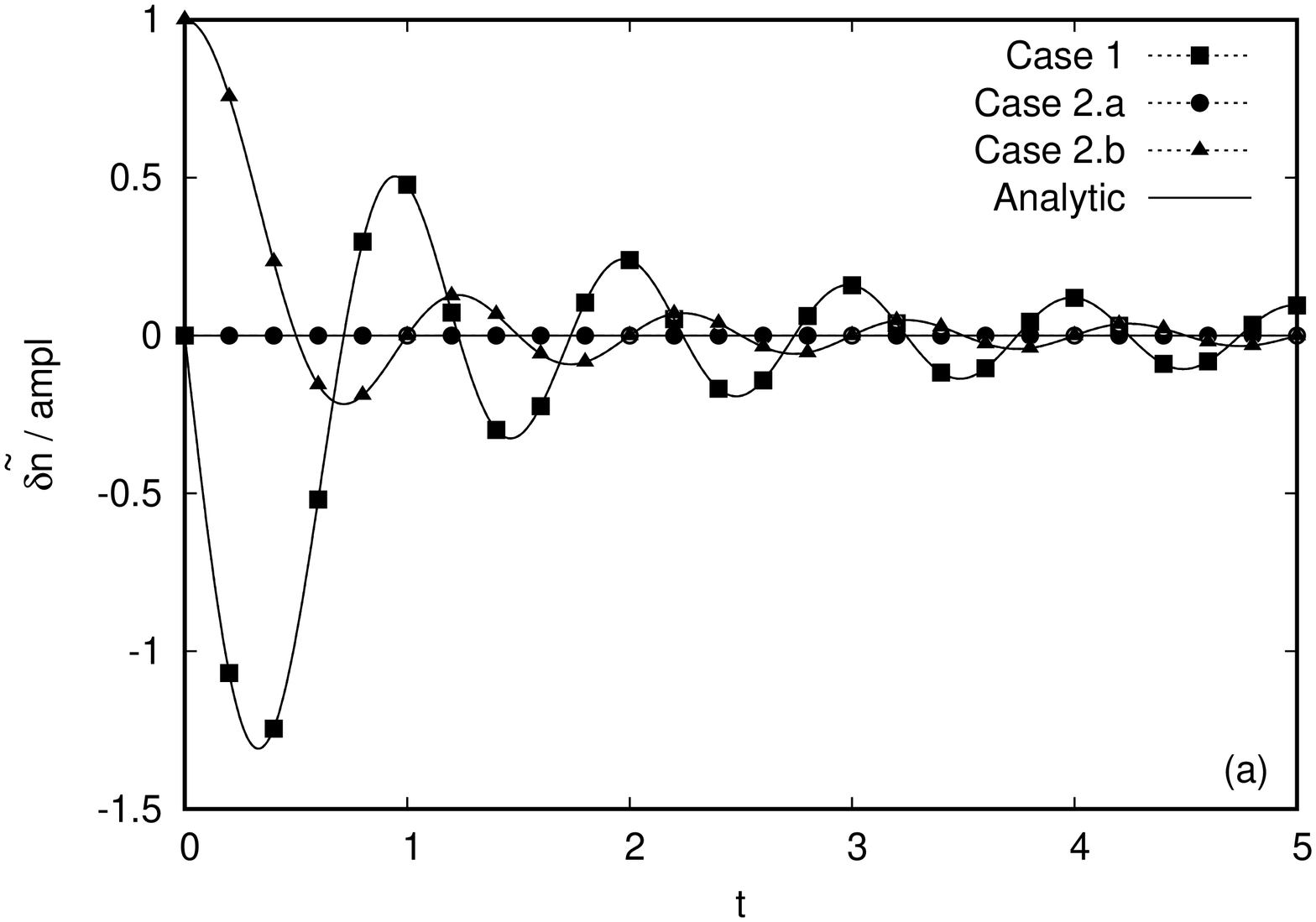} &
\includegraphics[angle=270,width=0.49\linewidth]{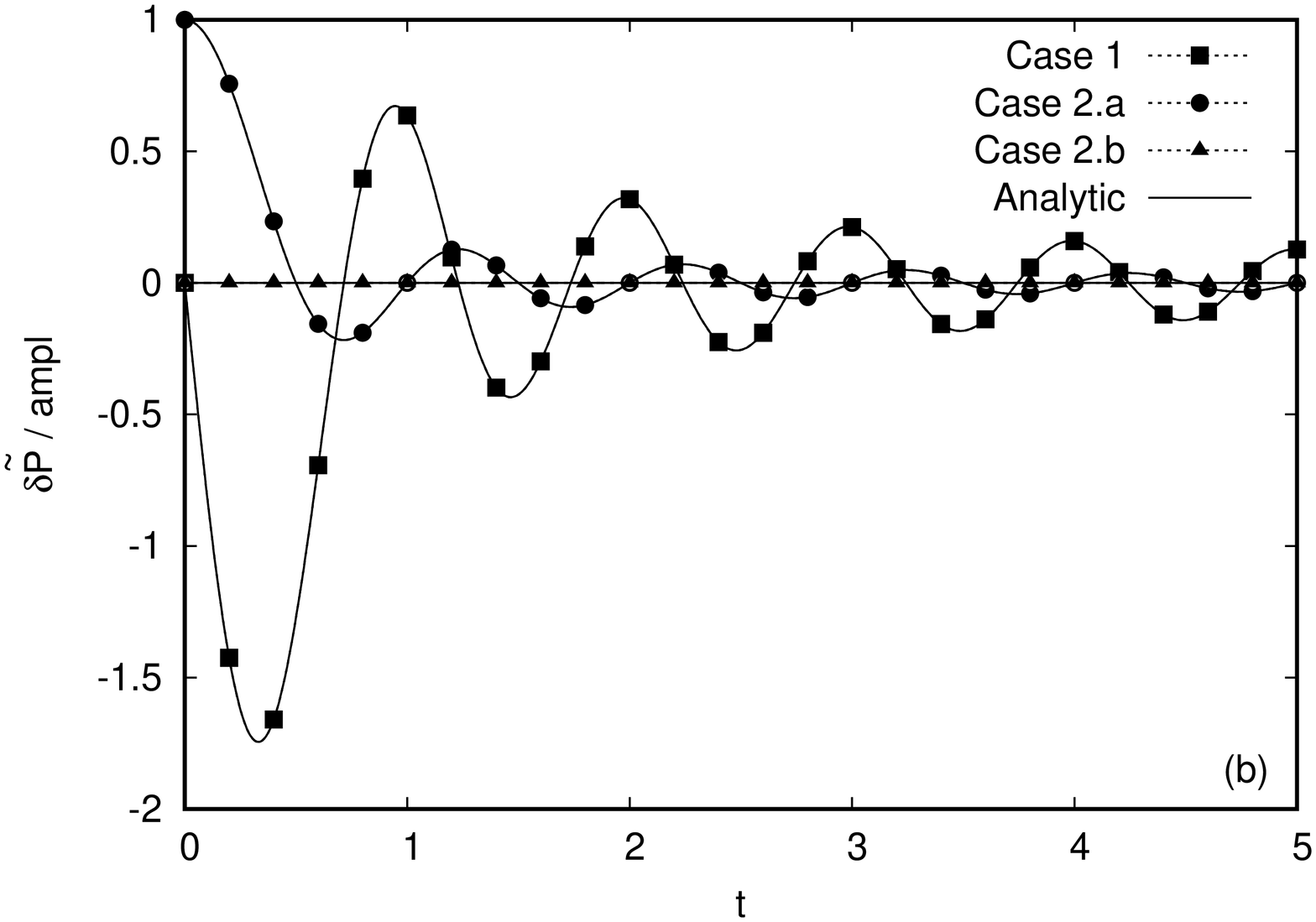} \\
\includegraphics[angle=270,width=0.49\linewidth]{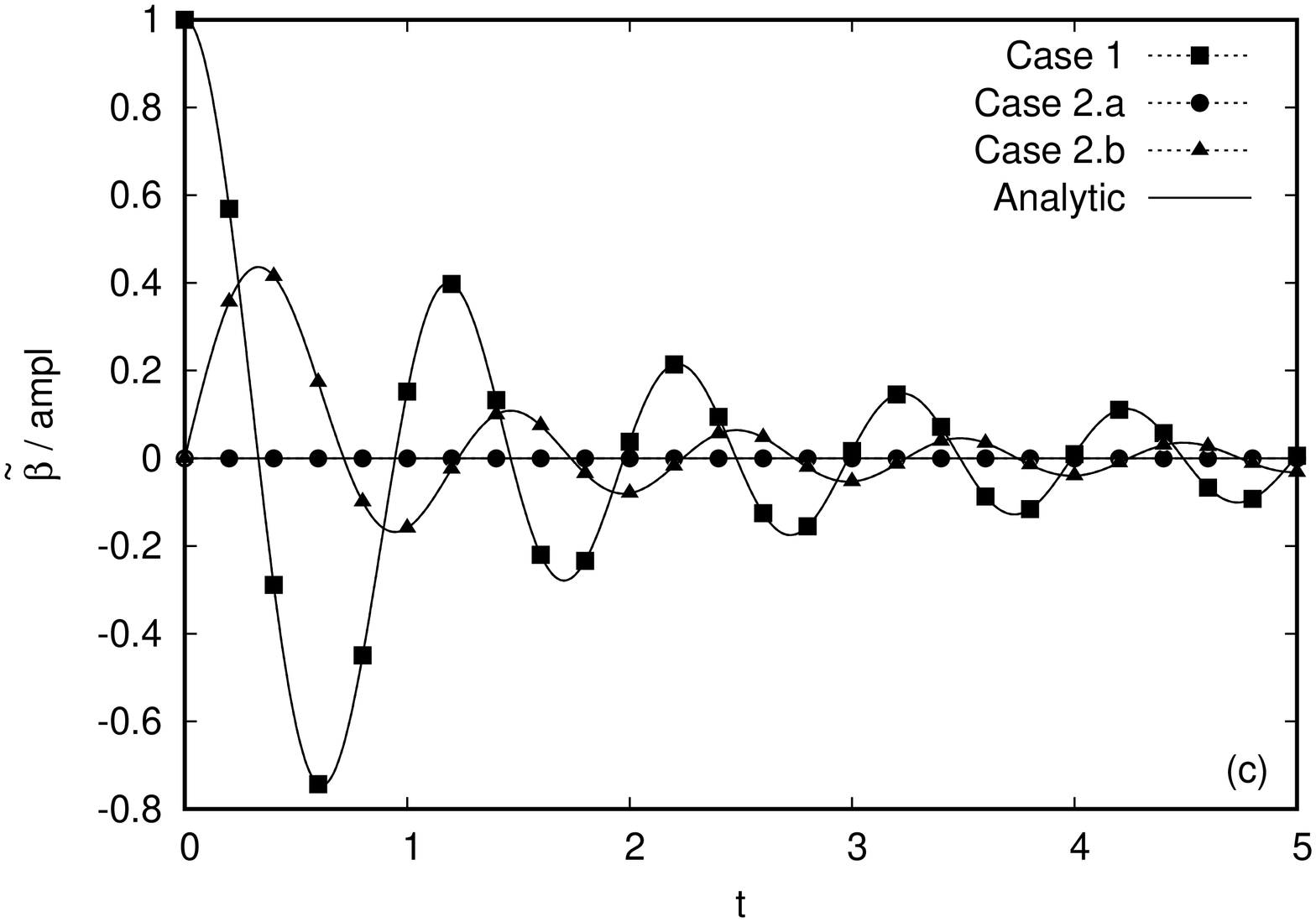} &
\includegraphics[angle=270,width=0.49\linewidth]{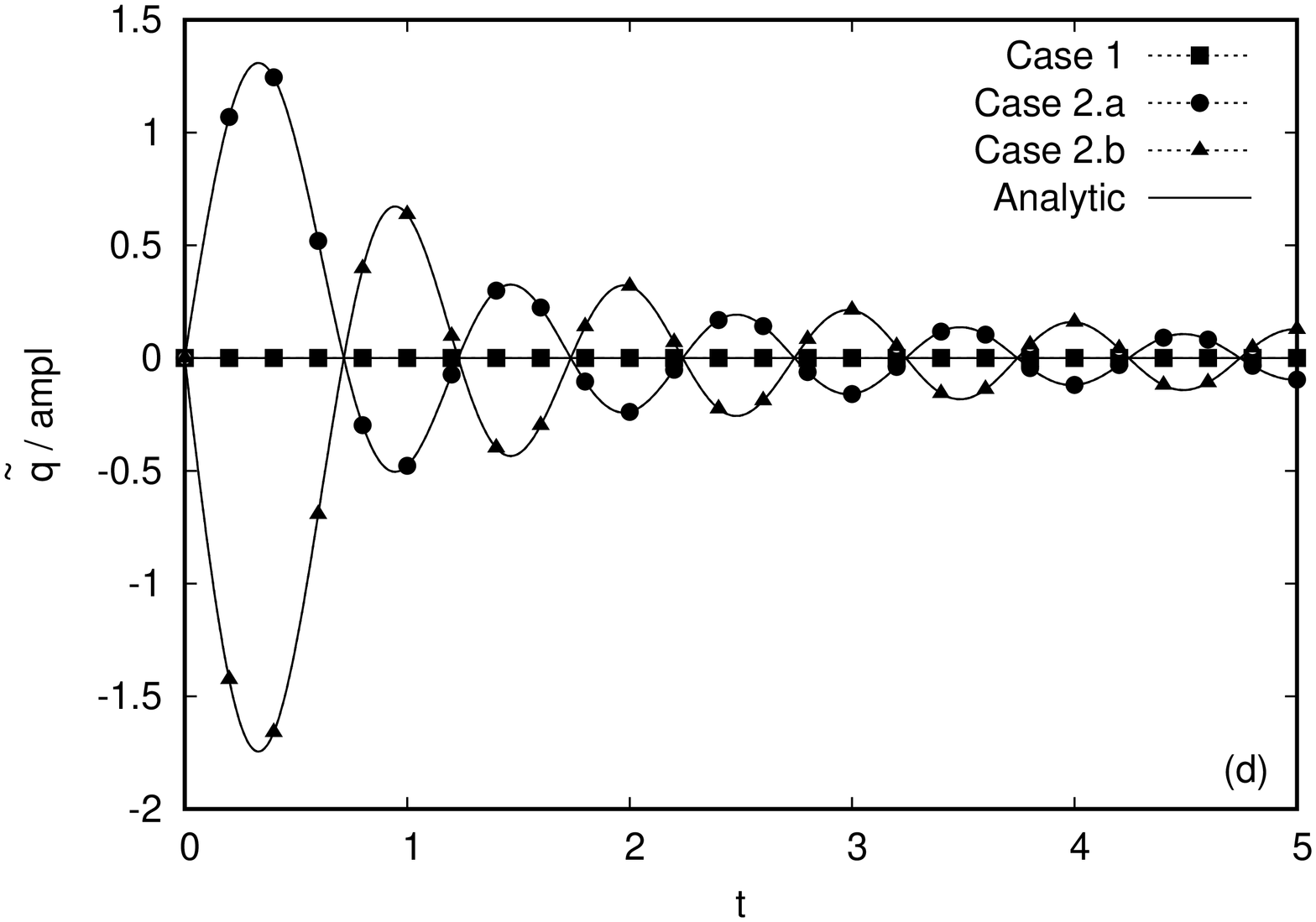}
\end{tabular}
\begin{tabular}{c}
\includegraphics[angle=270,width=0.49\linewidth]{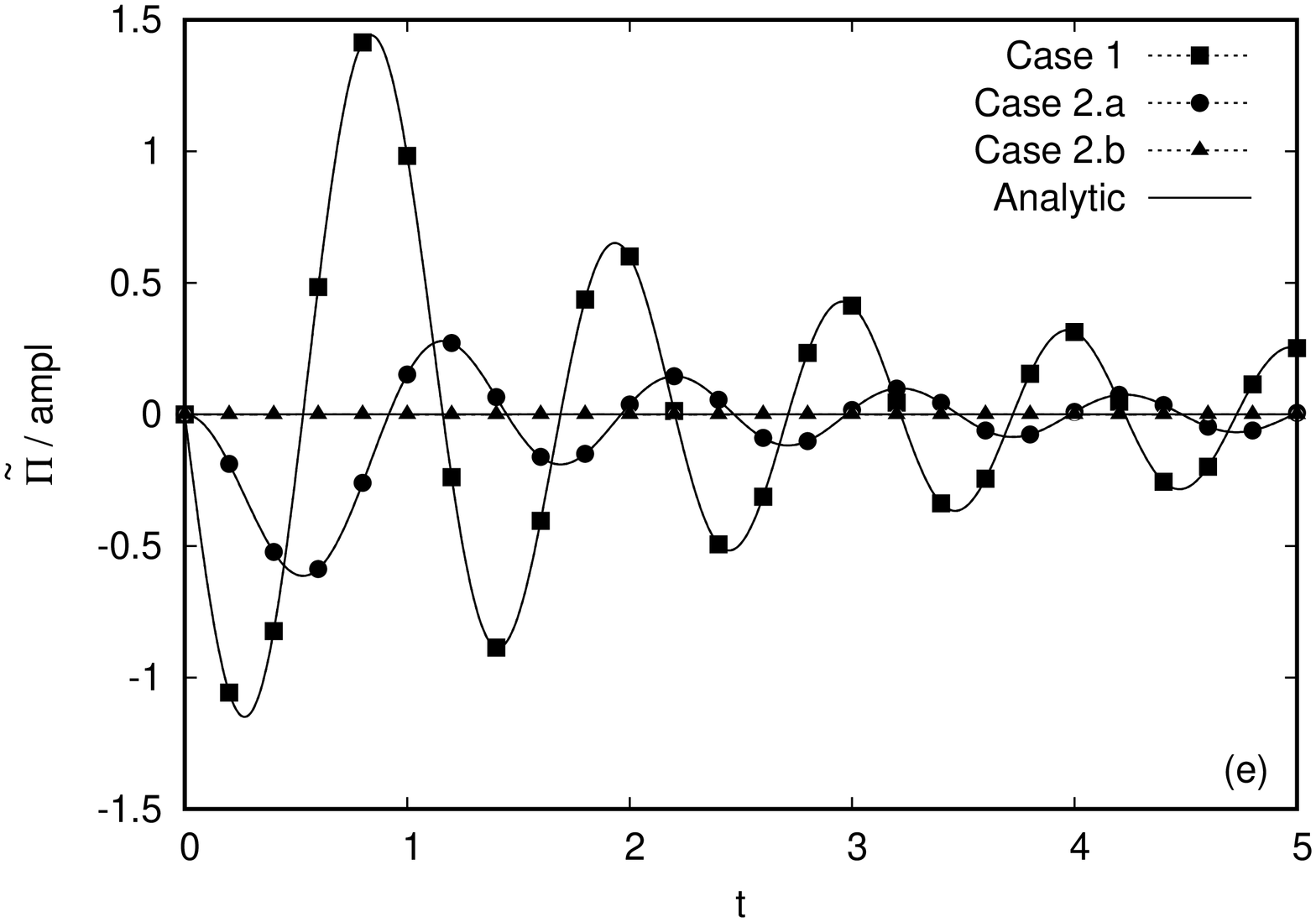}
\end{tabular}
\end{center}
\caption{
Time evolution of (a) $\widetilde{\delta n}$, (b) $\widetilde{\delta P}$,
(c) $\widetilde{\beta}$, (d) $\widetilde{q}$ and (e) $\widetilde{\Pi}$ 
in the free-streaming regime, divided by the wave amplitude.
The system is initialised with the following initial conditions:
{\em Case 1}: $\delta n_0 = \delta P_0 = 0$ and the wave amplitude is $\beta_0 = 10^{-3}$; 
{\em Case 2(a)}: $\delta n_0 = \beta_0 = 0$ and the wave amplitude is $\delta P_0 = 10^{-3}$;
{\em Case 2(b)}: $\delta P_0 = \beta_0 = 0$ and the wave amplitude is $\delta n_0 = 10^{-3}$,
such that the wave amplitude is always $10^{-3}$.
The numerical results are represented with dashed lines and points, while the analytic 
results corresponding to Eq.~\eqref{eq:bal} are represented using solid lines. The 
analytic and numerical curves are indistinguishable.
}
\label{fig:bal}
\end{figure*}

The time evolution of the macroscopic quantities
$n$, $P$, $\beta$, $q$ and $\Pi$ can be obtained from $N^\mu$ and $T^{\mu\nu}$
\eqref{eq:NT_def}, which reduce to:
\begin{align}
 N^\mu(t, z) =& \int_0^\infty dp\,p^2 \int d\Omega\, v^\mu\, f(z - \xi t), \nonumber\\
 T^{\mu\nu}(t, z) =& \int_0^\infty dp\,p^3 \int d\Omega\, v^\mu v^\nu\, f(z - \xi t),
\end{align}
where $v^\mu = p^\mu / p = (1, \sin \theta \cos\varphi, \sin\theta \sin\varphi, \cos\theta)$ and 
$\cos\theta = \xi$. The result is:
\begin{align}
 \widetilde{\delta n} =& \delta n_0 \frac{\sin kt}{kt} + 3\beta_0 n_0 
 \left[\frac{\cos kt}{kt} - \frac{\sin kt}{(kt)^2}\right], \nonumber\\
 \widetilde{\delta P} =& \delta P_0 \frac{\sin kt}{kt} + 
 4P_0 \beta_0 \left[\frac{\cos kt}{kt} - \frac{\sin kt}{(kt)^2} \right], \nonumber\\
 \widetilde{\beta} =& -\frac{\delta n_0}{n_0} \left[\frac{\cos kt}{kt} - \frac{\sin kt}{(kt)^2}\right] \nonumber\\
 &+ 3\beta_0 \left[\frac{\sin kt}{kt} + \frac{2\cos kt}{(kt)^2} - \frac{2\sin kt}{(kt)^3}\right],\nonumber\\
 \widetilde{q} =& P_0 \left(\frac{4\delta n_0}{n_0} - \frac{3\delta P_0}{P_0} \right) 
 \left[\frac{\cos kt}{kt} - \frac{\sin kt}{(kt)^2}\right],\nonumber\\
 \widetilde{\Pi} =& 2\delta P_0 \left[\frac{\sin kt}{kt} + \frac{3 \cos kt}{(kt)^2} - 
 \frac{3 \sin kt}{(kt)^3} \right] \nonumber\\
 &+ 8P_0 \beta_0 \left[ \frac{\cos kt}{kt} - \frac{4 \sin kt}{(kt)^2} - \frac{9 \cos kt}{(kt)^3} 
 + \frac{9 \sin kt}{(kt)^4} \right].
 \label{eq:bal}
\end{align}
The leading order term in all of the above expressions is damped 
according to a factor of $t^{-1}$.

Figure~\ref{fig:bal} illustrates the close agreement between the numerical results and the 
analytic solution \eqref{eq:bal} when $k = 2\pi / L$ and $L = 1$. 
Each plot in Fig.~\ref{fig:bal} contains three pairs of curves, each pair corresponding 
to the initial conditions described in {\it Cases 1} 
($\delta n_0 = \delta P_0 = 0$, $\beta_0 = 10^{-3}$), 
{\it 2a} ($\delta n_0 = \beta_0 = 0$, $\delta P_0 = 10^{-3}$) and 
{\it 2b} ($\delta P_0 = \beta_0 = 0$, $\delta n_0 = 10^{-3}$). 
The plots illustrate the time evolution 
of $\widetilde{\delta n}$, $\widetilde{\delta P}$, $\widetilde{\beta}$, $\widetilde{q}$ 
and $\widetilde{\Pi}$, where the numerical results are represented with dashed 
lines and points, while the 
analytic expressions \eqref{eq:bal} are represented using solid lines. 
The quantities on the vertical axis
are divided by the amplitude of the perturbation, namely $\beta_0$ for {\it Case 1}, 
$\delta P_0$ for {\it Case 2a} and $\delta n_0$ for {\it Case 2b}. For simplicity, the 
amplitude was taken equal to $10^{-3}$ in all cases. It can be seen that the 
agreement between the numerical results and the analytic expressions is excellent. 

Each plot in Fig.~\ref{fig:bal} displays two non-trivial curves and a line corresponding to 
a vanishing value. This is because all of the expressions in Eq.~\eqref{eq:bal} 
have on the right hand side only two terms, e.g. $\widetilde{\delta n}$ vanishes
when $\delta n_0 = \beta_0 = 0$ for all values 
of $\delta P_0$, etc. It is worth pointing out that $\widetilde{\delta P}$ and 
$\widetilde{\Pi}$ vanish when $\beta_0 = \delta P_0 / P_0 = 0$, as also predicted 
by the second-order hydrodynamics theory discussed in Sec.~\ref{sec:hydro2}, as 
well as by the moment method presented in Sec.~\ref{sec:mom}. 

A fundamental difference between the hydrodynamic and the free-streaming regimes is that the 
attenuation of the wave perturbation in the former case is exponential (dissipative), 
while in the latter case, it is of the form $t^{-1}$ (dispersive). 

\section{Conclusion and outlook}\label{sec:conc}

In this paper, the attenuation of a longitudinal wave in a medium formed 
of ultrarelativistic (massless) particles was studied from the following
perspectives: the first- and second-order hydrodynamics equations, 
the moment method, the free-streaming regime and by employing the numerical 
method introduced in Ref.~\cite{blaga17prc}. These investigations were carried out
by considering the linearised limit of the hydrodynamics equations, 
which can be solved analytically.
The analytic solutions were confronted with the numerical results in order to highlight 
the properties of the transport coefficients (and relaxation times in the 
second-order hydrodynamics and moment-method cases) in this system, for 
three particular cases: in {\it Case 1}, the initial density and pressure perturbations 
vanish ($\delta n_0 / n_0 = \delta P_0 / P_0 = 0$); in {\it Case 2a}, the initial density 
and velocity 
perturbations vanish ($\delta n_0 / n_0 = \beta_0 = 0$); finally, in {\it Case 2b},
the initial pressure and velocity perturbations vanish ($\delta P_0 / P_0 = \beta_0 = 0$).
Since in {\it Case 1}, the flow is adiabatic (i.e. the heat flux vanishes at all times) 
for all tested values of the initial wave perturbation $\beta_0$, this case was considered 
for the study of the shear viscosity $\eta$ by following the attenuation of the 
density, velocity, pressure and shear pressure perturbations. Cases 
{\it 2a} and {\it 2b} were considered in order to study the heat conductivity $\lambda$ 
by following the attenuation of the heat flux.

Throughout this paper, two types of tests were performed: (a) comparisons of the time 
evolution of the amplitudes $\widetilde{\delta n}$, $\widetilde{\delta P}$, 
$\widetilde{\beta}$, $\widetilde{q}$ and $\widetilde{\Pi}$ of the wave perturbations
obtained from the numerical simulations with the analytic predictions corresponding to 
the Chapman-Enskog and Grad expressions for the transport coefficients;
and (b), nonlinear numerical fits of these analytic expressions to the 
numerical results by considering the transport coefficients as fitting parameters.
The analytic solutions for the evolution of the amplitudes 
were considered in terms of the modes allowed by the hydrodynamic 
equations. 

In the first-order theory, three modes were highlighted: an evanescent mode 
(corresponding to the dampening coefficient $\alpha_\lambda$) and two modes undergoing 
oscillatory attenuation described by the dampening coefficient $\alpha_d$ and the 
oscillation angular frequency $\alpha_o$. Since there is no contribution to the heat flux 
from the oscillatory modes, the evanescent mode can be regarded as describing the 
heat flux sector (also, $\alpha_\lambda$ is determined exclusively in terms of 
the heat conductivity $\lambda$). Conversely, the oscillatory modes describe the 
shear pressure sector, since $\alpha_d$ and $\alpha_o$ depend only on the 
shear viscosity $\eta$. Thus, the numerical fits of $\widetilde{\delta n}$, 
$\widetilde{\beta}$, $\widetilde{\delta P}$ and $\widetilde{\Pi}$ were performed 
by considering $\alpha_d$ and $\alpha_o$ as free parameters, while $\alpha_\lambda$ 
was considered as a free parameter during the nonlinear fit of the expression for 
$\widetilde{q}$ to the numerical data. 

In the analysis based on the second-order hydrodynamics 
equations, the heat flux sector is described by two modes ($\alpha_{\lambda, +}$ and 
$\alpha_{\lambda, -}$) which are evanescent for $\tau$ smaller than some 
value $\tau_{\lambda,{\rm lim}}$, while 
for $\tau > \tau_{\lambda, {\rm lim}}$, their attenuation is oscillatory. The coefficients 
$\alpha_{\lambda, d}$ and $\alpha_{\lambda, o}$ now depend on $\lambda$ and also on 
the heat flux relaxation time $\tau_q$. The shear pressure 
sector is also enlarged by the addition of an evanescent mode corresponding to the
dampening coefficient $\alpha_{\eta, r}$, while the other two modes describe an oscillatory 
attenuation with dampening coefficient $\alpha_{\eta, d}$ and angular frequency 
$\alpha_{\eta, o}$. These three coefficients depend on $\eta$, as well as on the 
shear pressure relaxation time $\tau_\Pi$. The nonlinear fit of the heat flux was performed 
by considering $\alpha_{\lambda, d}$ and $\alpha_{\lambda, o}$ as free parameters with 
the expression of $\widetilde{q}$ written in both the evanescent (overdamped, OD) and in the 
oscillatory (underdamped, UD) forms. The nonlinear fits of the other amplitudes was performed 
by considering $\alpha_{\eta, *}$ ($* \in \{r, d, o\}$) as free parameters, as well as a 
fourth parameter ($\tau_\Pi$ for $\widetilde{\delta n}$, $\widetilde{\delta P}$ and 
$\widetilde{\beta}$ and the ratio $\eta / \tau_\Pi$ for $\widetilde{\Pi}$), 
which was considered as a free parameter due to the mathematical form of the analytic 
solution. 

In the case of the moment-based method, the shear pressure sector was found to be 
identical to that 
obtained within the second-order hydrodynamics approach. Due to the addition of a sixth 
(non-hydrodynamic) mode, the heat flux sector was enlarged by the addition of a 
purely evanescent mode damped by the coefficient $\alpha_{\lambda, r}$, while the 
other two modes are of oscillatory type, damped by the coefficient $\alpha_{\lambda, d}$ and 
having oscillation frequency $\alpha_{\lambda, o}$. In comparison to the 
second-order hydrodynamics result, this solution behaves as nearly evanescent at small values 
of $\tau$, since $\alpha_{\lambda, d} \sim \tau^{-1}$ quickly suppresses the oscillatory 
contributions. At larger values of $\tau$, the oscillatory modes contribute significantly 
to the time evolution of $\widetilde{q}$, which explains the better agreement to the numerical 
data observed at large values of $\tau$. The nonlinear fit of $\widetilde{q}$ 
in the case of the moment method was performed by considering $\alpha_{\lambda, *}$
($* \in \{r, d, o\}$) and $\tau$ as free parameters. The nonlinear fit of 
$\widetilde{\Pi}$ was performed by considering $\alpha_{\eta, *}$
($* \in \{r, d, o\}$) and $\tau$ as free parameters.

Both tests described above support the conclusion that, at small values of the 
Anderson-Witting relaxation time $\tau$ (typically, $\tau \lesssim 0.05$), the 
expressions for the first-order transport coefficients 
($\lambda$ and $\eta$) corresponding to the Anderson-Witting collision term are 
those predicted through 
the Chapman-Enskog procedure, which differ from the expressions obtained using 
Grad's 14 moment approach. 
This conclusion is supported by various evidence in the literature
\cite{florkowski13a,florkowski13b,florkowski15,ryblewski15,denicol14,bhalerao14,
blaga17prc,gabbana17b}. Also in the limit $\tau \lesssim 0.05$, the above 
tests confirmed that 
the relaxtion times for the shear pressure $\tau_\Pi$ and heat flux $\tau_q$ are 
$\simeq \tau$. These results were also obtained analytically when a moment-based method
was employed in order to construct the solution of the AWB equation.
As remarked in Ref.~\cite{denicol12}, this moment-based method is capable of reproducing 
the Chapman-Enskog transport coefficients since the distribution function is 
expanded with respect to orthogonal polynomials, while in the standard Grad 
method, the expansion is performed with respect to the nonorthogonal basis 
consisting of powers of the particle momentum $p^\mu$. 

By performing the test (a), it was highlighted that the analytic solution 
obtained using the first-order hydrodynamics theory loses applicability when 
$\tau \gtrsim 0.05$.
Since the constitutive equations for the heat flux and shear pressure tensor do not 
allow initial conditions to be specified for these fields, the first-order approximation is 
always inaccurate for a time scale $t \simeq 5\tau$. On this interval, the solution of 
the second-order 
hydrodynamics equations reproduce with good accuracy the numerical 
results for $\tau \lesssim 0.1$.
While the evolution of the amplitude of the shear pressure $\widetilde{\Pi}$ is the same 
within the frames of the second-order hydrodynamics and the moment-based method considered 
in this paper, numerical experiments show that the evolution of the heat flux amplitude 
$\widetilde{q}$ is better captured by the moment method, which offers a reasonable 
agreement with the numerical results up to $\tau \simeq 0.22$.

Furthermore, the viability of the functional form of the analytic solutions 
obtained using the various hydrodynamic theories described above was 
considered. To this end, the nonlinear fits corresponding to test (b) 
were performed and the results were analysed in three ways, as 
described below.

First, a comparison was considered between the numerical results for 
$\widetilde{q}$ and $\widetilde{\Pi}$ and their analytic expressions corresponding 
to the best-fit values of the free parameters. At $\tau = 0.26$, this analysis 
was used to highlight that the analytic expression for $\widetilde{\Pi}$ obtained 
using the second-order hydrodynamics theory 
(also from the moment-based method) was indistinguishable from the numerical results when 
the best-fit values of $\alpha_{\eta, *}$ and $\tau_\Pi$ were used, compared to the 
analytic prediction for these coefficients. The improvement of the analytic expression 
obtained using the first-order hydrodynamics equations was not significant, since 
this expression does not permit the value of $\widetilde{\Pi}$ to be fixed at $t = 0$.
In the case of the heat flux, the moment-based method provided a much more robust 
analytic expression for $\widetilde{q}$, which could be fitted remarkably well to the 
numerical results even at $\tau = 1$, while the solution obtained within the second-order 
hydrodynamics formulation corresponding to the best fit parameters was in visible disagreement 
compared to the numerical result, although the overall evolution was still in 
reasonable agreement with the numerical data.

Furthermore, the dependence on $\tau$ of the free parameters used in the 
nonlinear fitting procedure was considered. In all approaches, the dampening 
coefficient of the 
oscillatory modes on the shear pressure sector ($\alpha_\eta$ in the first-order theory, 
$\alpha_{\eta, d}$ in the second-order theory and in the moment-based approach) was predicted 
analytically to grow (almost) linearly with $\tau$. However, the numerical fits indicate that 
this coefficient increases at a much slower rate when $\tau \gtrsim 0.05$. A similar 
behaviour was highlighted for the coefficient governing the evanescent mode on the 
heat flux sector ($\alpha_\lambda$ in the first-order theory, $\alpha_{\lambda, -}$ 
in the second-order theory and $\alpha_{\lambda, r}$ in the moment-based method).
Thus, the above analysis indicates that the hydrodynamic theories considered in 
this paper break down when $\tau \gtrsim 0.05$.

A similar analysis of the dependence of the best fit parameters with respect 
to the wave amplitude at fixed $\tau = 0.0083$ was performed. 
Significant deviations from the analytic predictions were found when the amplitude 
($\beta_0$ for {\it Case 1}, $\delta P_0 / P_0$ for {\it Case 2a} and 
$\delta n_0 / n_0$ for {\it Case 2b}) exceeded $\sim 0.05$, 
indicating the inapplicability of the analysis in the linearised approximation 
at wave amplitudes larger than this value.

In order to gain some insight on the reason for the failure of the hydrodynamic
theories to describe the attenuation of the longitudinal wave at larger values 
of $\tau$, the ballistic (free molecular flow) limit of this problem was investigated.
In this case, the analytic solution for the distribution function $f$
indicates that the attenuation of the longitudinal wave is dispersive (i.e. the dampening 
is polynomial in $t^{-1}$) instead of dissipative (i.e. there is no exponential dampening).
This behaviour was exactly recovered numerically, confirming the applicability of 
the numerical method in this regime.
Thus, the solution of the hydrodynamics equations cannot describe correctly the attenuation 
of the longitudinal wave in the transition regime, where the dispersive component 
becomes important,
since the functional form of these solutions does not include terms which are polynomial in 
$t^{-1}$.

It is worth presenting the particular case when 
at initial time, the velocity and pressure perturbations of the wave vanish, i.e. 
$\beta_0 = \delta P_0 / P_0 =0$, while $\delta n_0 = 10^{-3}$. In this case,
the second-order hydrodynamics theory and the moment method predict that 
the pressure and shear pressure remain constant in time and the attenuation 
of the density, velocity and heat flux is purely evanescent (i.e. non-oscillatory).
This prediction is confirmed by the numerical simulations.
However, the first-order theory always predicts 
an oscillatory attenuation of all variables (including the pressure and 
shear pressure, but excluding the heat flux), 
where the amplitude of the oscillations is of the same order 
of magnitude as the evanescent component. The above behaviour persists at small 
values of $\tau$ (the tests were performed at $\tau =0.0083$), indicating 
a fundamental flaw of the first-order hydrodynamics equations.

It is worth noting that the generalisation of the conclusions 
presented in this paper to higher-order extensions of the Chapman-Enskog 
procedure or of the moment methods is not straightforward, since 
Ref.~\cite{cercignani02} warns that the higher orders in the Chapman-Enskog 
expansion can introduce spurious steady-state solutions, and in certain 
circumstances, the Chapman-Enskog series may exhibit a divergent 
behaviour \cite{denicol16}.

The present work can be naturally extended to the analysis of dissipative 
phenomena in fluids composed of massive particles \cite{gabbana17a,gabbana17b} 
or which obey quantum statistics \cite{coelho17}. Another extension can be 
made towards the analysis of dissipation in flows on curved spaces, such as 
the Bjorken flow in the Milne universe, where the solution of the Boltzmann 
equation is known semi-analytically \cite{florkowski13a,florkowski13b}, as 
well as the homogeneous and isotropically expanding flow on a 
background Friedmann-Lema\^itre-Robertson-Walker (FLRW) space, which
was studied analytically and numerically in Refs.~\cite{bazow16} and 
\cite{tindall17}, respectively.

\section*{Acknowledgements}
The author is grateful to Prof. Amaresh Jaiswal for useful discussions, as 
well as to an anonymous Phys. Rev. C referee for carefully reading 
the manuscript and for many useful comments.
This work was supported by a grant of the Romanian National Authority 
for Scientific Research and Innovation,
CNCS-UEFISCDI, project number PN-II-RU-TE-2014-4-2910. 

\appendix

\section{Non-dimensionalisation convention} \label{app:nondim}

\begin{table}
\begin{center}
\begin{tabular}{|l|l|}
\hline
 Reference quantity  & Conventional value \\\hline\hline
 $\hat{L}_{\rm ref}$ & $\hat{L} = 2\pi / \hat{k}$\\
 $\hat{v}_{\rm ref}$ & $\hat{c}$ (speed of light in vacuum) \\
 $\hat{n}_{\rm ref}$ & $\hat{n}_0$\\
 $\hat{T}_{\rm ref}$ & $\hat{P}_0 / \hat{K}_B \hat{n}_0$\\\hline
 $\hat{t}_{\rm ref}$ & $\hat{L}_{\rm ref} / \hat{c}_{\rm ref}$\\
 $\hat{p}_{\rm ref}$ & $\hat{K_B} \hat{T}_{\rm ref} / \hat{c}$ \\ 
 $\hat{f}_{\rm ref}$ & $\hat{n}_{\rm ref} / \hat{p}_{\rm ref}^3$\\\hline
\end{tabular}
\caption{Reference quantities employed for the non-dimensionalisation 
procedure employed in this paper.\label{tab:nondim}}
\end{center}
\end{table}

The dimensional form of the Boltzmann equation \eqref{eq:boltz} is:
\begin{gather}
 \hat{p}^\mu \hat{\partial}_\mu \hat{f} = 
 \frac{\hat{p} \cdot \hat{u}_L}{\hat{c}^2 \hat{\tau}} 
 (\hat{f} - \hat{f}^{(\mathrm{eq})}_L), \nonumber\\
 \hat{f}^{(\mathrm{eq})}_L = \frac{\hat{n}_L}{8\pi (\hat{K}_B \hat{T}_L / c)^3}
 \exp\left(\frac{\hat{p}\cdot \hat{u}_L}{\hat{K}_B \hat{T}_L}\right),
 \label{eq:boltz_dim}
\end{gather}
where the convention that dimensional quantities are written 
using a hat was employed.
The above equation can be non-dimensionalised using fundamental reference 
quantities, which are chosen as:
\begin{gather}
 \hat{L}_{\rm ref} = \hat{L},\qquad 
 \hat{v}_{\rm ref} = \hat{c}, \qquad 
 \hat{n}_{\rm ref} = \hat{n}_0, \nonumber\\
 \hat{T}_{\rm ref} = \hat{T}_0 = \hat{P}_0 / \hat{K}_B \hat{n}_0, 
\end{gather}
where $\hat{L}$ is the wavelength of the longitudinal wave, $\hat{c}$ is the 
speed of light in vacuum and $\hat{n}_0$ and $\hat{P}_0$ represent the average 
density and pressure of the medium. From the above fundamental reference quantities, 
the reference time $\hat{t}_{\rm ref}$, reference momentum $\hat{p}_{\rm ref}$ 
and reference particle distribution 
function $\hat{f}_{\rm ref}$ can be derived, as summarised in Tab.~\ref{tab:nondim}.
The non-dimensional form \eqref{eq:boltz} of the Boltzmann equation 
can be obtained by multiplying \eqref{eq:boltz_dim} by 
$\hat{t}_{\rm ref} / \hat{f}_{\rm ref}$, while the non-dimensional relaxation time 
$\tau$ is given by
\begin{equation}
 \tau = \frac{\hat{\tau}}{\hat{t}_{\rm ref}} = \frac{\hat{c} \hat{\tau}}{\hat{L}}.
\end{equation}

\section{Numerical method} \label{app:num}

In order to solve the AWB equation \eqref{eq:boltz}, we employ the relativistic 
spherical lattice Boltzmann (R-SLB) models introduced in Ref.~\cite{blaga17prc} 
as an extension of the non-relativistic spherical lattice Boltzmann (SLB) models 
introduced in Ref.~\cite{ambrus12}. 

A number of $N=100$ nodes are chosen along the $z$ axis, where periodic 
boundary conditions apply, while the flow is assumed to be homogeneous along the 
$x$ and $y$ directions. The advection and time evolution are performed using the 
fifth-order weighted essentially non-oscillatory (WENO-5) \cite{rezzolla13}
and third-order TVD Runge-Kutta (RK-3) \cite{trangenstein07} schemes, as presented 
in Ref.~\cite{blaga17prc}.
The lattice spacing is $\delta z = 10^{-2}$. The time step was set to 
$\delta t = 10^{-3}$ for the analysis performed within the frame of the 
first-order hydrodynamics theory in Sec.~\ref{sec:hydro1}. 
In the case of the second-order hydrodynamics theory and moment-based method 
considered in Secs.~\ref{sec:hydro2} and \ref{sec:mom}, a time 
step $\delta t = 10^{-4}$ was employed to allow an increased temporal resolution (i.e. more 
data points) for the study of the early time evolution of the longitudinal wave.

The momentum space is factorised using spherical coordinates $p$, $\theta$ 
and $\varphi$, which are discretised using $Q_L$, $Q_\xi$ and $Q_\varphi$ quadrature 
points, respectively. The quadrature order along the $p$ direction is set to 
$Q_L = 2$, while the azimuthal quadrature order is $Q_\varphi = 1$. The model
thus employs $Q_L \times Q_\xi \times Q_\varphi = 2Q_\xi$ velocities.
The value of $Q_\xi$ is chosen depending on the value of the relaxation time $\tau$. 
As discussed in Ref.~\cite{blaga17prc}, $Q_\xi = 6$ is sufficient to obtain accurate 
results at $\tau < 0.01$. For values of $\tau$ between $0.01$ and $0.1$,
$Q_\xi = 20$ was employed, while for $\tau \ge 0.1$, $Q_\xi$ was set to 
$Q_\xi = 200$.

The system is initialised with an equilibrium distribution $\feq$ \eqref{eq:feq}
at each point $z_\ell = -0.5 + (\ell - \frac{1}{2}) \delta z$, truncated 
to $N_L = 1$ and $N_\Omega = 5$ with respect to $p$ and $\xi = \cos\theta$, 
as explained in Ref.~\cite{blaga17prc}.

At a later time $t_s = s \, \delta t$ ($s = 1, 2, \dots T$), the quantities 
with tilde defined in Eqs.~\eqref{eq:ansatz} are obtained as:
\begin{align}
 \begin{pmatrix}
  \widetilde{\beta}_s\\
  \widetilde{q}_s
 \end{pmatrix} =& 2 \delta z \sum_{\ell = 1}^N
 \begin{pmatrix}
  \beta_{s,\ell}\\
  q_{s,\ell}
 \end{pmatrix}
 \sin k z_\ell, \nonumber\\
 \begin{pmatrix}
  \widetilde{\delta n}_s\\
  \widetilde{\delta P}_s\\
  \widetilde{\Pi}_s
 \end{pmatrix} =& 2 \delta z \sum_{\ell = 1}^N
 \begin{pmatrix}
  n_{s,\ell} - n_0\\
  P_{s,\ell} - P_0\\
  \Pi_{s,\ell}
 \end{pmatrix}
 \cos k z_\ell.
 \label{eq:tilde_num}
\end{align}

In the case of the analysis using the first-order theory performed in Sec.~\ref{sec:hydro1},
$T = 20,000$ and the resulting values $\widetilde{\beta}_s$, etc.~are stored at intervals of 
$10\delta t = 0.01$ resulting in a number of $2,000$ values which are further processed using 
Mathematica\textsuperscript{TM} to obtain a nonlinear fit, based on the 
analytic solution for the flow, as described in Secs.~\ref{sec:hydro1:case1} and 
\ref{sec:hydro1:case2}. Since in the first order theory, $q$ and $\Pi$ cannot be 
imposed at initial time, the analytic expressions for the evolutions of $\widetilde{q}$ 
and $\widetilde{\Pi}$ are not accurate at small values of $t$. Thus, 
the first $50$ points (i.e. up to $t = 0.5$) in the data sets were
always ignored when performing the nonlinear fits for these quantities.

In the case of the analysis using the second-order theory and the moment method, performed 
in Secs.~\ref{sec:hydro2} and \ref{sec:mom}, respectively, the values of the field amplitudes 
$\widetilde{\beta}_s$, etc.~were stored at intervals of $\delta t = 10^{-4}$. 
The value of $T$ was obtained using the following algorithm. For 
$0.001 \le \tau \le 0.1$, a number of time steps equal to 
$T = 100 (\tau / \delta t)$ was considered (i.e.~up to $t = 100 \tau$). This 
ensured a balanced coverage of the initial stage corresponding to the relaxation 
of the nonequilibrium parameters $\widetilde{q}$ and $\widetilde{\Pi}$ 
from their initial vanishing values towards the values predicted by the first 
order theory (up to $t \sim 5\tau$), as well as of the later stage of the 
wave evolution, where the attenuation effects dominate. 
For $0.1 < \tau < 0.32$, a number of $T = 1 / (\tau \delta t)$ points 
was chosen (i.e. corresponding to $t \simeq \tau^{-1}$), while for all 
$\tau \ge 0.32$, $T = 32,000$ time steps were performed (i.e. up to $t = 3.2$).

\end{document}